\documentclass[usenatbib]{mnras}
\RequirePackage{rotating}
\usepackage{dcolumn}
\usepackage{bm}
\usepackage{lscape}
\usepackage{graphicx}
\usepackage{amsmath}
\usepackage{hyperref}

\usepackage{siunitx}
\usepackage{braket}
\usepackage{color}
\usepackage{epsfig}
\usepackage{longtable}
\usepackage{makecell}
\usepackage{array}
\usepackage{subcaption}
\usepackage{amsmath,amssymb}
\usepackage[thinlines]{easytable}
\usepackage{multicol}
\usepackage{float}
\usepackage{xcolor}
\usepackage{soul}
\usepackage{lastpage}

 at 13 truept

\begin{document}
\twocolumn

\title[SKYSURF IX]{SKYSURF IX - The Cosmic Optical  Background from Integrated Galaxy Light Measurements}
\author[Tompkins et al.] 
{Scott A.\,Tompkins$^{2}$, 
Simon P.\,Driver$^{2}$,  
Aaron S.\,G.\,Robotham$^2$,
Rogier A.\,Windhorst$^1$,\newauthor
Delondrae Carter$^1$, 
Timothy Carleton$^1$,
Zak Goisman$^1$,
Daniel Henningsen$^1$,
Luke J.\,Davies$^{2}$,\newauthor
Sabine Bellstedt$^{2}$,
Jordan C. J. D'Silva$^{2}$,
Juno Li$^{2}$,
Seth H. Cohen$^1$,
Rolf A. Jansen$^1$,
Rosalia O'Brien$^1$,\newauthor
Anton M. Koekemoer$^3$,
Norman Grogin$^3$,
John MacKenty$^3$\\
\\
$^1$ School of Earth and Space Exploration, Arizona State University,
Tempe, AZ 85287-1404 \\
$^2$ International Centre for Radio Astronomy Research (ICRAR), University of Western Australia, Crawley, WA 6009, Australia \\
$^3$ Space Telescope Science Institute, 3700 San Martin Drive, Baltimore, MD 21218, USA \\
}

%\pubyear{2026} \volume{000}

%\input{psfig}

\maketitle
\label{firstpage}

\vspace{-2.0cm}

\begin{abstract}
As part of the SKYSURF Hubble Space Telescope (HST) Legacy Archival program we present galaxy number counts which yield measurements of the extragalactic background light (EBL) at 15 different wavelengths. We have processed 82,752 HST images across 23 filters into 16,686 mosaics using the same software and processing pipeline throughout. Using 17/23 filters that give reliable galaxy counts, we constrain the integrated galaxy light (IGL) with a 1.1-9\% error between 0.3 and 1.6 $\mu$m in combination with 8 bands from WAVES (Wide Area VISTA Extragalactic Survey) and DEVILS (Deep Extragalactic Visible Legacy Survey). While HST was never intended to undertake large area surveys, through extensive quality control and filtering, we were able to extract a reliable and representative sample of fields distributed across the sky. Our final catalogs cover a combined $\approx 19.6 \deg^2$, with individual filters covering areas ranging from $\approx 0.16-7.0 \deg^2$. The combination of numerous independent sight-lines and area coverage allows us to reduce cosmic variance uncertainties in deep number counts to 0.21\%-1.8\%. For the first time we are able to establish a measurement of the IGL, $\mathrm{8.90 \pm 0.29 nW m^{-2} sr^{-1}}$, at 0.59 $\mu$m using HST data. We obtain a cosmic optical background value of $\mathrm{25.15 \pm 0.49 nW m^{-2} sr^{-1}}$. Different techniques used to measure the COB, both directly and indirectly, have recently converged indicating that the COB arises almost exclusively from processes within galaxies. This in combination with the recent values reported from New Horizons and very high energy (VHE) constraints leaves very little room for any diffuse emission coming from outside the Milky Way.

\end{abstract}


\section{Introduction}
%$\mathrm{nW m^{-2} sr^{-1}}$
The extragalactic background light (EBL) is composed of all radiation originating from outside of our galaxy \citep{Hill_2018, Driver_2021}. The EBL is typically broken down into wavelength regimes such as the cosmic optical background (COB $\mathrm{0.1 -8 \mu m}$) , cosmic infrared background (CIB $\mathrm{8 -1000\mu m}$) and cosmic radio background (CRB); e.g. \citep{Hill_2018}. The cosmic microwave background (CMB) is distinct from other astrophysical backgrounds as it is a relic from when the early Universe cooled to a temperature where neutral atoms could form. The remaining EBL from $\gamma$-ray to radio wavelengths contains information about all physical processes that have produced photons since recombination and thus can be considered separate from the CMB. The COB and CIB dominate the EBL and provide roughly equal energy contributions \citep{Driver_2016, Koushan_2021}, though they both respectively fall a factor of $\approx$40x below the intensity of the CMB. The COB is perhaps the most important component of the EBL due to its intimate relationship with the cosmic star formation history (CSFH). The CSFH is defined as the volume averaged rate at which stars form as a function of either look-back time or redshift, for a thorough review see \citep{Madau_2014}. The COB can be directly predicted from the CSFH assuming an implementation of a stellar initial mass function (IMF), dust attenuation model, metallicity evolution history, and stellar population model which has been demonstrated in works such as \cite{Andrews_2018} and \cite{Koushan_2021}. Thus, empirical constraints of the COB can be used to test these models. The contribution from AGN is also important; especially at UV and mid-IR wavelengths, and can be modelled. The cosmic AGN luminosity history (CAGNH) is known to co-evolve with the CSFH \citep{D'Silva_2023}. Despite its importance, the COB remains one of the most difficult portions of the EBL to study in large part due to the presence of foregrounds which outshine the EBL, see \citep{Bernstein_2002, Windhorst_2022, Caddy_2022, Carleton_2022, Windhorst_2023}. 

Chief among these foregrounds is the Zodiacal light (ZL) that originates from sunlight scattered off of diffuse dust grains within the inner solar system. The ZL is the brightest of the optical and NIR foregrounds and constitutes as much as $95\%$ of the sky brightness at $\mathrm{ \lambda = 1.25 \mu m}$ \citep{Windhorst_2022}. It is also time variable, which requires observations of diffuse light over long periods of time from outside of the Earth's atmosphere as done in \cite{Kelsall_1998}. More recent studies of the ZL such as \cite{Korngut_2022} show that the theoretical model posed in \cite{Kelsall_1998} still provides an accurate description of the ZL when allowing for a free offset parameter. Part of the SKYSURF project was designed to improve the models of the ZL and to attempt to place upper limits on the EBL at NIR wavelengths using HST, see, in particular, \cite{O'Brien_2023}. However, from low Earth orbit (LEO), stray-light from the Sun and reflections from the Earth and moon pose significant challenges to surface brightness studies using HST. In \cite{Caddy_2022}, the authors show that at optical wavelengths scattered light can be even more intense than the ZL when HST is pointing close to the Earth's limb, as it approaches the daylight portion of its orbit. The final diffuse foreground is known as the diffuse galactic light (DGL). The DGL is produced from starlight scattered by the interstellar medium of the Milky Way and cannot be avoided when attempting a direct measurement of the EBL. Subtracting these foreground sources of diffuse light is difficult but necessary to obtain a direct measurement of the EBL \citep{Bernstein_2002, Bernstein_2007}. 

Efforts to disentangle the DGL from other foregrounds include the dark cloud method employed by works such as \cite{Mattila_2017}, in which an optically thick molecular cloud is used to establish lines of sight which end at the cloud or continue into extragalactic space. However, results may be impacted by scattered starlight from the cloud. Recent attempts to measure the EBL directly from the outer Solar System have also been performed, e.g., \cite{Lauer_2022} and \cite{Lauer_2024} in which the New Horizons mission was able to take optical images data using the LORRI camera at distances greater than 50 AU from the Sun, which should place the spacecraft well beyond the interference produced by the Zodiacal light, Earth, and Moon. Efforts were made to avoid bright stars and correct for both DGL and scattered starlight off of the spacecraft.

Indirect detection techniques have become possible with advancements in the ability to study high energy $\gamma$-ray sources. The interaction between photons from distant $\gamma$-ray sources, typically blazars, and optical EBL photons results in an attenuation of TeV energy $\gamma$-rays through particle/anti-particle pair production interactions as discussed in \cite{Desai_2017}. Using blazar spectra with established redshifts, they are able to estimate the attenuation of $\gamma$-ray spectra by EBL photons along the line of sight between the source and detectors. While difficult, this method (hereafter VHE) is less susceptible to foregrounds and is also unaffected by issues such as cosmic variance and zero point uncertainties, which plague the IGL. However VHE techniques typically apply a universal spectral shape to the $\gamma$-ray sources and infer deviations from this spectrum as entirely due to interactions with EBL photons or intergalactic magnetic fields. The technique typically requires an existing COB model and just constraints the normalization of the model. Recent developments in VHE studies of the EBL, \citep[e.g.][]{Gréaux_2023, Gréaux_2024} have adopted a Bayesian framework to avoid limitations introduced by assuming a model and spectral shape for both the EBL and gamma ray source. Using the observed $\gamma$-ray spectra of blazars they present preliminary results in which they now extract the COB shape and amplitude. Uncertainties in measurements of the COB established by VHE studies to 15\% and will be developed further in the coming years. 

A commonly employed method to place constraints on the EBL without accounting for foregrounds is to use galaxy number counts and measure the total intensity of all resolved galaxies in works such as \citep[e.g.][Carter et al. 2025(submitted)]{Madau_2000, Totani_2001, Xu_2005, Dole_2006, Hopwood_2010, Keenan_2010, Berta_2011, Bethermin_2012, Driver_2016, Koushan_2021, Windhorst_2022, Windhorst_2023}. The IGL can be well constrained by surveying a large enough volume to obtain a representative sample of the Universe. Galaxy number counts, or rather the galaxy flux density, can be integrated to provide a measurement of their combined intensity. The IGL is insensitive to diffuse foregrounds as positive sky values across the image are subtracted during processing and/or source detection. In particular, for nearly 30 years, standard HST data processing has removed foreground sky levels by subtracting a 2D model fit to the image after initially detected discrete objects are masked out \citep{Windhorst_2022}. The sky subtraction includes signals from the zodiacal light, diffuse galactic light, and scattered sun and/or moonlight. Similar processing is done for ground-based observations which must also deal with atmospheric effects such as airglow. 

IGL based measurements of the COB are not sensitive to proposed diffuse sources of the EBL such as intra-cluster light (ICL), diffuse light from galaxy halos, or ultra-diffuse galaxies (UDGs) in galaxy clusters. Extended low surface brightness (SB) features may be lost during sky subtraction or remain undetected by automated source finding tools. In clusters, ICL may contribute as much as $\approx10-15\%$ of cluster luminosity at SB levels fainter than $\mathrm{26\ mag/arcsec^2}$ \citep{Montes_2019}. Similar fractions of $\approx 15\%$ from the extended halos of individual galaxies have been proposed in works such as \citep{Ashcraft_2018, Borlaff_2019, Cheng_2021}. While missed by IGL studies, though as direct, indirect, and source-count based measurements improve tighter constraints will be imposed on these diffuse contributions to the COB.

While establishing robust lower limits, the IGL is prone to uncertainties introduced by cosmic variance and magnitude zero-point differences present between the facilities providing the number counts. Cosmic variance and zero point errors remain the dominant sources of uncertainty in measuring the IGL. In works such as \cite{Driver_2016} these contribute uncertainties of 10\% and 5\%, respectively. More recent studies such as \cite{Koushan_2021}, through the use of larger area surveys and updated zero point calibrations, were able to bring these uncertainties down to 4\%. However, \cite{Driver_2016} and \cite{Koushan_2021} relied only on a small number of deep HST fields. Work such as \cite{Windhorst_2011} and \cite{Rafelski_2015} provide the faintest optical galaxy number counts and have only been surpassed by JWST recently in the NIR \citep{Windhorst_2023}. The two HST surveys included in \cite{Driver_2016} and \cite{Koushan_2021} cover a combined 40-60 arcmin$^2$, and thus cosmic variance remains the dominant source of uncertainty in these surveys at 10-15\% as estimated using the formula developed in \cite{Driver_2010}. Uncertainties in the IGL have also been reduced to between by Carter et al. 2025 (submitted). SKYSURF aims to use a significant fraction of the area covered by HST over 20 years of observations to further reduce uncertainties in measurements of the IGL primarily by reducing the cosmic variance errors of individual HST deep surveys used in previous works.

\begin{figure*}
\vspace*{0.000cm}
\includegraphics[width=18cm, height = 11.33cm]{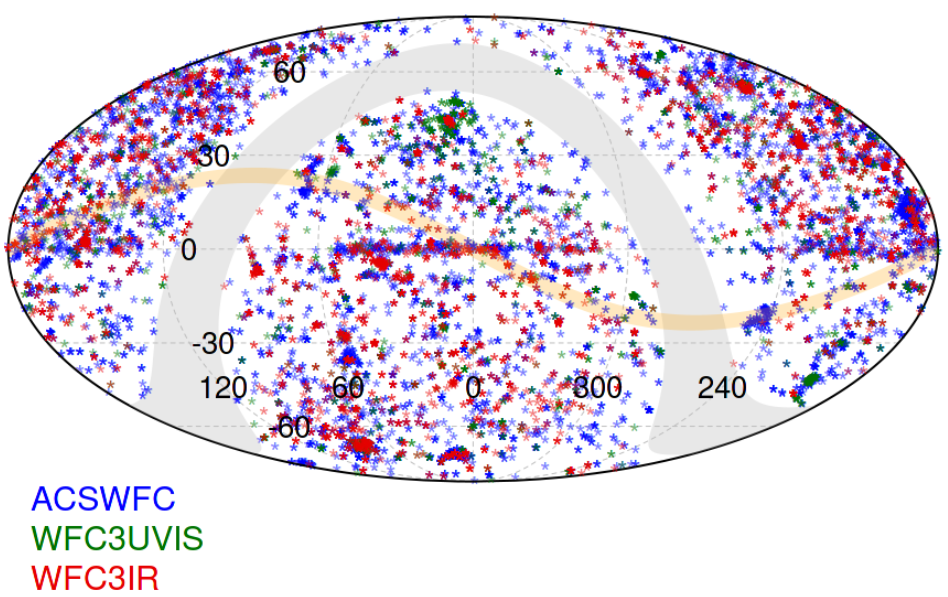}
\caption{ An equatorial projection showing the location of the 15,339 mosaics produced during this work. This projection displays all SKYSURF fields before any filtering or field rejection steps are taken. The galactic plane, $\mathrm{\pm 10}$ degrees, is shown as as gray overlay, with the zodiacal plane, $\mathrm{\pm 5}$ degrees, shown in orange. 
\label{fig:fig_projection}}
\end{figure*}

Despite efforts to measure the COB both directly and through the IGL, disagreements between the upper limits established by work of \cite{Carleton_2022} \cite{O'Brien_2023}, \cite{Mattila_2017}, \cite{Lauer_2022} and lower limits established by the IGL in \cite{Driver_2016} and \cite{Koushan_2021} remained factors of $\mathrm{2-4\times}$ apart. The newest direct measurements of \cite{Lauer_2024} from New Horizons suggest that the tension between the IGL and direct measurements has finally been resolved. Again using the LORRI instrument aboard New Horizons, measurements have yielded a significant detection of the EBL at the LORRI pivot wavelength of $\mathrm{ \lambda = 0.596 \mu m}$, with the main change in \cite{Lauer_2024} being an improved method for estimating the DGL. Within their respective errors, the measurements of \cite{Lauer_2024}, \cite{Koushan_2021}, and preliminary VHE measurements from \cite{Gréaux_2023} have finally converged. The explanation posed in \cite{Lauer_2024} is that the IGL composes the most significant component of the COB and the majority of extragalactic optical flux originates from within galaxies. The JWST work of \cite{Windhorst_2023} suggests that no more than $\approx$20\% of the COB can originate from outside of galaxies. As such, efforts to reduce uncertainties in the IGL can better constrain models and provide a near complete census of photon production from  baryon-photon decoupling until the present day.

In this paper, we begin by introducing the WAVES (Wide Area Vista Extragalactic Survey) photometric data, though for a complete discussion of the WAVES survey see the WAVES overview paper (Bellstedt et al. (in prep)). For our other supplementary data set we provide a brief overview of the DEVILS (Deep Extragalactic VIsible Legacy Survey) photometric data \citep{Davies_2021}. We then detail the processing of the SKYSURF database from assembly, reduction, and data quality checks. The production of galaxy number counts and the necessary steps to reduce HST object catalogs to usable number counts is discussed in Section \ref{Section_3}. In Section \ref{Section_4} we discuss the most significant issue posed by using HST to produce wide area counts given HST's bias towards observing cluster fields. In Section \ref{section_5.0} we present measurements of the IGL, sources of uncertainty, and how they can be improved. We then discuss the difficulties imposed by remaining bias in our HST number counts. Finally, we compare our measurements to existing measurements of the COB from IGL studies, VHE constraints, and direct measurements in Section \ref{Section_6}. Throughout we assume a standard 7-3-7 cosmology with H0 = 70 $\mathrm{km   s^{-1} Mpc^{-1}}$, $\Omega_{m} = 0.3$ and $\Omega_{\lambda} = 0.7$.

\section{Data and Data Reduction}

The data presented in this work is derived from catalogs and/or galaxy number counts as part of the WAVES (Bellstedt et al. in prep), DEVILS \citep{Davies_2021},
and SKYSURF \citep{Windhorst_2022} survey programs. Here we introduce the relevant surveys and present our SKYSURF HST data processing pipeline.

\subsection{The WAVES Survey}

WAVES is a spectroscopic redshift survey to be conducted using the European Southern Observatory’s 4-Metre-Multi-Object Spectroscopic Telescope (4MOST) \citep{de_Jong_2011}. In this work, bright galaxy number counts are produced in preparation for the WAVES survey \citep[][Bellstedt et al. in prep]{Driver_2016b, Driver_2019} which covers a $\mathrm{1150 deg^2}$ region of the sky. The galaxy catalogs used for WAVES target selection  (Bellstedt et al. in prep) are produced from imaging taken from the VISTA KIlo-degree Infrared Galaxy \citep[VIKING]{Edge_2013} and the VST Kilo-Degree Survey \citep[VST-KiDS]{Kuijken_2019}, both of which are ESO public data sets. The data was WCS aligned using \textsc{ProPane} \citep{Robotham_2023},  and object detection is done using an \textit{rizY} inverse variance stack using \textsc{ProFound} \citep{Robotham_2018}. All bright galaxies were inspected for fragmentation following the detection phase and fixed. Following the detection phase, fluxes in each band were determined using forced aperture photometry from the detection stack. Stars brighter than GAIA g$\mathrm{_{AB}}=16$ mag were masked. Star-galaxy separation was achieved using an unsupervised machine learning technique developed in \cite{Cook_2024}. WAVES galaxy number counts also include an object classification uncertainty. Objects are designated as either a star, galaxy, or ambiguous. Galaxy number counts were constructed both with and without the inclusion of ambiguous objects. The final galaxy counts are defined as the mean between the two and the uncertainty spans the range between the galaxy and galaxy + ambiguous counts. See Bellstedt et al. (in prep) for full details on these steps. In this work we use WAVES input catalog v1.2 with 7 bands (\textit{ugrizYJH}) that overlap with the HST filter system. The completeness limit of each band is shown in Table \ref{tab:tab1} where completeness limits were defined by stepping back by one half-magnitude bin from where the counts reach a maximum. 
 
\begin{figure*}
\vspace*{0.000cm}
\includegraphics[width=13cm, height = 18.47cm]{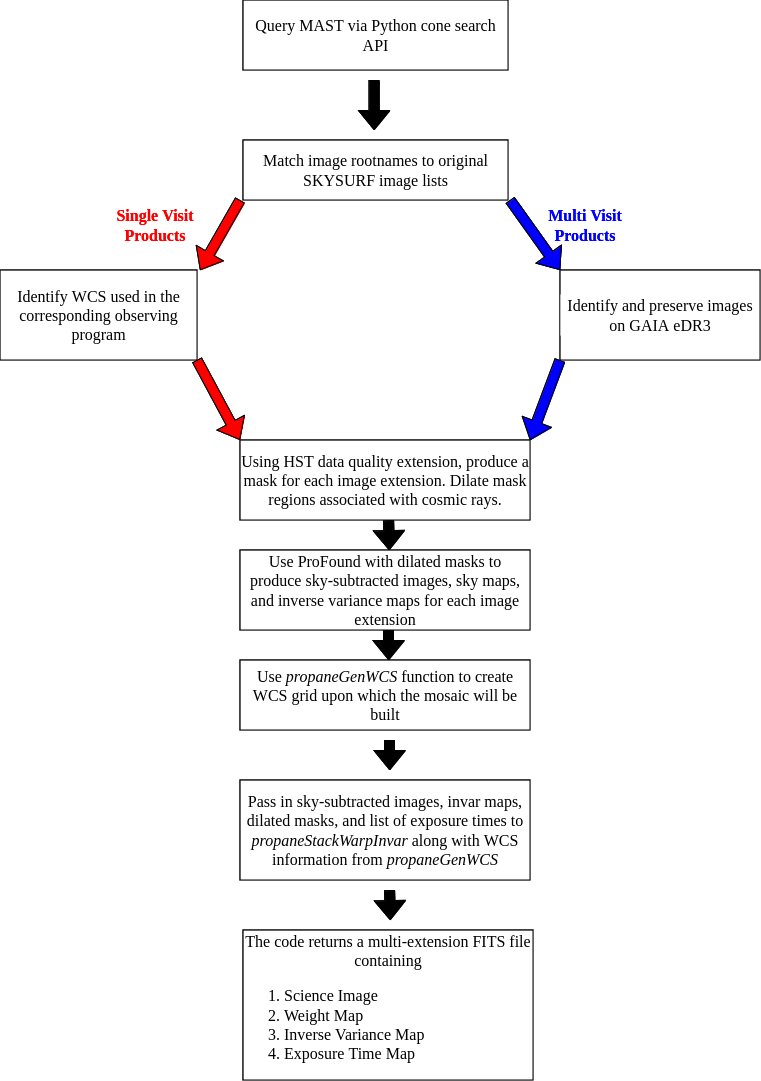}
\caption{ A flow chart illustrating the processing steps from MAST query to science ready multi-extension fits files used in this work.
\label{fig:chart}}
\end{figure*}

\begin{figure*}
    \centering
    \subfloat[\centering An example of a complex observing pattern from a WFC3IR F125W SKYSURF multi-visit product.]{{\includegraphics[width=8cm]{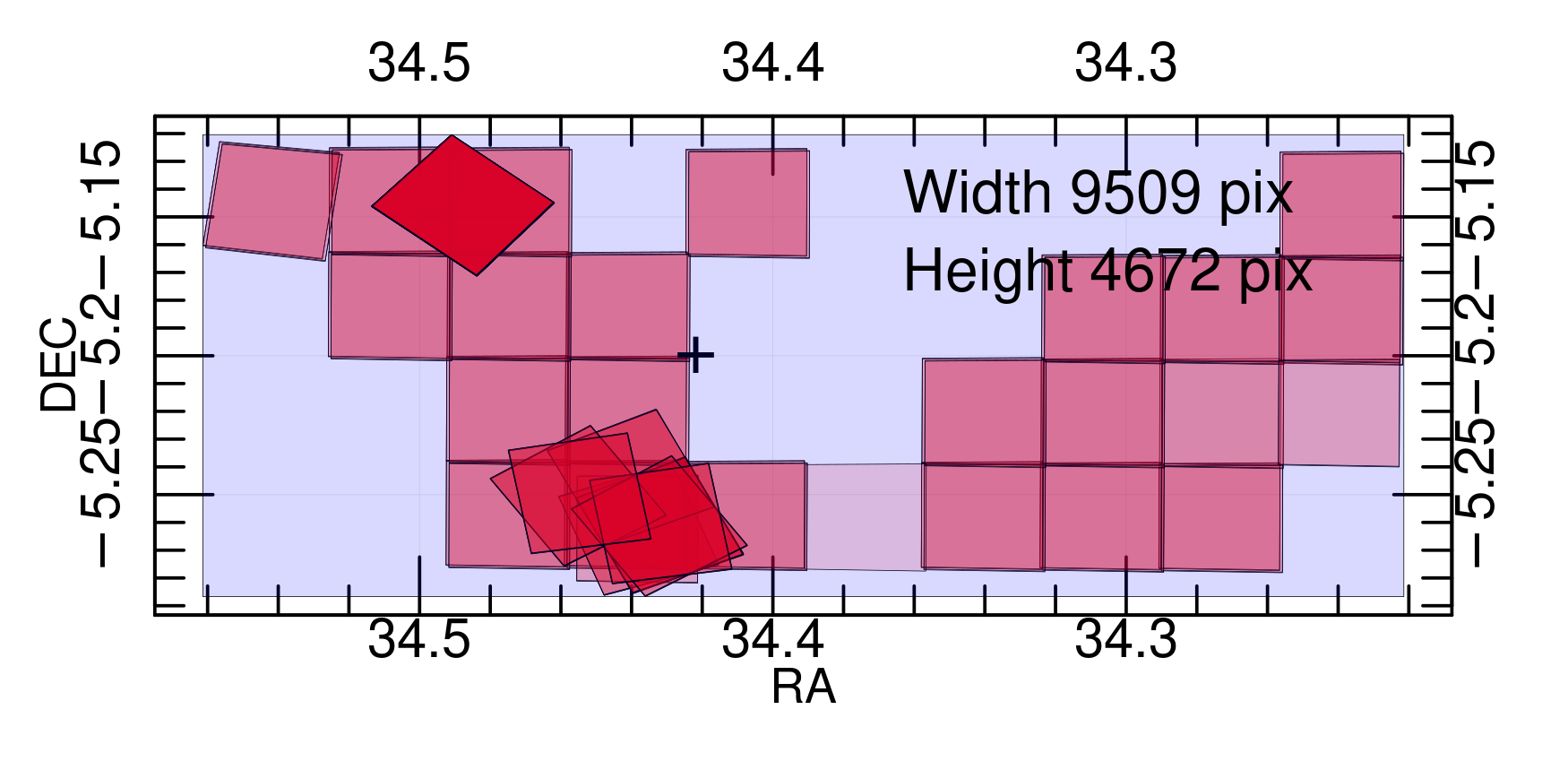} }}%
    \qquad
    \subfloat[\centering An example of a complex observing pattern from a WFC3IR F125W SKYSURF single-visit product.]{{\includegraphics[width=8cm]{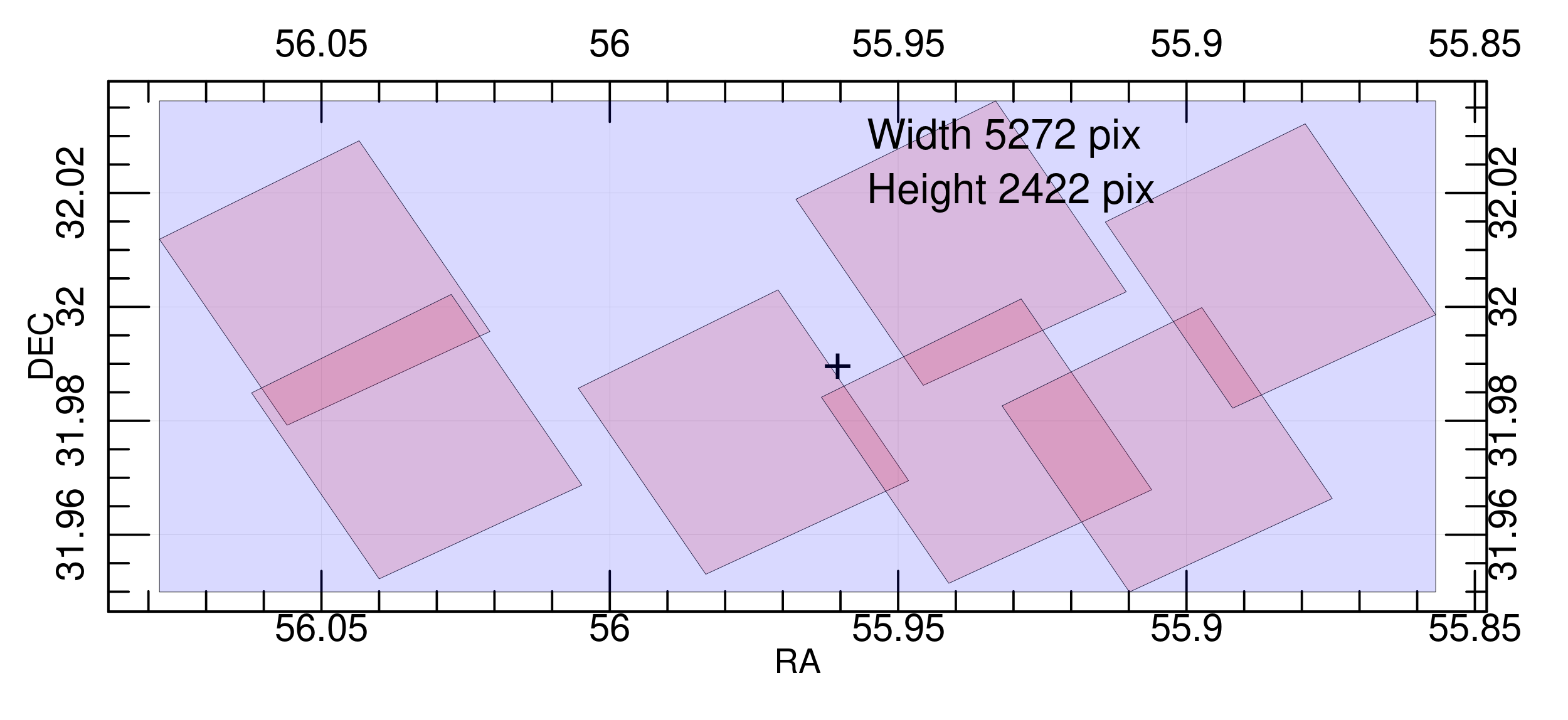} }}%

    \subfloat[\centering An example of a simple observing pattern from a WFC3IR F125W SKYSURF multi-visit product.]{{\includegraphics[width=7cm]{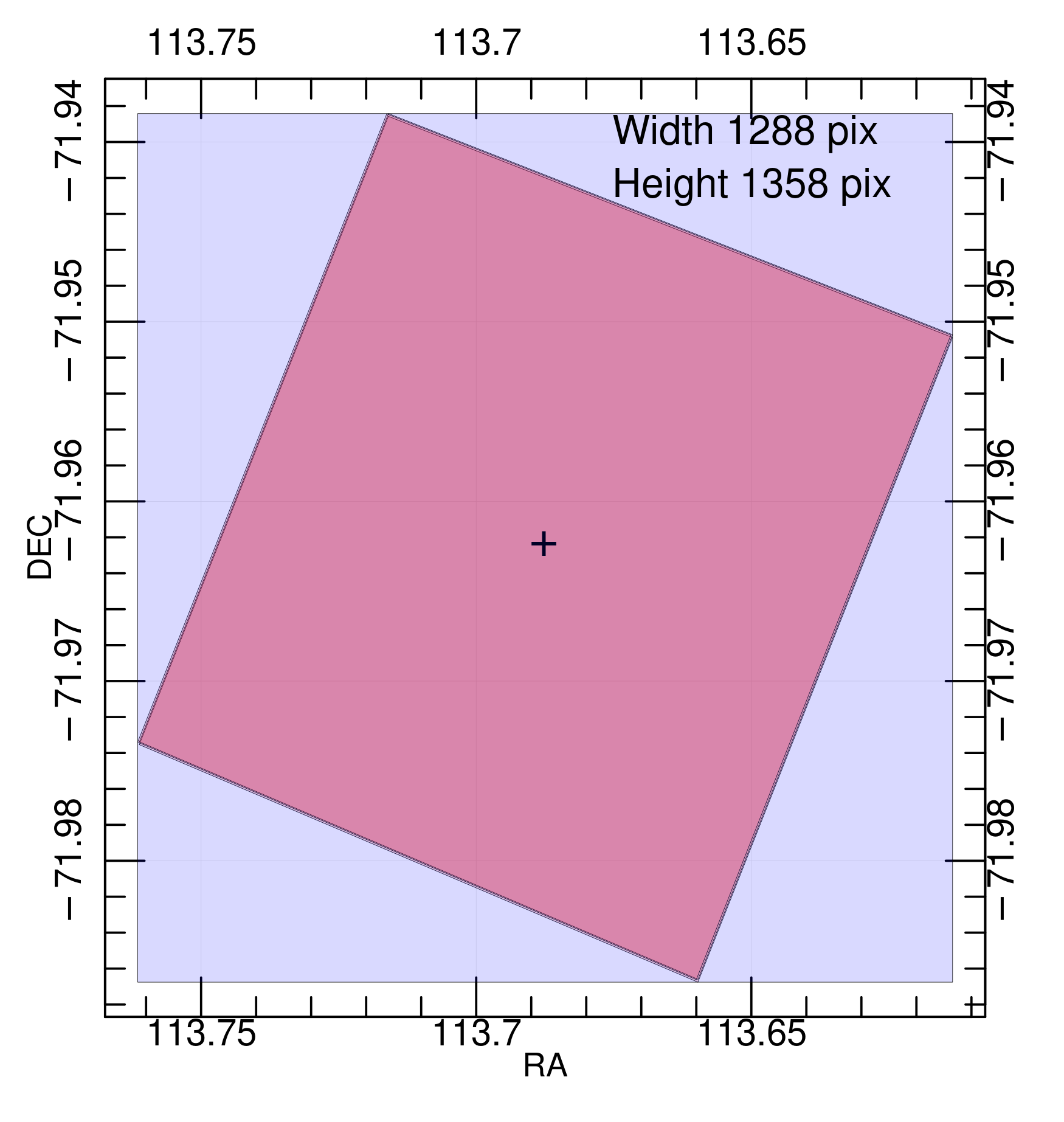} }}%
    \qquad
    \subfloat[\centering An example of a simple observing pattern from a WFC3IR F125W SKYSURF single-visit product.]{{\includegraphics[width=7cm]{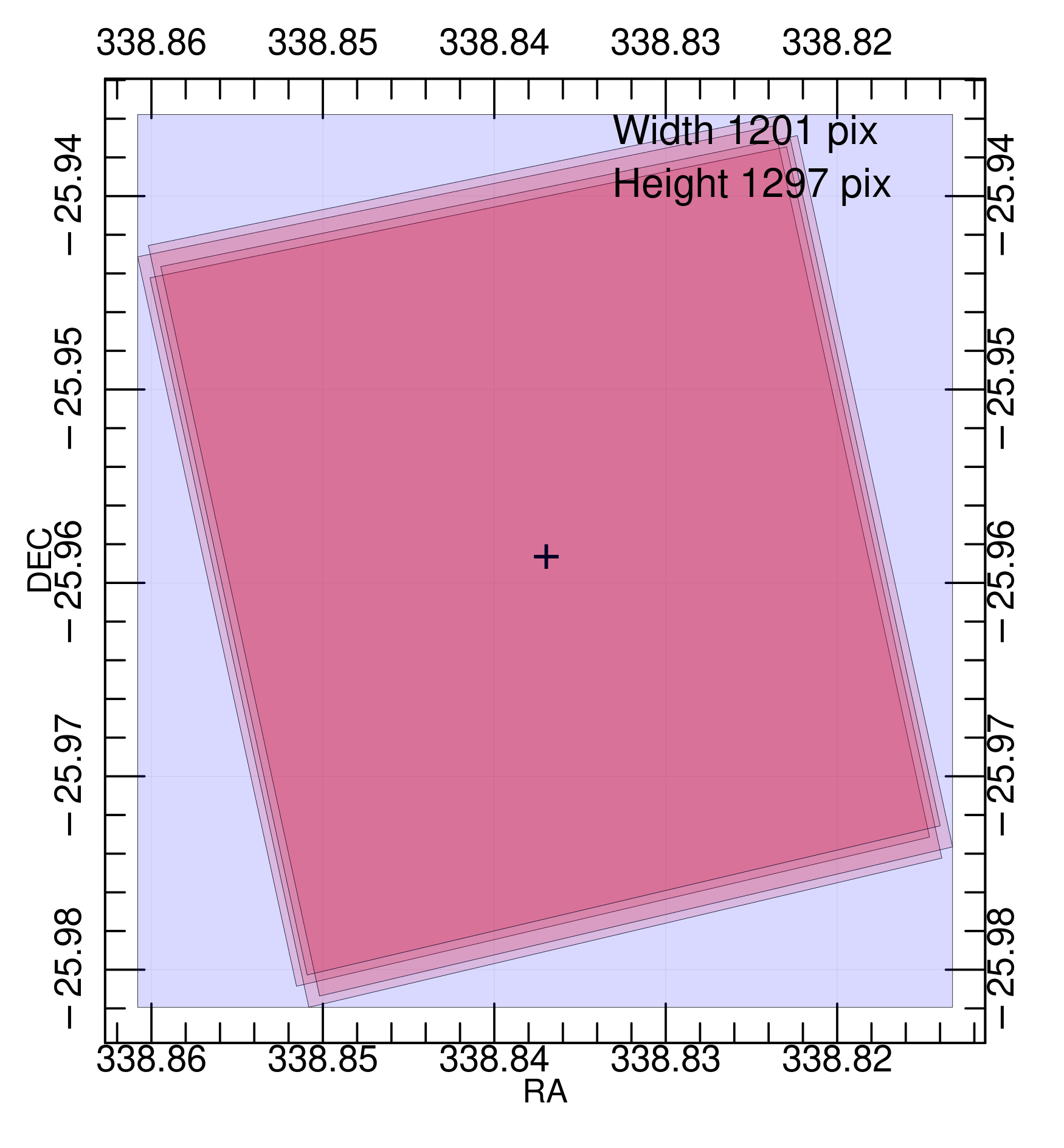} }}%
    \caption{Footprints from two complex and two simple SKYSURF mosaics.}%
    \label{fig:fig_footprints}%
\end{figure*}

\subsection{The DEVILS Survey}

 For intermediate magnitude number counts we turn to DEVILS \citep{Davies_2021}, using all three DEVILS fields, D02, D03, and D10. The three fields are centred on well-studied regions, XMM-LSS, ECDFS, and COSMOS, respectively. The DEVILS galaxy number counts are complete to $\approx$ AB 24 mag across all 3 fields and cover a combined area of 6 $\mathrm{deg^2}$. A brief summary of areas and completeness limits are provided in Table \ref{tab:tab1}. We elect to use the same 7 bands chosen from the WAVES survey, i.e. , $ugrizYJH$. We note that D02 and D10 were observed by the Hyper Suprime Cam (HSC) \citep{Hiroaki_2017} in \textit{griz} with the \textit{YJH} bands observed by VISTA. In D03 the \textit{griz} imaging was obtained by the VST. Finally, \textit{u}-band imaging was obtained from the Canada-France-Hawaii-Telescope (CFHT) and VST.
 
 Source detection was done using \textsc{ProFound} on an inverse variance weighted stack of three photometric bands, \textit{YJH}. This was done to maximize depth and to avoid any potential issues caused by artifacts in a single band. Source detection was performed on this stack and the resulting object contours were used for flux extraction in all bands. Star-galaxy separation was done using the relationship between half-light radius and Y band magnitude, as well as a relationship between magnitudes from the Y, J, H, and K bands. For a complete discussion of the DEVILS survey, see \cite{Davies_2021} and DEVILS data-release (Davies et al. 2025, Submitted).

\begin{table*}
\caption{Basic information for the ground-based data sets used in conjunction with SKYSURF.   \label{tab:tab1}}
    \centering
    \begin{tabular}{cccccc}
    \hline
        Band & WAVES Area & WAVES Completeness Limits & DEVILS Area & DEVILS Completeness Limits\\ \hline
           & $\mathrm{deg^2}$ & ABmag & $\mathrm{deg^2}$ & ABmag\\ \hline
        u   & 1058.30 & 11.75 - 23.25 & 5.96 & 15 - 25.5 \\ %\hline   
        g   & 1058.30 & 10.75 - 23.25 & 5.96 & 15 - 25 \\ %\hline
        r   & 1058.30 & 10.75 - 23.25 & 5.96 & 15 - 25 \\ %\hline
        i   & 1058.30 & 10.25 - 22.25 & 5.96 & 15 - 24.5 \\ %\hline
        z   & 1058.30 & 10.75 - 22.25 &  4.48 & 14.5 - 24.5 \\ %\hline
        Y   & 1058.30 & 9.75 - 21.75 &  5.97 & 15 - 24.5 \\ %\hline
        J   & 1058.30 & 10.25 - 21.25 &  5.96 & 15 - 24 \\ %\hline
        H   & 1058.30 & 10.75 - 21.25 &  5.96 & 14.5 - 24 \\ \hline
    \end{tabular}
\end{table*}

\subsection{The SKYSURF HST Archival Program}
\label{sec:sec2}
%\citealt[][Carter et al. in prep]{O'Brien_2023} 
SKYSURF is an HST cycle 27-29 Legacy Archival program (AR-09955 \& AR-15810) designed to use a significant amount of HST imaging data to measure the sky-surface brightness (sky-SB) across the wavelength range covered by HST, i.e., $\mathrm{ 0.2-1.6 \mu m}$. This requires the reprocessing of archived HST data using consistent software pipelines throughout. During the last two years, the SKYSURF project has measured the sky-SB and divided it into its components, the ZL, DGL, and EBL,  \citep[e.g.][Carter et al. 2025 (Submitted)]{O'Brien_2023, Carleton_2022, Windhorst_2022}. The SKYSURF project is split into two separate teams using independent processing pipelines and methodologies to obtain mosaics, sky-SB values, and object catalogs. This paper constitutes the portion of the project led from the University of Western Australia focused on integrated galaxy light (IGL) measurements using object catalogs generated using the \textsc{ProPane} and \textsc{ProFound} software packages,  \citep{Robotham_2023, Robotham_2018} as was used by WAVES and DEVILS. 
In Carter et al. 2025 (submitted), an independent processing of the SKYSURF archive was performed using the \textsc{AstroDrizzle} \citep{Gonzaga_2012} for image alignment and stacking. Source detection was done using \textsc{SourceExtractor} \citep{Bertin_1996}. Our overriding objective is to produce reliable galaxy number counts across as broad a magnitude and wavelength range as possible. We begin our SKYSURF selection by only including HST imaging with an exposure time of 200 seconds or longer and at galactic latitude $\mathrm{|b| > 10}$. A projection showing the location of SKYSURF data used in this work is shown in Figure \ref{fig:fig_projection}. All data was downloaded from the Mikulski Archive for Space Telescopes (MAST)\footnote{https://mast.stsci.edu/portal/Mashup/Clients/Mast/Portal.html}.

A number of tests and findings from earlier SKYSURF papers have contributed to the final decisions and methodology used in this work. Our science goals require an ability to demonstrate reliable sky surface brightness estimation and subtraction. In SKYSURF I \cite{Windhorst_2022}, we investigated nine different sky background estimation algorithms. We produced simulated WFC3IR images with known sky values either with or without an input gradient across the image. A number of different observing conditions such as sun angle, moon angle, or off-axis bright stars can introduce a non-uniform sky value which varies across the image.

We found that the \textsc{ProFound} based sky estimation was able to recover input sky values within $\approx0.1$ \& $\approx 0.8$\% for images with and without sky gradients, respectively. SKYSURF IV \citep{O'Brien_2023} implemented different sky estimation techniques on SKYSURF data, demonstrating that sky background levels increase as ecliptic latitude and/or sun angle decrease. This is not a surprising result, but reinforces the need to reliably subtract the sky background dominated by zodiacal light.

\begin{table}
\caption{A summary of the final HST data used in our analysis after 
all filtering described in Section \ref{Section_4} is performed.
 \label{tab:tab2}}
    \begin{tabular}{ccc}
    \hline
        Filter & \# of Fields & Final Area ($\mathrm{deg^2}$) \\ \hline
        ACSWFC F435W & 186 & 0.564 \\ %\hline
        ACSWFC F475W & 335 & 0.992 \\ %\hline
        ACSWFC F555W &  58 & 0.175 \\ %\hline
        ACSWFC F606W & 978 & 2.940 \\ %\hline
        ACSWFC F625W &  188 & 0.587 \\ %\hline
        ACSWFC F775W &  424 & 1.350 \\ %\hline
        ACSWFC F814W &  2201 & 7.04 \\ %\hline
        ACSWFC F850LP &  299 & 1.04 \\ %\hline
        WFC3UVIS F336W &  97 & 0.190 \\% \hline
        WFC3UVIS F390W &  80 & 0.160 \\ %\hline
      %  WFC3UVIS F438W &  ~ & ~ \\ \hline
      %  WFC3UVIS F475W &  ~ & ~ \\ \hline
      %  WFC3UVIS F555W &  ~ & ~ \\ \hline
        WFC3UVIS F606W &  279 & 0.529 \\ %\hline
       % WFC3UVIS F625W & ~ & ~ \\ \hline
        WFC3UVIS F814W & 479 & 0.880 \\ %\hline
        WFC3IR F105W &  345 & 0.460 \\ %\hline
        WFC3IR F110W &  245 & 0.329 \\ %\hline
        WFC3IR F125W &  344 & 0.531 \\ %\hline
        WFC3IR F140W & 288 & 0.466 \\ %\hline
        WFC3IR F160W &  1007 & 1.372 \\ \hline
    \end{tabular} 
\end{table}

\begin{figure*}
\includegraphics[width=15cm, height = 9cm]{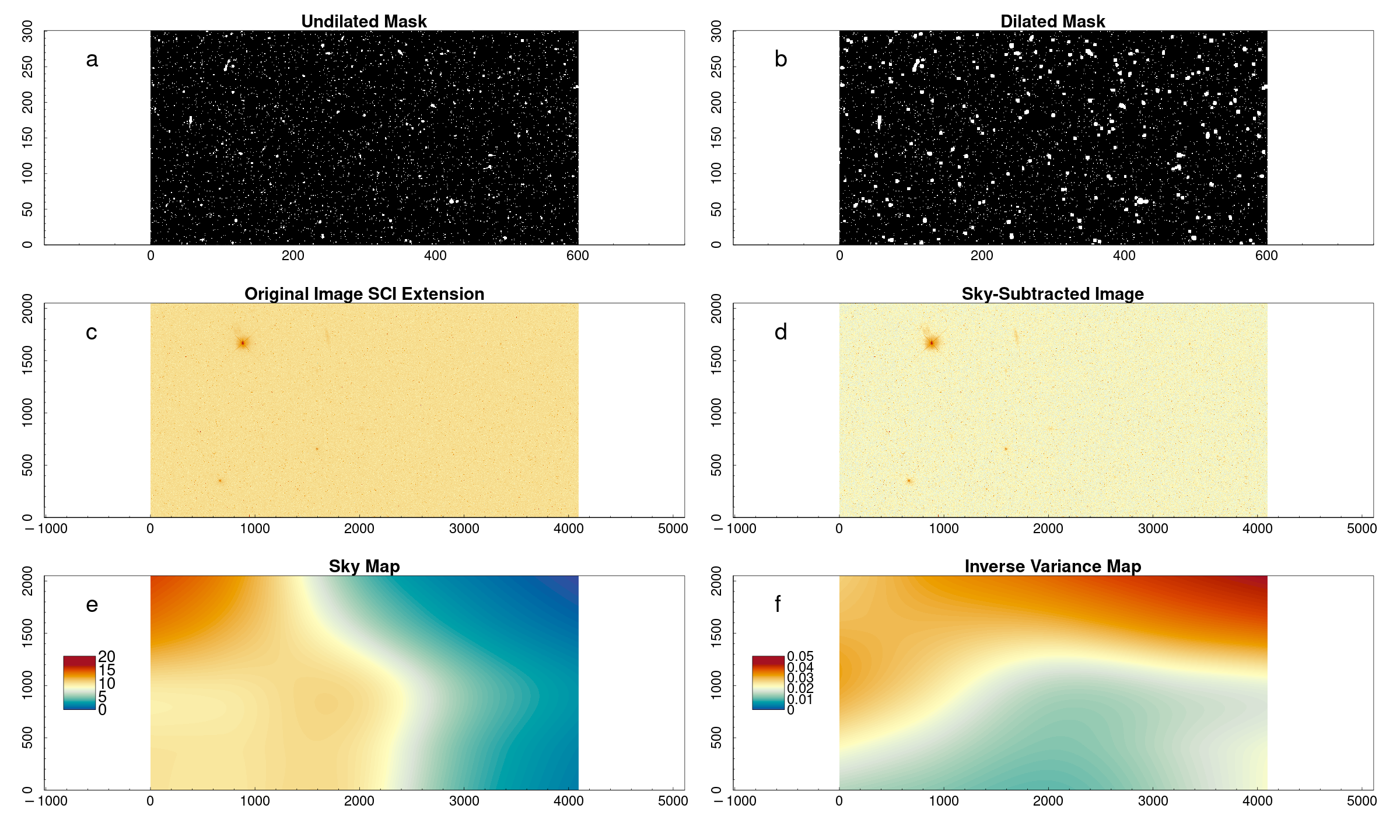}
\caption{ An example showing an output produced by a single image from WFC3UVIS. All labeled image dimensions are in units of pixels. 
\textbf{(a.)} Zoomed-in cut out of the undilated mask produced from the HST DQA extension.
\textbf{(b.)} The same zoomed in region after dilation.
\textbf{(c.)} The original HST 'SCI' image extension before processing.
\textbf{(d.)} The same image after sky-subtraction is performed using \textsc{ProFound}.
\textbf{(e.)} The sky map generated from the original image extension.
\textbf{(f.)} The inverse variance map generated from the original extension.
These maps are used as input for \textsc{propaneStackWarpInvar} as described in Figure \ref{fig:chart}.
\label{fig:fig_extensions}}
\end{figure*}

\begin{figure*}
\vspace*{0.000cm}
\includegraphics[width=15cm, height = 13.225cm]{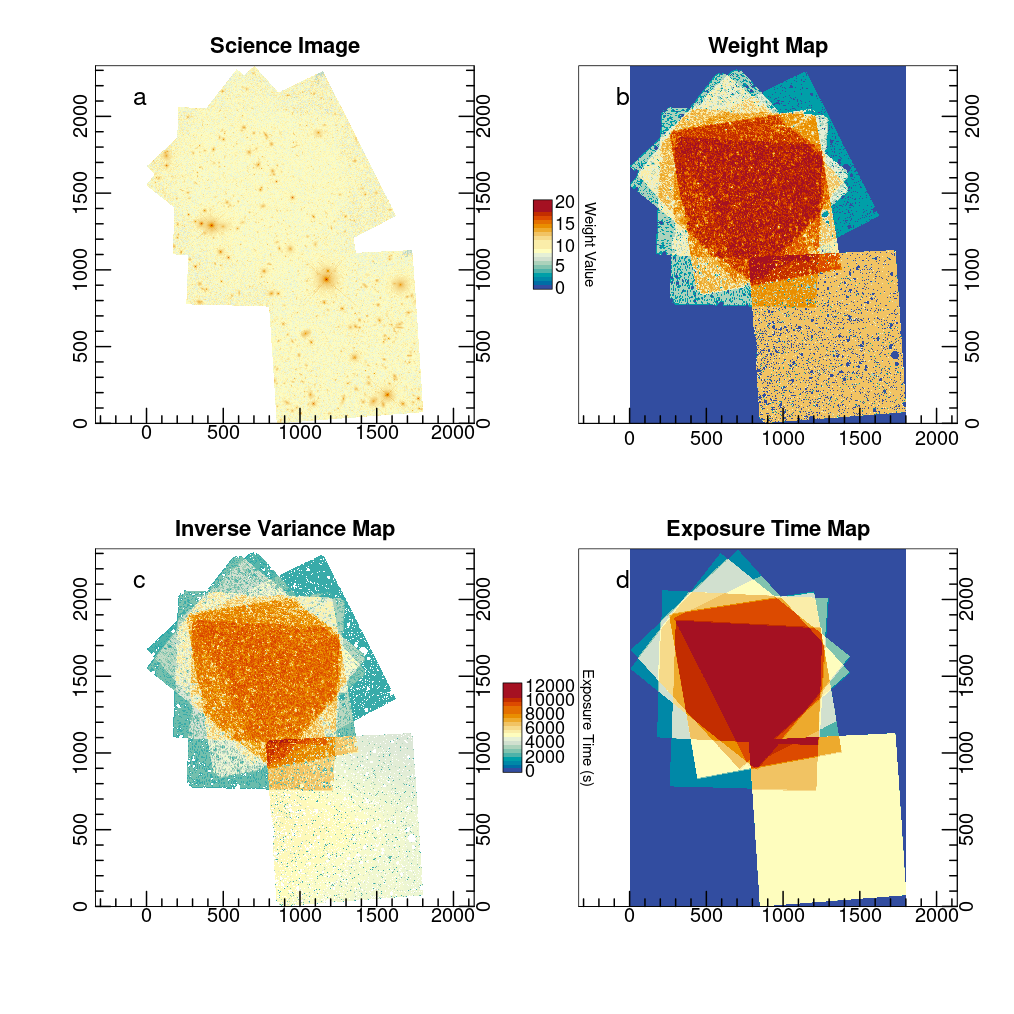}
\caption{ The final multi-extension FITS image with all extensions displayed. All labeled image dimensions are in units of pixels. \textbf{(a.)} Sky-Subtracted and \textsc{ProPane}-stacked image extension.
\textbf{(b.)} The weight map for the image. Weight values represent the number of unmasked pixels in the stack at each location.
\textbf{(c.)} The inverse variance map used as input to \textsc{ProFound} during the object detection stage.
\textbf{(d.)} The exposure time map for the field.
\label{fig:fits_final}}
\end{figure*}

\subsection{SKYSURF Data Reduction}

We use the data assembled in SKYSURF-1 \cite{Windhorst_2022} initially filtered as described in Section \ref{sec:sec2} and construct the mosaics summarized in 
Table \ref{tab:tab2} as follows. Starting with calibration level 3 products from MAST we produce sky-subtracted science images, inverse variance maps, and masks for each image extension using the \textsc{ProFound} image processing software \citep{Robotham_2018}. A flow chart summarizing the data processing steps is shown in Figure \ref{fig:chart}. In this work, we use the term "mosaic" to refer to any processed data regardless of how many exposures or observing programs are included. For example, some SKYSURF final data products consist of a single exposure while others are built from hundreds of exposures. SKYSURF mosaics comprised of multiple overlapping observing programs are referred to as multi-visit products. Mosaics produced from single observing programs are referred to as single-visit products. A series of footprints showing both complex and simple mosaic tiling patterns are displayed in Figure \ref{fig:fig_footprints}.

To ensure accurate alignment during stacking of multi visit data we further restrict our input images to those recalibrated to GAIA DR3 WCS. For mosaics composed only of a single observing program, we choose the data with the original WCS system associated with the program at the time of observation in order to be maximally inclusive.  An example intermediate data product is shown in Figure \ref{fig:fig_extensions} with a final mosaic shown in Figure \ref{fig:fits_final}. Mask generation is done using the HST data quality (DQ) FITS extension. We mask pixels associated with potential cosmic rays, and dilate these regions by three pixels to be conservative in removing potentially contaminated pixels not flagged in the original DQ extension. All other bad detector pixels and readout columns are masked but not dilated. Each individual sky subtracted image, inverse variance map, and mask are used to produce inverse variance weighted mosaics using the \textsc{propaneStackWarpInvar} function in \textsc{ProPane} (see Figure \ref{fig:fig_extensions}). The science mosaics are in flux units of electrons per second. Each mosaic contains an image extension, inverse variance map, weight map, and exposure time map (see Figure \ref{fig:fits_final}). The processing software pipeline used for the project is open source and publicly available on github\footnote{\url{github.com/stompkins7192/HST_Builder}}. 

\section{Object Catalog and Counts Construction}
\label{Section_3}
Here we discuss the methods used for the SKYSURF catalog construction. Initial object detection was achieved using \textsc{ProFound} \citep{Robotham_2018}. We first used a subset of images from the database with a range of filters and exposure times to optimize the settings in \textsc{ProFound} for object detection via an iterative process of varying settings and inspecting the resulting diagnostics, checking for completeness, reliability, and optimal deblending. The final \textsc{profoundProFound} object detection settings are shown below.

\texttt{profoundProFound     (                image = sci, \\
                                              header = sci[,]header, \\
                                              mask = mask, \\
                                              plot = FALSE,  \\
                                              magzero = magzero, \\
                                              pixscale = ps, \\
                                              rem\_mask = TRUE, \\
                                              keepim = FALSE,  \\
                                              skyRMS = (invar)$^{-0.5}$, \\ 
                                              \#Use inverse variance map as skyRMS  \\
                                              skycut = 1.5,   \\
                                              pixcut = 4.0,   \\
                                              ext = 4,  \\
                                              smooth = TRUE,  \\
                                              sigma = 0.8,  \\
                                              tolerance = 8,  \\
                                              reltol = 1,  \\
                                              cliptol = 300,  \\
                                              size = 5,    \\
                                              iters = 7,  \\
                                              box = c(((dim(mask)[1])/3),\\
                                              ((dim(mask)[2])/3)),   
                                              boxiters = 2,  \\
                                              roughpedestal = FALSE \\
                                              pixelcov = FALSE \\
                                              boundstats = TRUE,  \\
                                              nearstats = TRUE,  \\
                                              redosky = TRUE,  \\
                                              redoskysize = 21,  \\
                                              SBdilate = 2  ) \\
                        }

We then applied \textsc{profoundProFound}, the top level source detection and cataloging function in \textsc{ProFound}, to all 16,686 mosaics including both single and multi-visit products and detected a total of 153 million objects (including stars, galaxies, and artifacts). All magnitude values for galaxy number counts used in this work are sky-subtracted apparent AB magnitudes as calculated by \textsc{profoundProFound}. After running source detection, we implement a four step process to go from our raw catalogs to galaxy number counts. These steps are galactic extinction corrections, star-galaxy separation, bright star masking, and completeness corrections as described in the following sections.

\begin{figure*}
\includegraphics[width=4.65cm, height = 3.57cm]{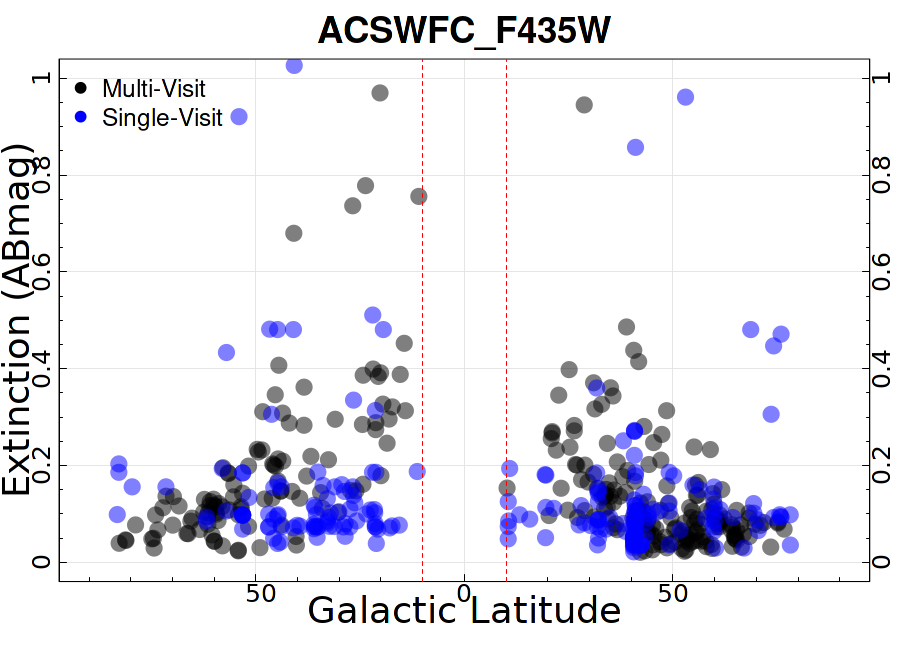}
\includegraphics[width=4.65cm, height = 3.57cm]{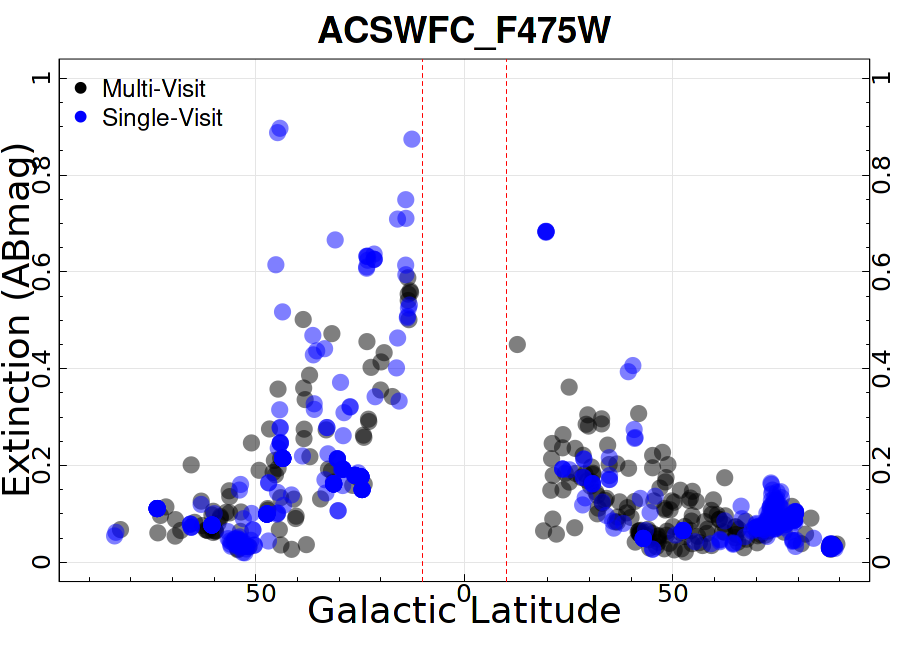}
\includegraphics[width=4.65cm, height = 3.57cm]{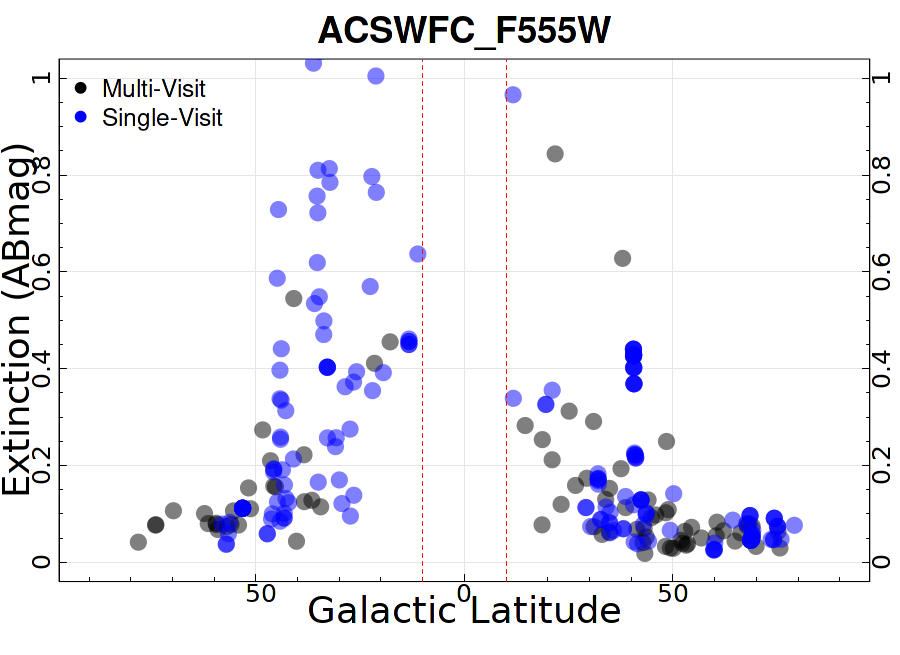}

\includegraphics[width=4.65cm, height = 3.57cm]{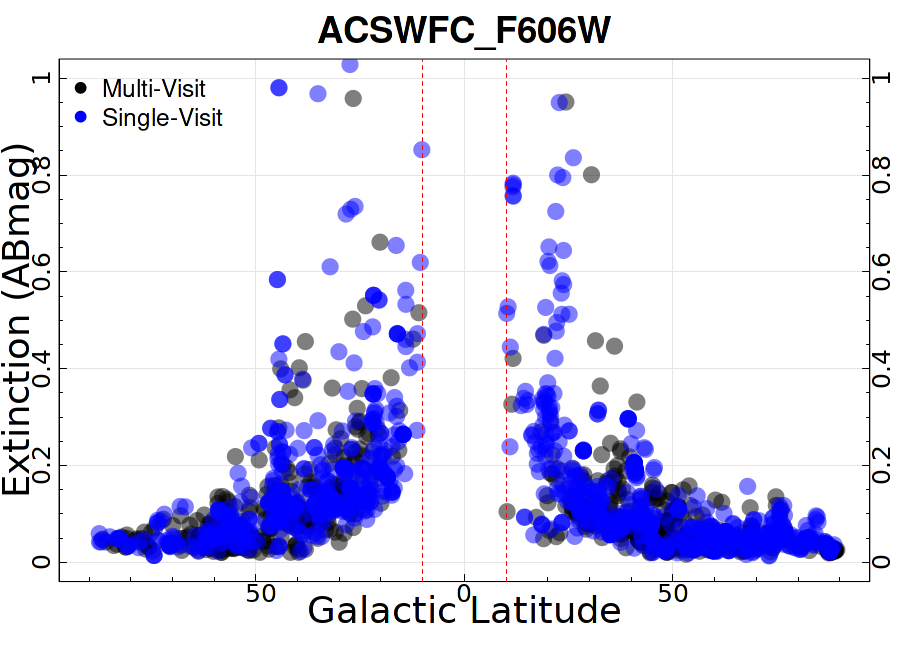}
\includegraphics[width=4.65cm, height = 3.57cm]{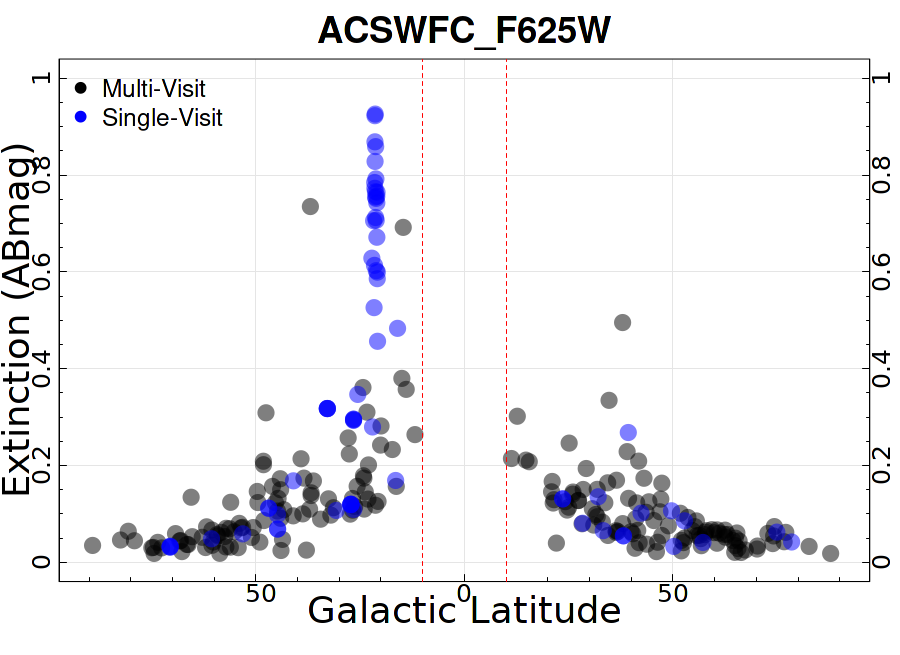}
\includegraphics[width=4.65cm, height = 3.57cm]{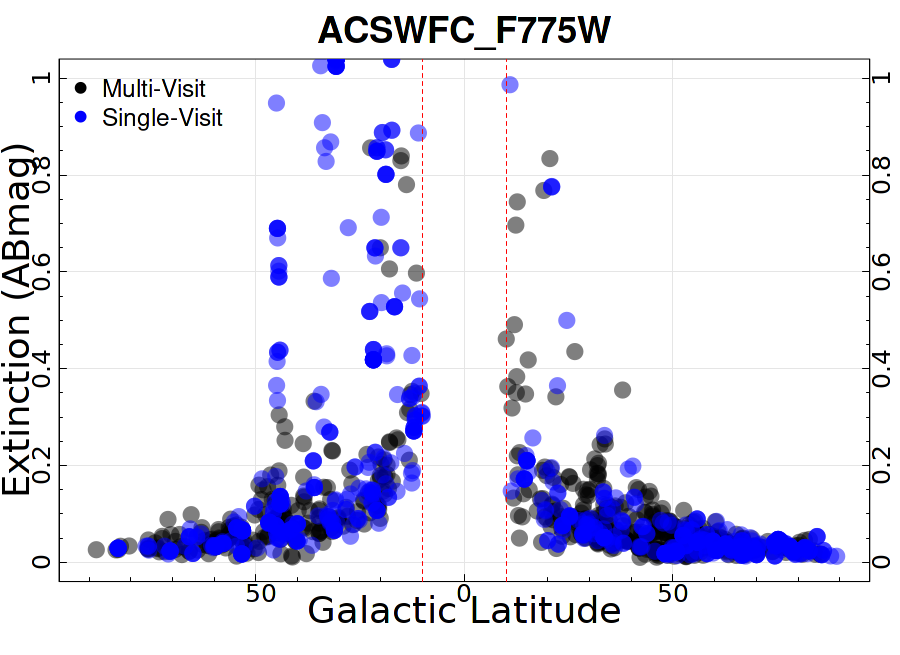}

\includegraphics[width=4.65cm, height = 3.57cm]{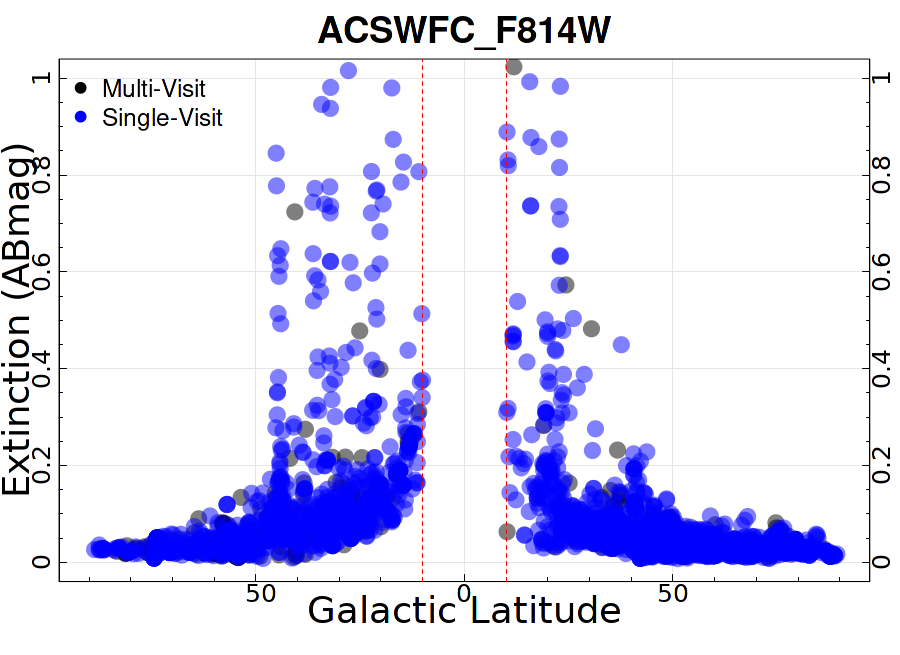}
\includegraphics[width=4.65cm, height = 3.57cm]{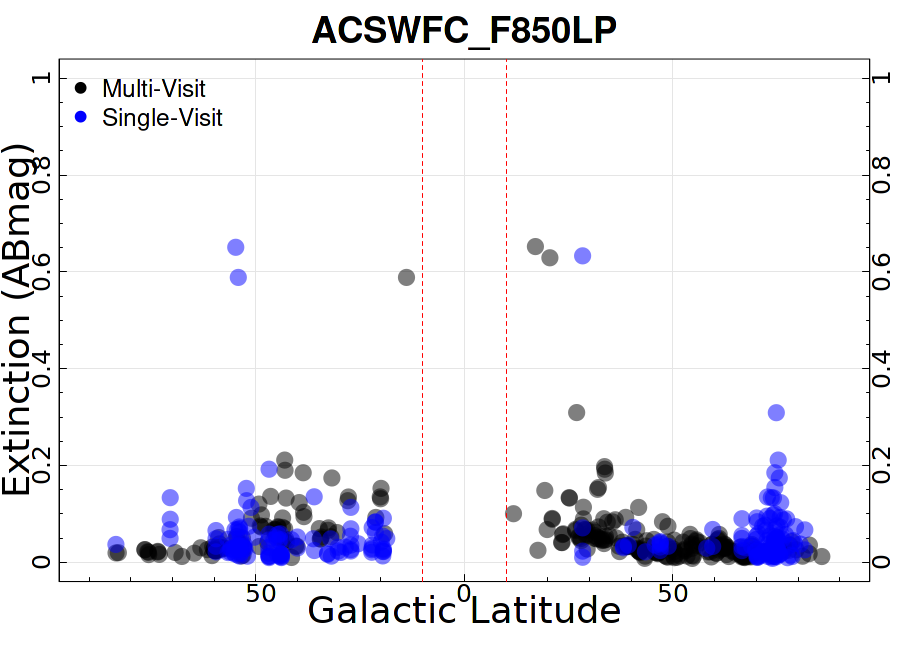}
\includegraphics[width=4.65cm, height = 3.57cm]{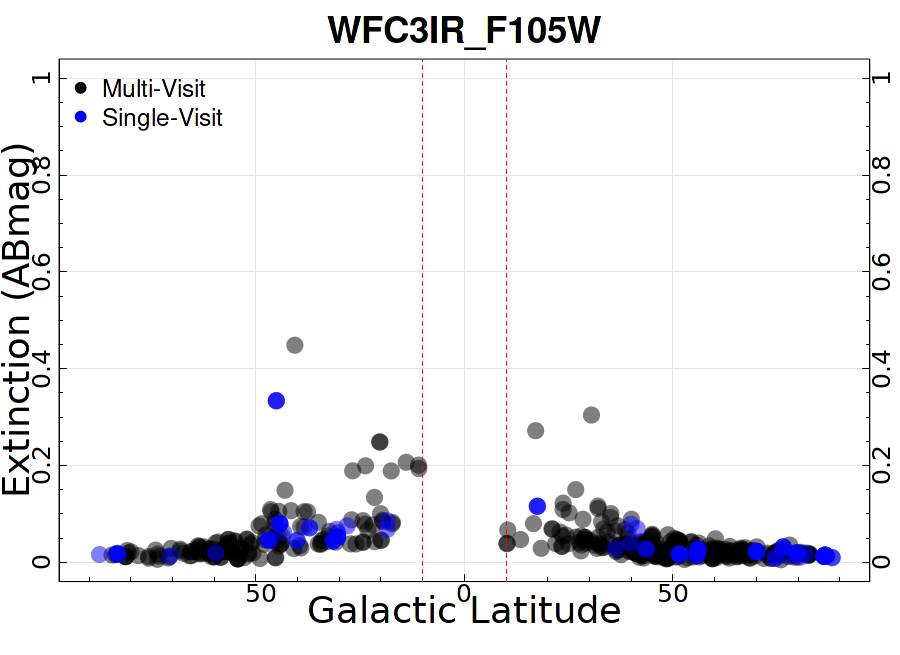}

\includegraphics[width=4.65cm, height = 3.57cm]{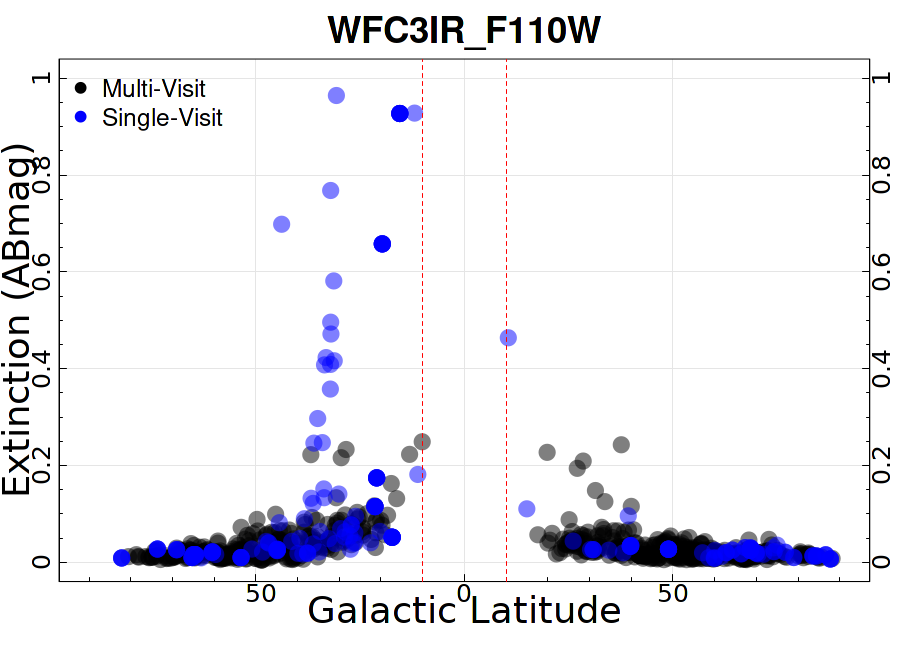}
\includegraphics[width=4.65cm, height = 3.57cm]{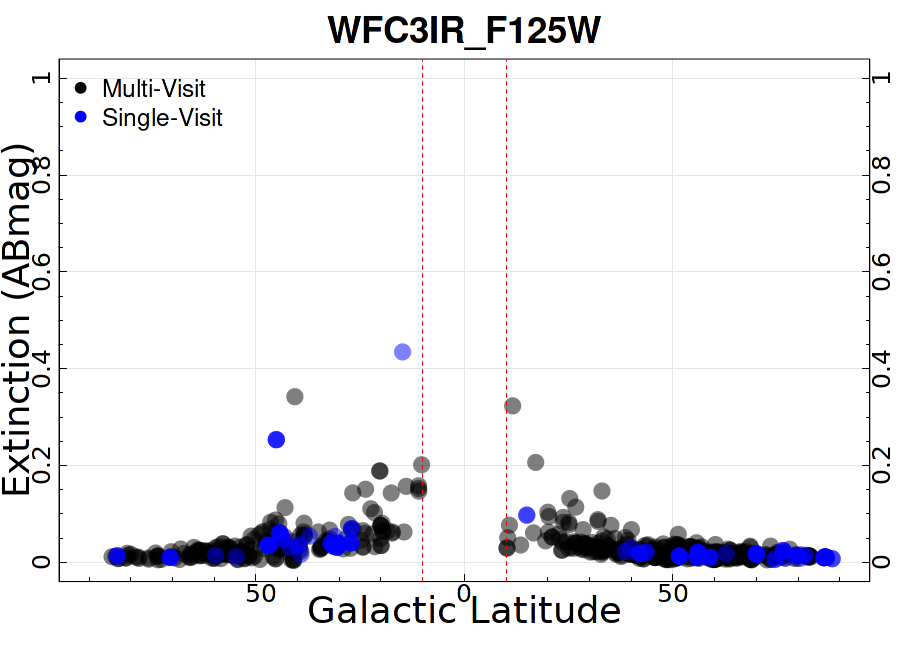}
\includegraphics[width=4.65cm, height = 3.57cm]{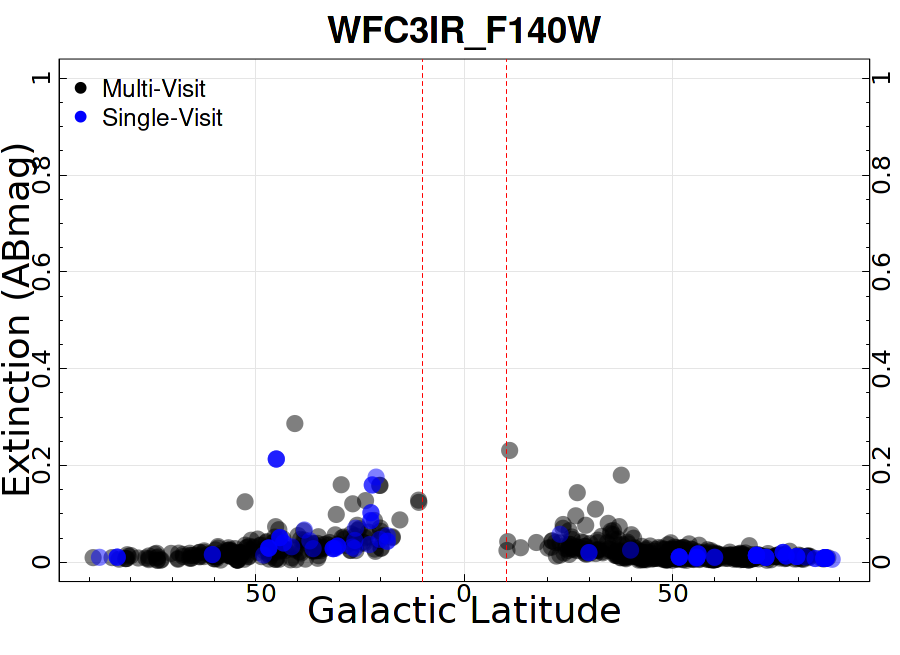}

\includegraphics[width=4.65cm, height = 3.57cm]{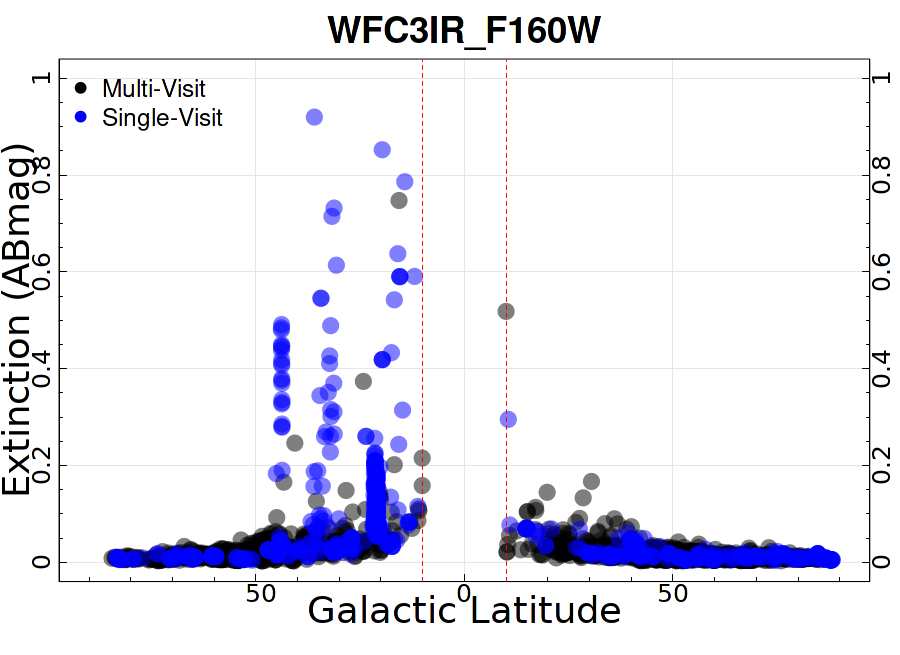}
\includegraphics[width=4.65cm, height = 3.57cm]{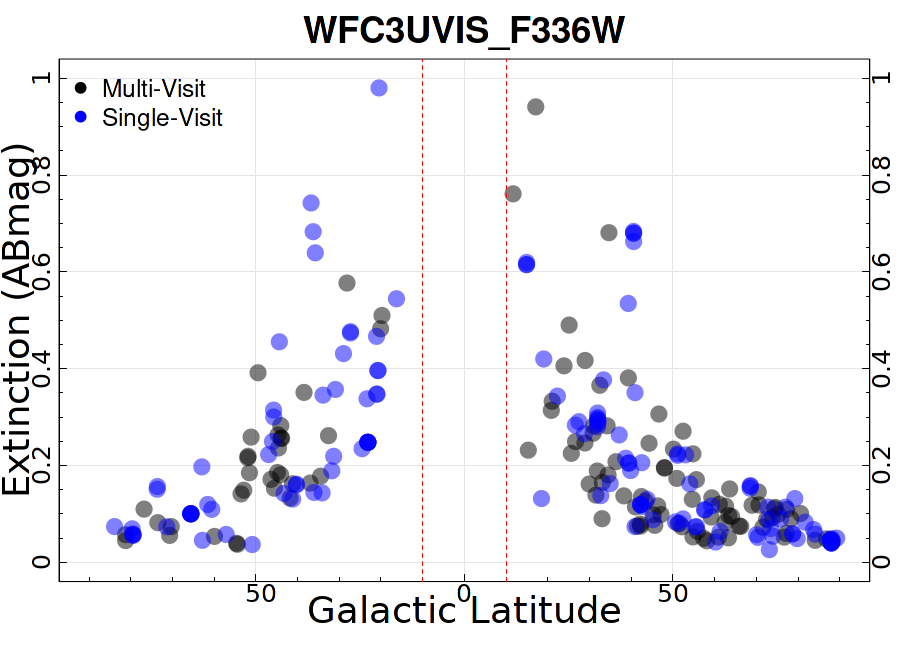}
\includegraphics[width=4.65cm, height = 3.57cm]{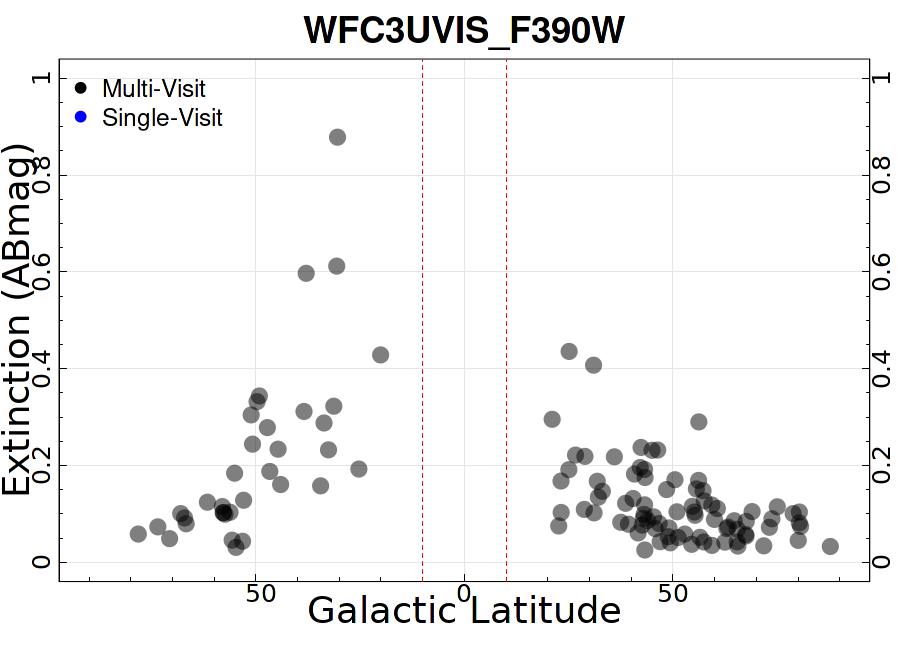}

\includegraphics[width=4.65cm, height = 3.57cm]{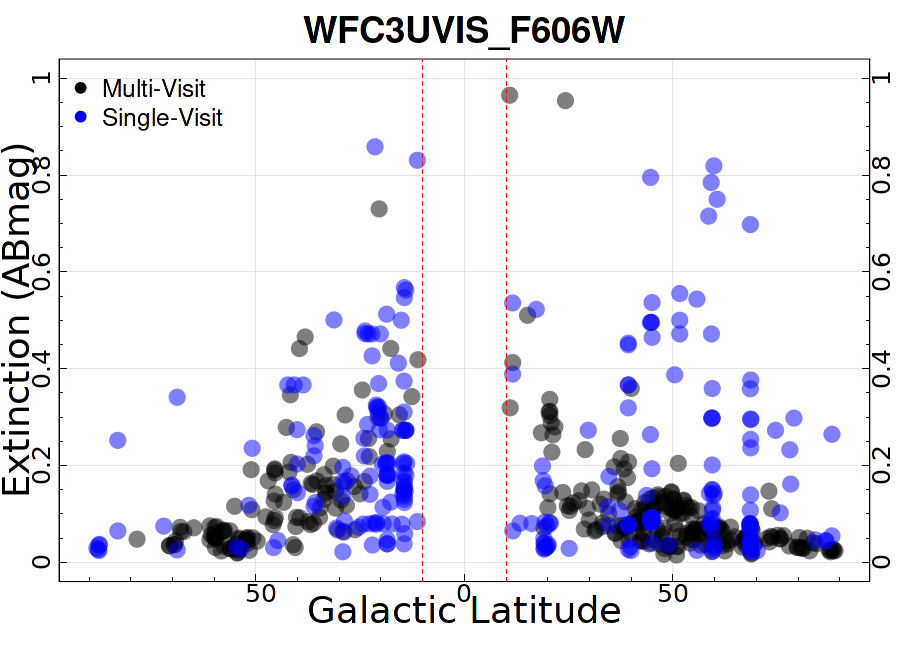}
\includegraphics[width=4.65cm, height = 3.57cm]{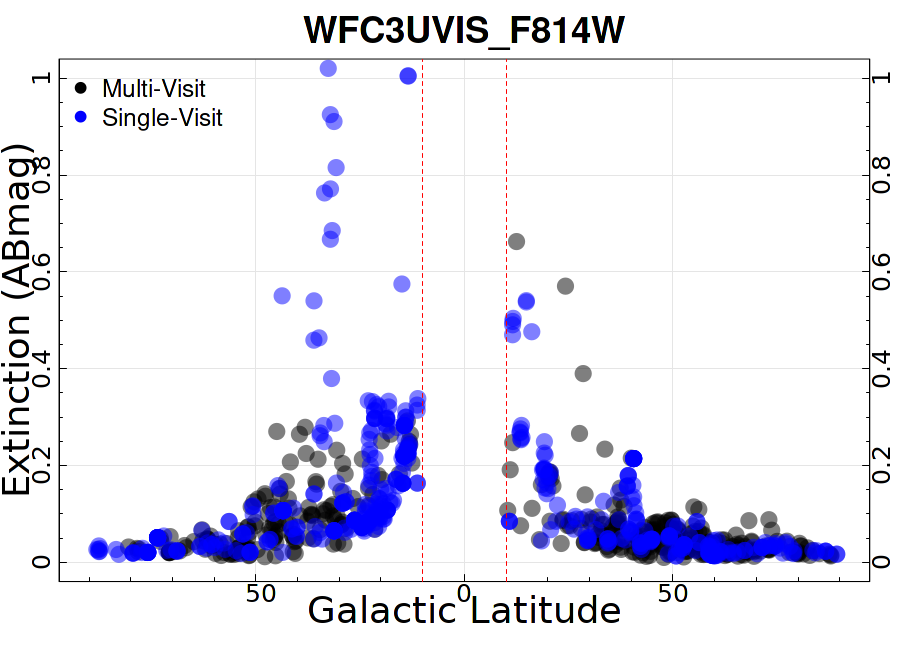}

\caption{
\label{fig:fig_extinct}
Galactic extinction correction versus galactic latitude for all 14 HST filters used in SKYSURF.}
\end{figure*}

\begin{figure*}
\vspace*{0.000cm}
\includegraphics[width=12cm, height = 10cm]{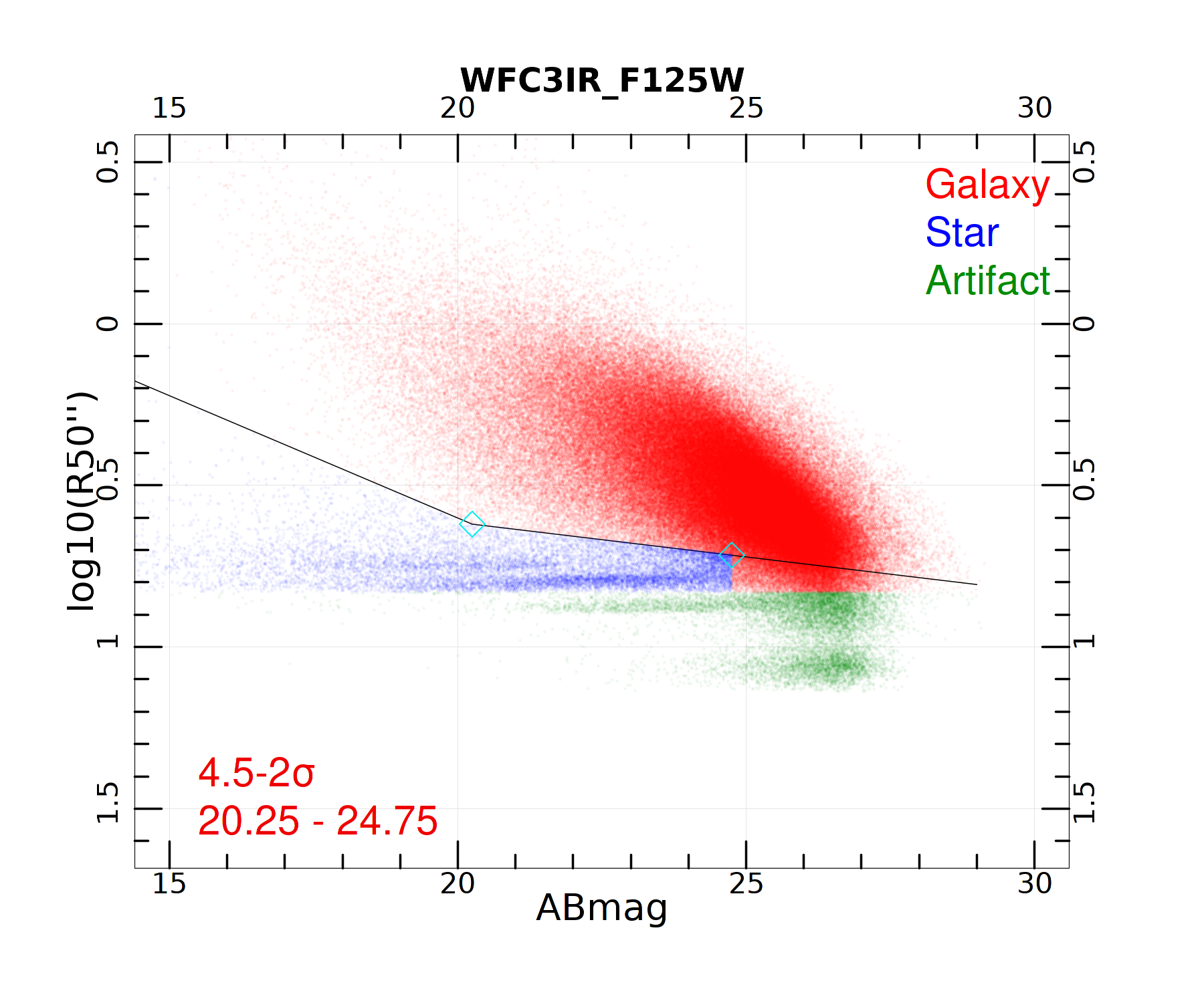}
\caption{ An example figure showing the star, galaxy, and artifact boundaries used in this work for the WFC3IR F125W filter. The numbers quoted in red represent the boundaries between which the star-galaxy separation boundary is drawn. See the text for details on how these values are obtained. The same parameter space and dividing lines are shown in the appendix for all other HST filters used in this work. A histogram showing the distribution of object sizes between AB 20-22 mag is shown in Figure \ref{fig:r50_hist}.
\label{fig:fig_sgs}}
\end{figure*}

\begin{figure}
\vspace*{0.000cm}
\includegraphics[width=7cm, height = 7cm]{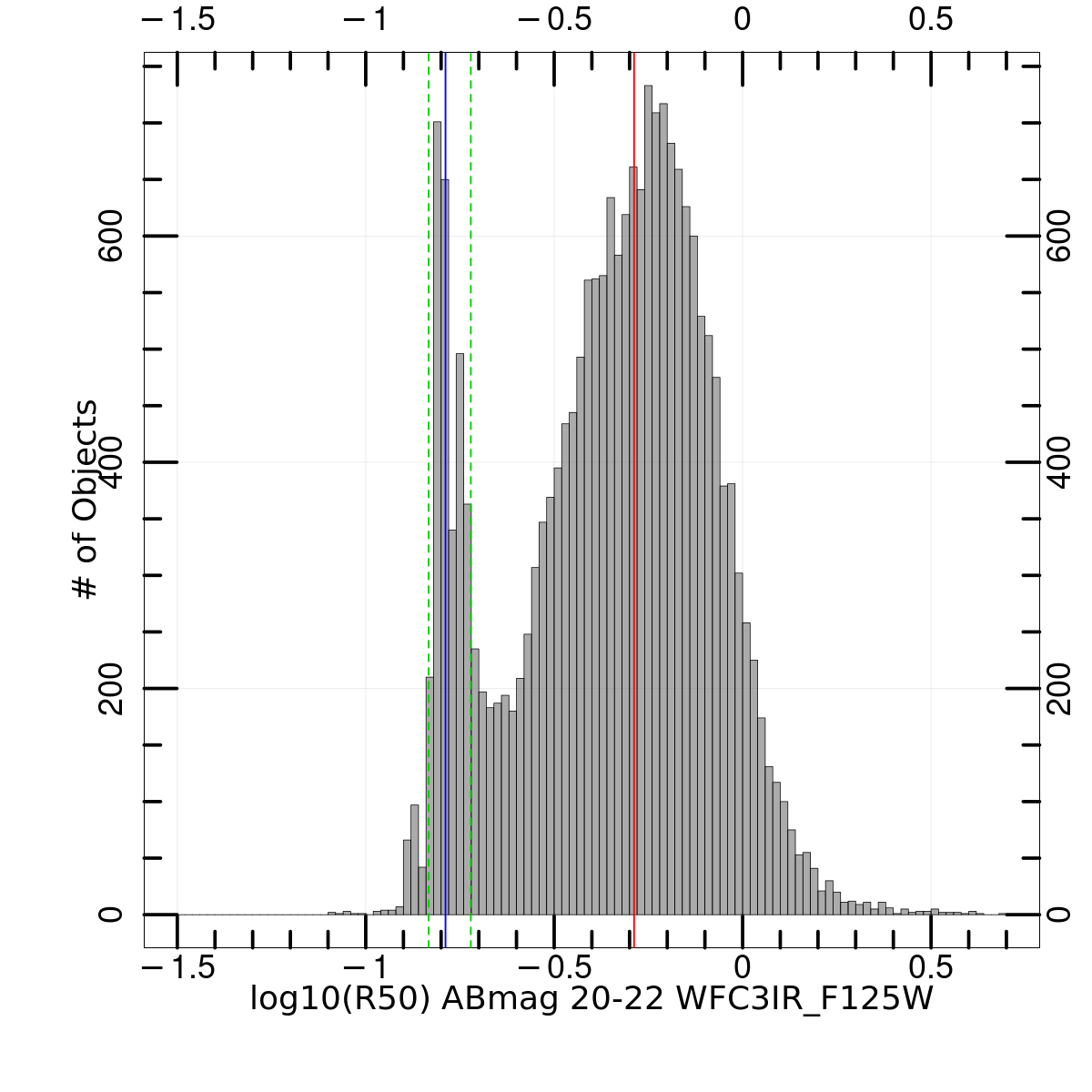}
\caption{ The distribution of $\mathrm{\log_{10} (R50)}$ for all objects in the WFC3IR F125W filter between AB 20-22 mag. The mean values from both Gaussians are marked by vertical blue lines for the star \& galaxy populations, respectively. The dividing boundaries at $\mathrm{\pm 1.5\sigma}$ are marked by the vertical green lines. }
\label{fig:r50_hist} 
\end{figure}

\begin{figure}
\vspace*{0.000cm}
\includegraphics[width=8cm, height = 8cm]{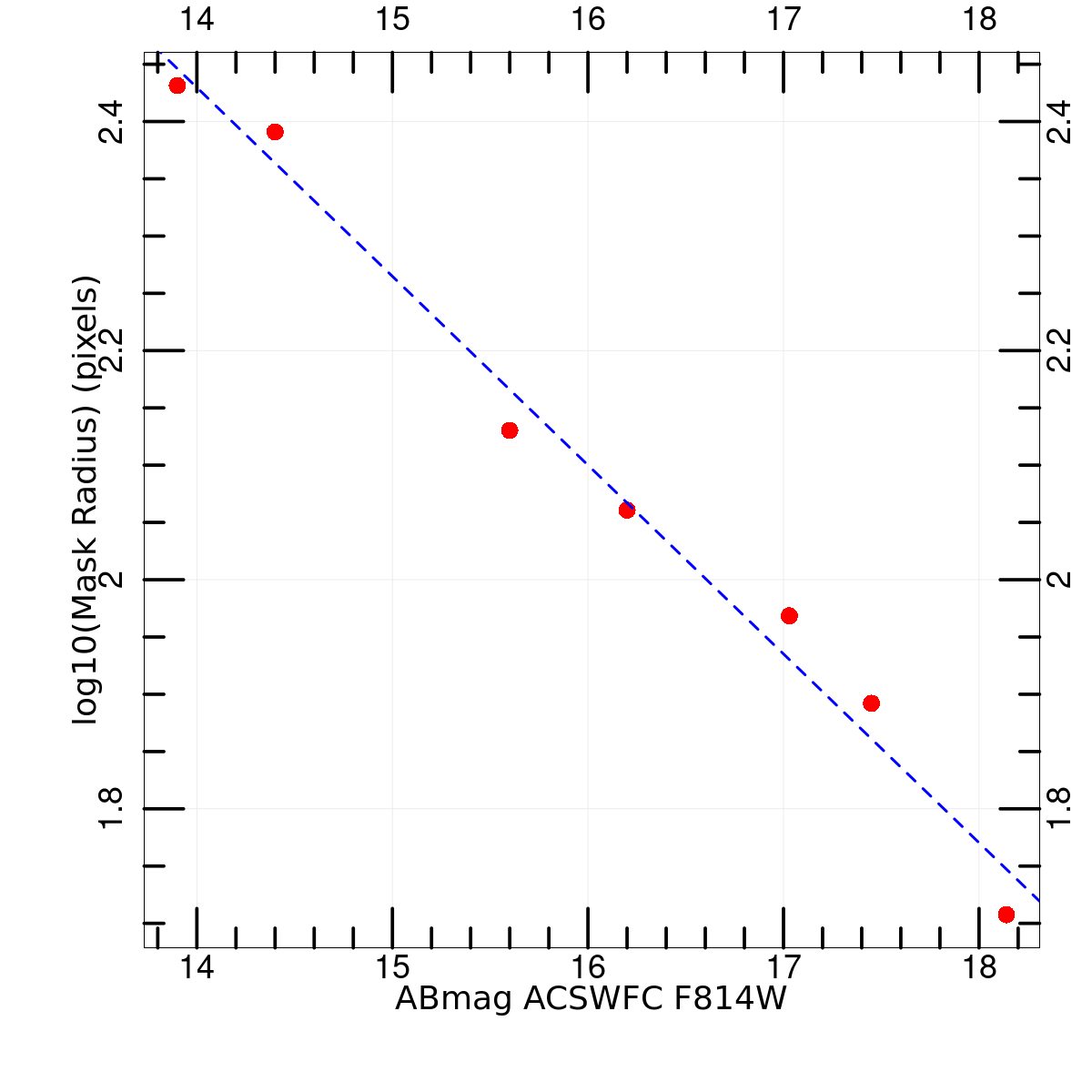}
\caption{ The relation between magnitude and mask radius in pixels for a set of stars in ACSWFC F814W. 
\label{fig:mask_radius}}
\end{figure}

\begin{figure*}
\vspace*{0.000cm}
\includegraphics[width=14.6cm, height = 12cm]{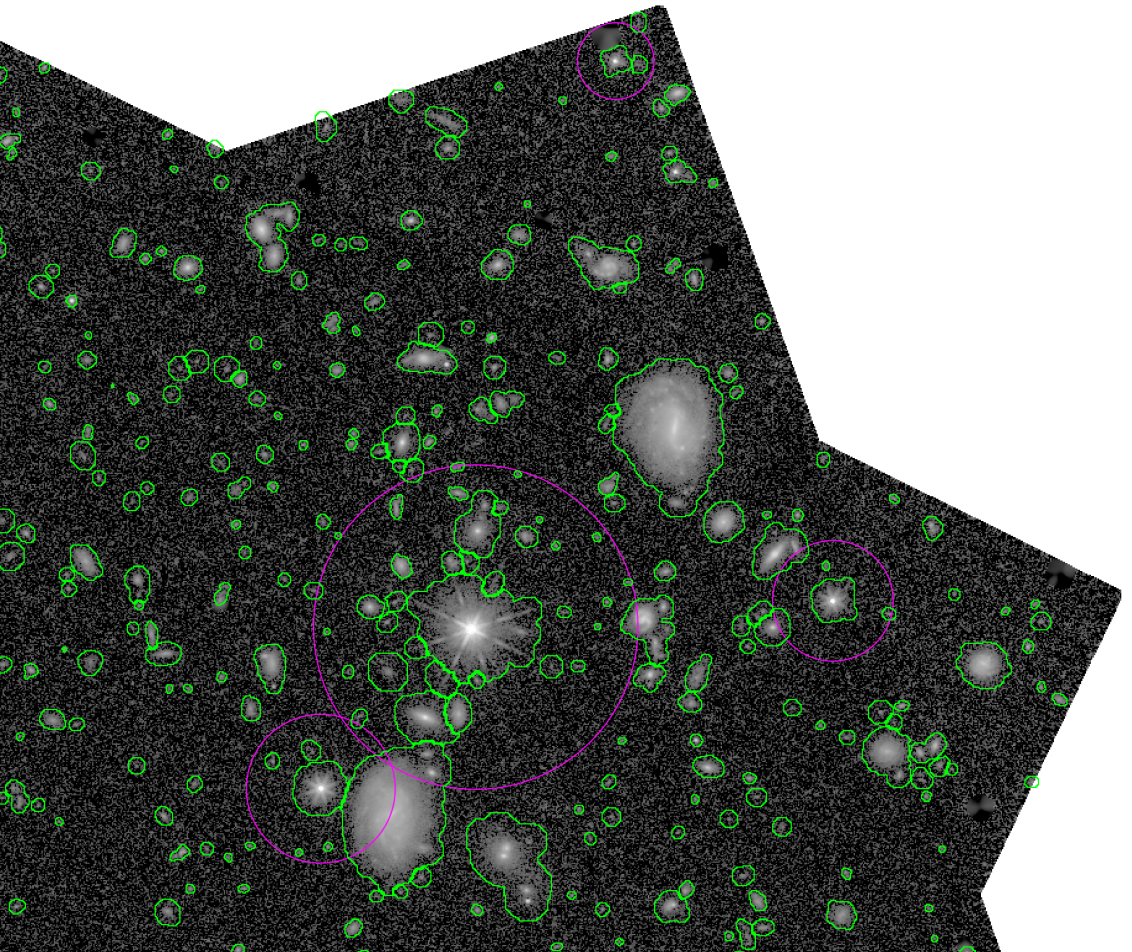}
\caption{An zoomed-in example showing successful GAIA catalog coordinate matches circled in magenta. While successful in identifying stars, the inclusion of galaxies such as the example seen in the bottom right in the GAIA catalog demonstrates it cannot be used as a delimiter between stars and galaxies. Stars in the central region also show the effects of stacking images taken at different position angles where multiple sets of diffraction spikes can be seen.
\label{fig:posang}}
\end{figure*}

\subsection{Galactic Extinction Corrections}
We correct for galactic extinction using the Planck satellite dust maps accessible through the
web tool\footnote{https://irsa.ipac.caltech.edu/applications/DUST/} with extinction estimates provided by \cite{Finkbeiner_1998}. Given the small angular size of SKYSURF fields compared to the resolution of the dust maps, we elect to use the centre of the field to obtain a single E(B-V) value and thus apply one extinction value to all objects in that field. Using the Spanish Virtual Observatory tool provided by \citep{Rodrigo_2020}\footnote{http://svo2.cab.inta-csic.es/theory/fps/}, we obtain $A_V$ values for each HST filter. The $\mathrm{A_{f} / A_{V}}$ values adopted for each HST filter used in this work along with the pivot wavelengths adopted in this work are listed in Table \ref{tab:tab3}. $\mathrm{A_{f} / A_{V}}$ is defined as the ratio between extinction at $\mathrm{\lambda_{ref}}$. and visual extinction. In Figure \ref{fig:fig_extinct} we show the Galactic extinction values for each mosaic as a function of Galactic latitude. By design, SKYSURF omitted all observations within 10 degrees of the galactic plane, as indicated by the vertical lines in Figure \ref{fig:fig_extinct}. 

\begin{table}
    \caption{Tabled values obtained from the SVO website \citet{Rodrigo_2020} used to perform galactic extinction corrections in this work. \label{tab:tab3}.}
        \centering %\frac{}{}A$_{\lambda}$$\frac{A_{\lambda}}{A_{\nu}}$
    \begin{tabular}{lllll} \hline 
        Filter & Pivot Wavelength (Å) & $\mathrm{A_f / A_{V}}$  Value  \\ \hline
        ACSWFC F435W & 4369.62 & 1.32  \\ 
        ACSWFC F475W & 4783.06 & 1.20  \\ 
        ACSWFC F555W & 5413.32 & 1.04 \\ 
        ACSWFC F606W & 5962.2 & 0.924  \\ 
        ACSWFC F625W & 6323.81 & 0.853  \\ 
        ACSWFC F775W & 7763.11 & 0.645  \\ 
        ACSWFC F814W & 8102.91 & 0.602  \\ 
        ACSWFC F850LP & 8945.58 & 0.505  \\ 
    %    WFC3UVIS F225W & 2354.44 & 2.68  \\ 
     %   WFC3UVIS F275W & 2725.91 & 2.03  \\ 
      %  WFC3UVIS F300X & 2763.84 & 1.94  \\ 
        WFC3UVIS F336W & 3371.04 & 1.69  \\ 
        WFC3UVIS F390W & 3955.42 & 1.47  \\ 
     %   WFC3UVIS F438W & 4325.41 & 1.32  \\ 
     %   WFC3UVIS F475W & 4788.5 & 1.19  \\ 
     %   WFC3UVIS F475X & 4950.34 & 1.15  \\ 
      %  WFC3UVIS F555W & 5307.2 & 1.05  \\ 
        WFC3UVIS F606W & 5946.48 & 0.93  \\ 
       % WFC3UVIS F625W & 6277.98 & 0.866  \\ 
       % WFC3UVIS F775W & 7734.8 & 0.65  \\ 
        WFC3UVIS F814W & 8107.52 & 0.603  \\ 
        %WFC3UVIS F850LP & 9121.74 & 0.493  \\ 
       % WFC3IR F098M & 9860.02 & 0.438  \\ 
        WFC3IR F105W & 10542.47 & 0.395  \\ 
        WFC3IR F110W & 11696.34 & 0.34 \\ 
        WFC3IR F125W & 12503.91 & 0.30  \\ 
        WFC3IR F140W & 13983.36 & 0.256 \\ 
        WFC3IR F160W & 15437.71 & 0.22  \\ \hline
    \end{tabular}
\end{table}

\subsection{Star-Galaxy Separation}
 Without color information for all SKYSURF fields, only object sizes and single filter magnitudes can be used for star-galaxy classification. For star-galaxy separation we use the apparent magnitude and \textit{R50} parameters provided by \textsc{ProFound}. \textit{R50} is defined as the radius in arcseconds within which half of an object's integrated flux is contained. Figure \ref{fig:fig_sgs} shows the apparent magnitude versus $\mathrm{\log_{10} (R50)}$ plane for the WFC3IR F125W filter. The stars are shown as blue points and follow a nearly constant $\mathrm{\log_{10} (R50)}$ versus magnitude relationship below $\approx$ AB 20 mag. Between AB 20-22 mag, we fit a double Gaussian model in $\mathrm{\log_{10} (R50)}$ to the histogram of $\mathrm{\log_{10} (R50)}$ as shown in Figure \ref{fig:r50_hist} with the dividing boundaries marked. The stars form a narrow locus until $\lesssim$ AB 20 mag when their PSF begins to over-saturate the detector and their measured sizes increase.
 We use the width of the normal distribution occupied by the stars to define a gentle slope starting from AB 20.25-21.75 mag and ending at the mag boundary at which the populations blend together at faint magnitudes. While the minimum $\mathrm{\log_{10} (R50)}$ size between the two populations shown in Figure \ref{fig:r50_hist} occurs at -0.8, visual inspections of objects fainter than AB 21.25 mag confirm that objects between the displayed boundary and the galaxy locus confirmed that these objects are galaxies. Star-galaxy separation is done before performing galactic extinction corrections. At bright magnitudes we define a steeper slope from the size-magnitude relationship of the stars that forks upwards when the stars begin to over-saturate the detector. This is done independently for each filter. We also use the measured width of the stellar population, $\sigma$, to define a boundary between real objects and artifacts, which are denoted by green points in Figure \ref{fig:fig_sgs}.
 
 Visual inspections suggest that objects with sizes less than $1\sigma$ below the median $\mathrm{\log_{10}(R50)}$ value of the stellar locus are artifacts and thus we exclude them for all purposes in this work. The truncation point of star galaxy separation is derived by attempting to fit a double-Gaussian model to the $\mathrm{\log_{10}(R50)}$ parameter in each half-magnitude bin. Once the double Gaussian model is no longer successful due to the blending of the populations we stop attempting to separate stars from galaxies. This typically occurs at AB $\mathrm{\gtrapprox} 25.5$ mag for ACSWFC and WFC3UVIS. This occurs at slightly brighter magnitudes, $\mathrm{\gtrapprox} 24.5$ mag, for WFC3IR given its lower resolution. Note that at this flux level the galaxy number density far outnumbers that of the stars. We also choose to mask all objects brighter than AB 15 mag given the risk of fragmentation during source detection. The results of these selections are shown in Figure \ref{fig:fig_sgs} with our galaxy sample shown in red, stars and objects brighter than AB 15 mag shown in blue, and artifacts shown in green.

\begin{figure*}
\vspace*{0.000cm}
\includegraphics[width=15cm, height = 15cm]{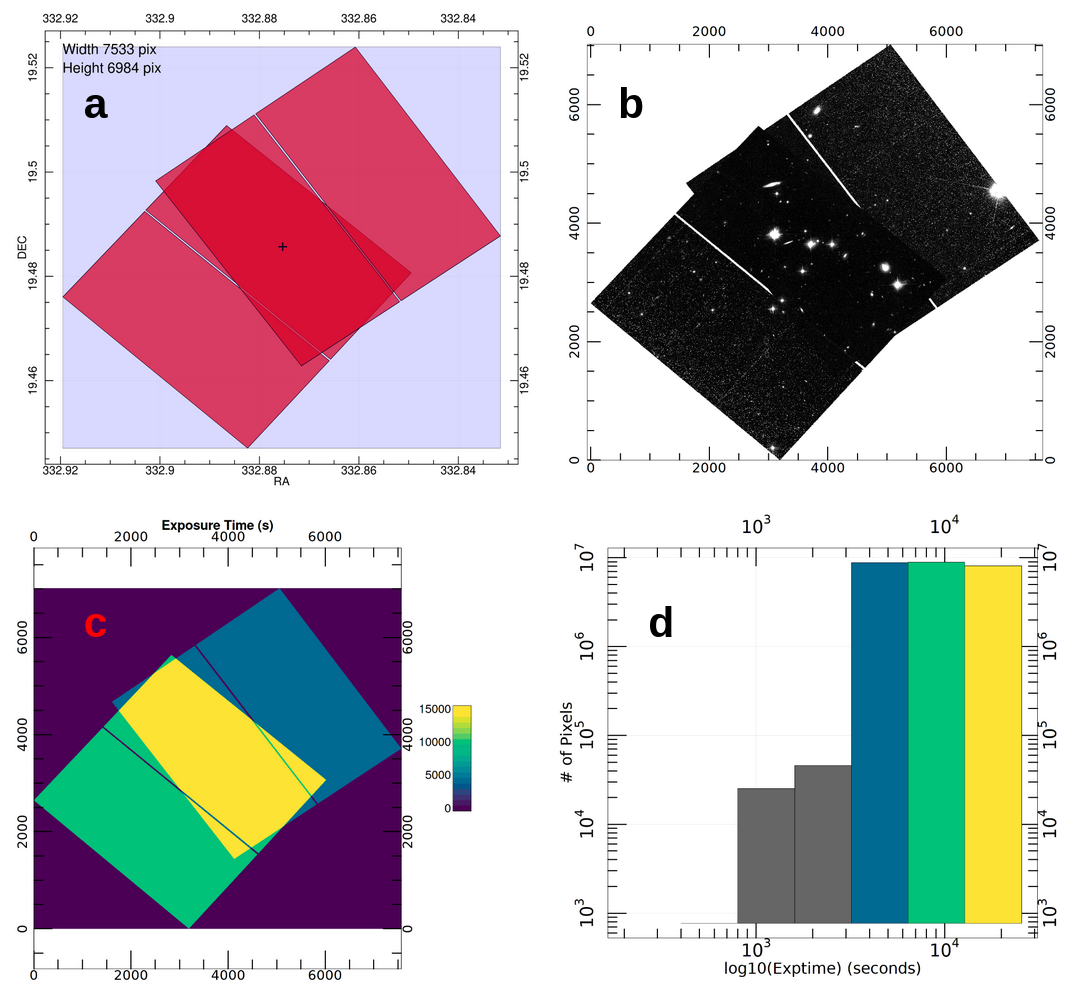}
\caption{ A four panel figure showing an example of how we split mosaics by exposure time. \textbf{(a.)} The footprint for the mosaic showing the outline of each image extension. \textbf{(b.)} The final mosaic. \textbf{(c.)} The exposure time map for the mosaic, with exposure time in seconds displayed on the color bar to the right. \textbf{(d.)} A histogram showing the number of pixels in each exposure time bin. The bars are color-coded to match the exposure time map. The exposure time ranges covering $\lessapprox$ 0.2\% of the area are shown in gray given they cannot be displayed at this resolution and are discarded, see Section \ref{sec_3.4} 
\label{fig:fig_expmap}}
\end{figure*}

\subsection{Bright Star Masking}

Manually inspecting fields and their associated segmentation maps also revealed that \textsc{ProFound} identifies diffraction spikes and over-saturated regions associated with bright stars as objects. To combat this problem and flag potential diffraction spikes, we define a circular mask around stars within which all objects are flagged. The radius of this mask was determined by inspecting the ProFound segmentation maps and measuring the sizes of diffraction spikes detected by ProFound for stars of different magnitudes.

The results are shown in Figure \ref{fig:mask_radius}, where our derived relationship is $\mathrm{\log_{10}(M)} = a\rm mag_{AB} + b$, where M is the $\mathrm{\log}$ size mask radius in pixels as a function of apparent magnitude with slope, a, and y-intercept, b. 

This formula is used to define a circular region around stars to mask and exclude. Objects falling within these regions are flagged in the catalog as they could be diffraction spikes or have their photometery affected by the presence of the nearby star. To adjust for this in post processing, we re-calculated the area of the image after drawing circular masks around flagged objects and removing the enclosed pixels from the area calculation for each frame. We use the apparent magnitudes as measured by ProFound in this process ($\mathrm{mag_{ProFound}}$). Stars with $\mathrm{mag_{ProFound} > 19}$ mag are faint enough that \textsc{ProFound} does not deblend diffraction spikes into secondary objects and the star does not significantly affect the photometry of adjacent objects. We performed this for the WFC3IR F160W, WFC3UVIS F814W, and ACSWFC F814W filters, with the values for slope, a, and y intercept, b, shown in Table \ref{tab:tab4}. The resulting relationships between mask size and magnitude are used for all other filters in the corresponding instrument (WFC3IR, WFC3UVIS, and ACSWFC). Using a well-studied and defined HST point-spread-function mask is not feasible for SKYSURF as in regions where frames taken at different orientations are stacked, multiple sets of diffraction spikes and other associated issues are seen as in Figure \ref{fig:posang}. 
\begin{table}

\caption{ The values defining the linear relationship between mask radius in pixels $\mathrm{log_{10}}(M)$ and $\mathrm{mag_{ProFound}}$.
 \label{tab:tab4}}
    \begin{tabular}{ccc}
    \hline
        Filter & Slope (a) & Y-intercept (b) \\ \hline
        ACSWFC F814W &  -0.165 & 4.74 \\ %\hline
        WFC3UVIS F814W & -0.146 & 4.59  \\ %\hline
        WFC3IR F160W &  -0.137 & 3.99 \\ \hline
    \end{tabular} 
\end{table}

\subsection{Completeness Corrections}
\label{sec_3.4}
The nature of SKYSURF means that different regions in the same mosaic reach different magnitude limits as exposure times can vary significantly within individual mosaics. Thus, completeness limits for object counts cannot be defined by selecting the peak of the combined number counts within a mosaic. Without accounting for exposure time, the counts would gradually roll over (flatten) as sub-regions within their mosaic reach their completeness limits. We therefore developed a strategy using exposure time maps for each mosaic as shown in Figure \ref{fig:fig_expmap}. We select exposure time bins which increase in width by a factor of 2 beginning with 200 seconds (the minimum exposure time for frames comprising the SKYSURF database). 

For each exposure time bin, we calculate the combined area across all single \& multi visit mosaics and generate combined object counts for each exposure time region. Only exposure time regions covering more than 10\% of the unmasked area in a mosaic are considered for object counting purposes. We perform this operation in catalog space by adding the exposure time for each object as an additional column along with the unmasked area associated with that specific sub-region. We then collate objects with similar exposure times in the catalog and sum the pixel area from the final catalog containing all fields. 
Figure \ref{fig:fig_expmap} shows an example mosaic demonstrating this process from the WFC3UVIS F814W filter. In Table \ref{tab:tab5} we show the corresponding completeness limits, area coverage, and completeness limits for the entire WFC3UVIS F814W filter mosaic set.

\begin{figure*}
\vspace*{0.000cm}
\includegraphics[width=14.6cm, height = 22cm]{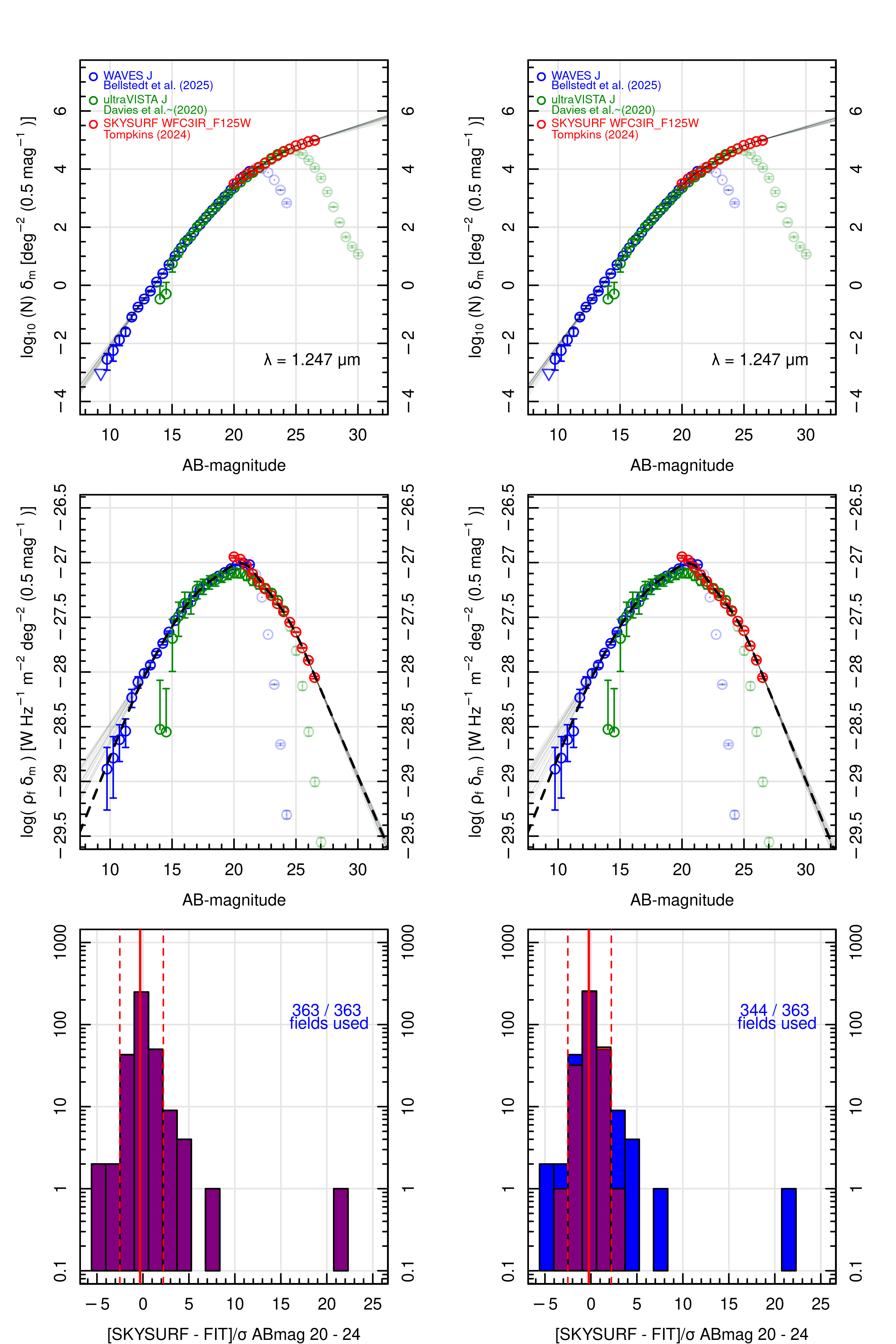}
\caption{ An example from the WFC3IR F125W filter demonstrating our methodology for removing over dense fields. On both sides the panels are as follows. (Top) The galaxy number counts shown in the traditional fashion before (left) and after (right) over-dense fields are trimmed. (Middle) The galaxy number counts adjusted to show flux density per magnitude bin as a function of magnitude before and after over dense fields are trimmed. (Bottom) A histogram showing the mean value of the ${(Data - Fit) / Error}$ quantity described in the text for each individual field. The original distribution is shown in blue while the distribution at the current iteration is overlain in purple. Transparent data points in existing data are affected by incompleteness and are not included in any fitting.}
\label{fig:fig_clip}
\end{figure*}

\begin{table*}
    \centering
                 \caption{A sample of measured completeness limits from different exposure bins from WFC3UVIS F814W. Regions with an exposure time of zero outside the borders of existing data are not included. The fraction column is simply defined as the fraction of the total area for all mosaics in WFC3UVIS F814W covered by each exposure time range.
    \label{tab:tab5}}
    \begin{tabular}{cccccc}
    \hline
        Exptime Range (s) & Median Exptime (s) & Area $\mathrm{deg^2}$ & Fraction & Completeness Limit (ABmag) \\ \hline
        400-800 & 780 & 0.154 & 0.159 & 24 \\ % \hline
        801-1600 & 1313 & 0.172 & 0.177 & 24 \\ %\hline
        1601-3200 & 2508 & 0.460 & 0.475 & 24.5 \\ %\hline
        3201-6400 & 4931 & 0.135 & 0.139 & 25 \\ %\hline
        6401-12800 & 7420 & 0.022 & 0.023 & 26 \\ %\hline
        12801-25600 & 18690 & 0.021 & 0.0214 & 26 \\ %\hline
        25601-51200 & 27250 & 0.006 & 0.006 & 27 \\ \hline
    \end{tabular}
\end{table*}

\begin{figure*}
\vspace*{0.000cm}
\includegraphics[width=1\textwidth]{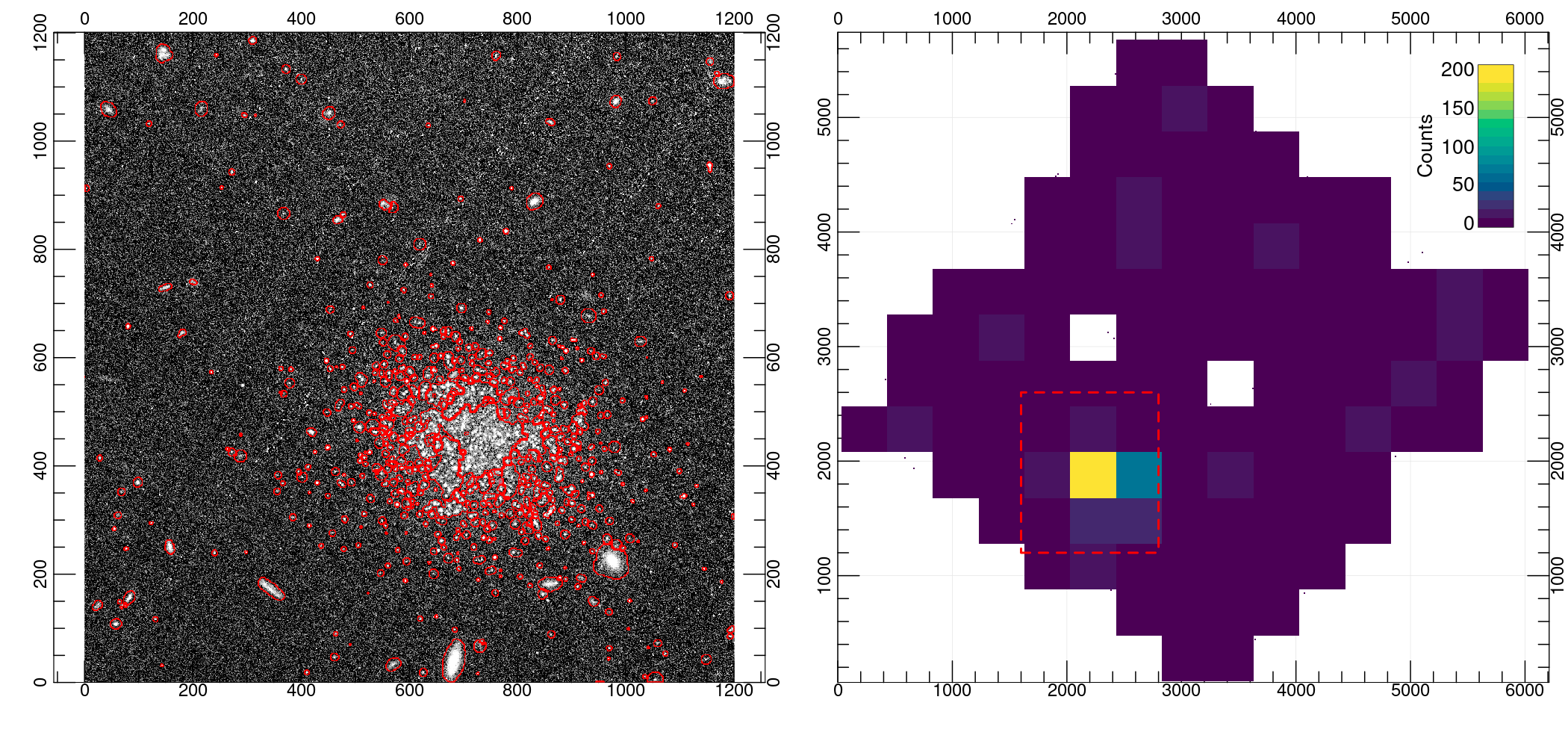}
\caption{ (Left) A zoomed in 1200x1200 pixel region selection showing a dwarf galaxy found in one of our ACSWFC F606W images. All objects outlined in red are classified as galaxies. Objects not outlined are classified as defects or stars. (Right) A 2-D histogram showing the number density of objects identified as galaxies in each 400x400 pixel box across the entire mosaic, with the corresponding dwarf galaxy clearly visible as the yellow region towards the bottom left. Regions outside of the north-aligned mosaic are left empty where no image data is present. The region highlighted in the selection shown on the left is enclosed in the red box shown on the right.
\label{fig:fig_dwarfgal}}
\end{figure*}

\section{Addressing HST's Observing Biases}
\label{Section_4}
    While the SKYSURF database covers a significant area, this area is not necessarily representative of the Universe at large. Cluster and nearby galaxies compose a significant fraction of the observations performed by HST and thus SKYSURF mosaics are prone to significant bias. We implement three strategies to address this. First we flag and ignore observing programs targeting known globular clusters or nearby galaxies by using the target name from the FITS header. Targets with names such as M31, 47TUC, and NGC were rejected. However, many proposals do not provide a target name and thus this is not the only method employed for sorting data. 

Secondly, we turn to statistical methods to remove dense cluster fields and under dense fields with either processing errors or due to observations of targets such as nebulae or nearby large galaxies that occupy a significant portion of the field of view. To remove such fields, we first create galaxy number counts from the catalog for each filter. We then perform a zero point correction to move both WAVES and DEVILS data to the chosen HST filter. This is done using the \textsc{FilterTranBands} function in \textsc{ProSpect} software package \citep{Robotham_2020}. For a detailed discussion of the magnitude zero-point corrections see Section \ref{section_5.3}.

From this we fit a spline to the number counts and the energy curve, see the dashed line fit in Figure \ref{fig:fig_clip} (centre). This represents our reference best fit line for that filter. The energy curve shows the flux density contained in each galaxy number count bin. From AB 20-24 mag we take the difference between the reference fit to the energy curve and the energy curve data points for each individual SKYSURF field. The mean value of this quantity \textit{for each field} is recorded and shown in the histograms in Figure \ref{fig:fig_clip} (bottom). A 3$\sigma$ clip is performed to remove outlier fields, and the entire process is repeated until the sigma clipping converges when no additional fields are clipped. Figure \ref{fig:fig_clip} (centre) shows the effectiveness of the sigma clipping. With the left panel showing a disagreement with the WAVES data but agreement after removing over-dense fields. We also discard SKYSURF number counts at magnitudes brighter than AB 19.75 mag as brighter counts are far more robust from WAVES. However, for the F140W filter the entire magnitude range is used given the significant magnitude zero-point corrections between these filters and the WAVES ground-based Y\&J band data.

HST data is also prone to over-densities caused by excessive fragmentation of nearby galaxies during source detection and especially nearby low surface brightness objects, dwarf galaxies, globular clusters, and in some instances faulty readouts. While most of the area in such fields remains usable, these fields may contain unrealistic object densities at faint magnitudes. We thus elect to perform a third filtering step. We measure the number of objects brighter than AB 26 mag classified as galaxies in subregions (400x400 pixels for ACSWFC and WFC3UVIS, 100x100 for WFC3IR. The highest object density subregion for each field is stored. We removed fields containing subregions more dense than that 99th percentile of the total subregion density distribution. An example of a problematic field which we are able to identify this way is shown in Figure \ref{fig:fig_dwarfgal}.

\section{Galaxy Number Counts and the IGL}
\label{section_5.0}
\begin{figure*}
\vspace*{0.000cm}
\includegraphics[width=1\textwidth]{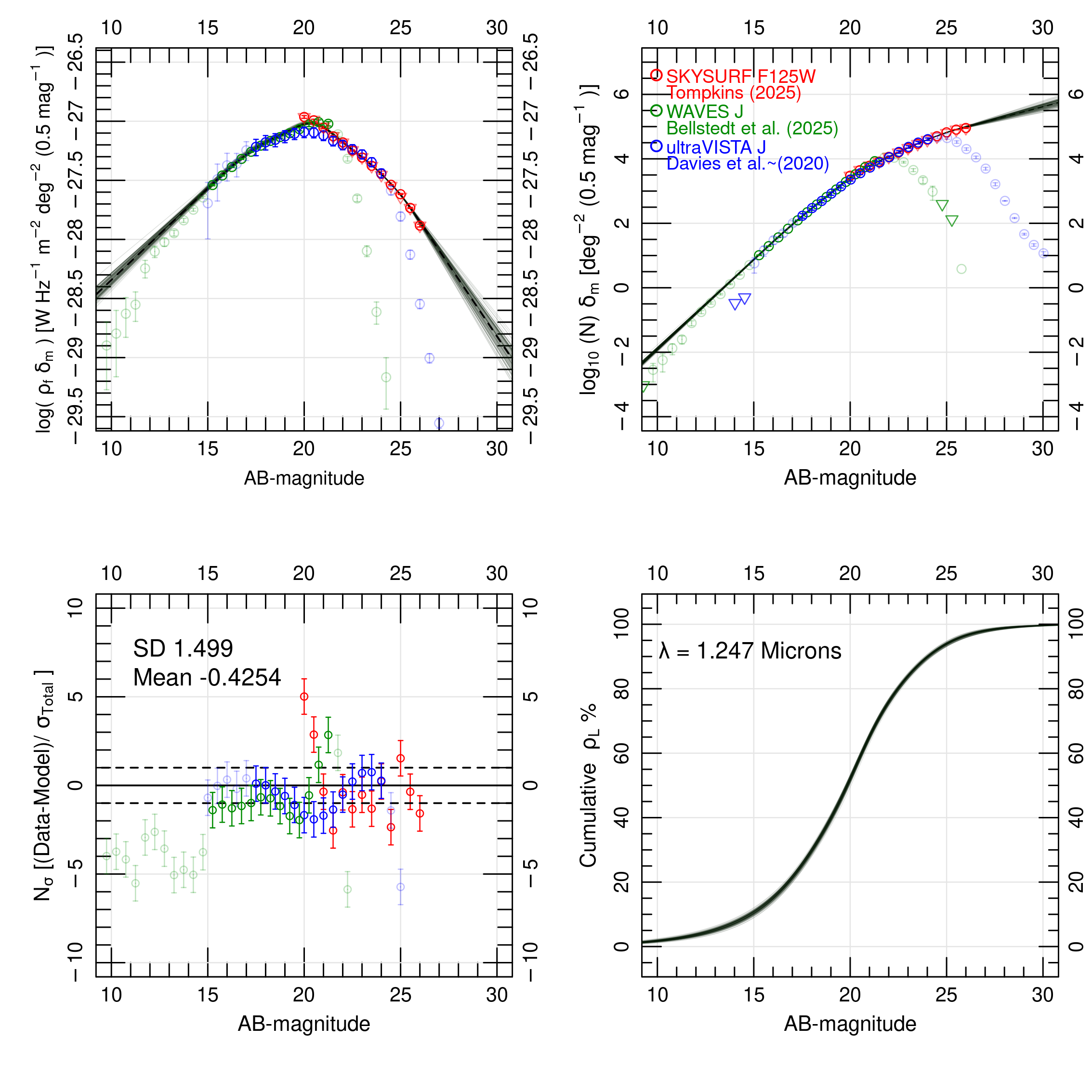}
\caption{(Top Left) The galaxy number counts adjusted to show the contribution of each
magnitude bin to the IGL. Transparent points are not included in the fitting but still displayed. Downward facing red triangles show where the unfiltered HST number counts lie with respect to the final data and are not used in fitting. The dashed line is the mean of all 10,000 Monte-Carlo iterations. (Top Right) The differential galaxy number counts. (Bottom Left) A goodness of fit metric showing $\mathrm{Data - Model}/\sigma_{Poisson}$, where $\mathrm{\sigma_{Total}}$ defines the weight on each data point. The spline fits to the data include the systematic uncertainties introduced by cosmic variance and magnitude zero point errors. See Sections \ref{Section_5.1} - \ref{sec:sec_5.5} for a discussion of error estimates.
\label{fig:f606w_igl}}
\end{figure*}

Here we discuss and present the SKYSURF galaxy number counts for the HST filter sets used in this analysis. In Figure \ref{fig:f606w_igl} we show the galaxy number counts (top right), integrated galaxy light (top left), goodness of fit (bottom left), and the normalized sum of the IGL (bottom right) for the F125W filter as an example. To convert from galaxy number counts to a measurement of the EBL, we use the following standard equations. These equations are used with the AB magnitude system. This mathematical derivation of the IGL from galaxy number counts follows the same methodology as \cite{Driver_2016, Koushan_2021}.

\begin{equation}
  IGL (W Hz^{-1} deg^{-2})\ \     \propto  \int_{-\infty}^{\infty} N_{m}\times(3631\times10^{-26})\times10^{-0.4m} \,dm\
\label{eqn:eqn1}
\end{equation}

The output of Equation \ref{eqn:eqn1} is then multiplied by the following constants to obtain the standard EBL units. 

\begin{equation}
  EBL ( nW m^{-2} sr^{-1}) = IGL \times \lambda(\frac{180}{\pi})^2 \times10^9 
\label{eqn:eqn2}
\end{equation}

where $\mathrm{N_m}$ is the number of galaxies in a magnitude bin between \textit{m} and $ m + \delta m$.

The best spline fit to the data in Figure \ref{fig:f606w_igl} is shown as a dashed black line alongside a subset of 100 Monte-Carlo spline fits to the data as described in Section \ref{Section_5.1}. The splines shown in Figure \ref{fig:f606w_igl} use 10 degrees of freedom. The weight for each data point is set solely by the Poisson uncertainty, $\mathrm{1/\sigma_{Pois}}$. These weights are maintained when testing different sources of uncertainty. In Figure \ref{fig:f606w_igl} Poisson, cosmic variance, and magnitude zero point uncertainties are tested simultaneously. The implementation of each is described in the following subsections. The same figures are shown for all other bands in the appendix, see Figures \ref{fig:uebl}, \ref{fig:gebl}, \ref{fig:r_ebl},
\ref{fig:ri_ebl}, \ref{fig:iz_EBL}, \ref{fig:y_EBL}, \& \ref{fig:jh_EBL}. We report an extrapolated IGL value where we integrate between AB $\mathrm{\pm}$ 100 mag, which is indistinguishable from integrating between $\mathrm{\pm} \infty$. These values are reported in Table \ref{tab:tab6}. We also report lower limits in Table \ref{tab:tab7} where we integrate to AB 15 mag. We do not include WAVES galaxy number counts brighter than AB 15 mag in the fitting to force a Euclidean behaviour of the bright number counts. This is maintained throughout all IGL measurements. When our galaxy number counts do not reach faint magnitudes, and/or the slope of the faint number counts is poorly defined, there can be significant differences between the extrapolated and non-extrapolated IGL. For example, in the F625W filter where the extrapolated IGL value of $\mathrm{9.07  \pm 0.40 nW m^{-2} sr^{-1}}$, and the lower limit ie $\mathrm{7.00 nW m^{-2} sr^{-1}}$. We thus elect to report a conservative mean value in with an uncertainty stretching between the $\mathrm{\pm 1\sigma}$ on the respective bounds. These values are given in Tables \ref{tab:tab8} \& \ref{tab:tab11}.

In this work we face four sources of uncertainty which affect the IGL. Poisson statistical error, cosmic variance (CV), magnitude zero point differences between bandpasses, and fitting errors. We now discuss each of these errors in detail below.
\subsection{Poisson Statistical Errors}
\label{Section_5.1}
 For all spline fitting we use a $\mathrm{\frac{1}{\sigma}}$ weight on each data point where $\sigma$ is the typical $\mathrm{\frac{1}{\sqrt{N}}}$ (where N is the number of galaxies in each half magnitude bin) Poisson statistical uncertainty on each data point. To the Poisson error we also add an error based on our field selection criteria to account for the significant sampling bias in our HST data. For each galaxy magnitude bin we take the difference between the untrimmed and trimmed galaxy number counts shown in Figure \ref{fig:fig_clip} left \& right, respectively. The square root of this value is added in quadrature to the Poisson error. The untrimmed HST galaxy number counts are shown in Figure \ref{fig:f606w_igl}, as downward facing red triangles, representing absolute upper limits. These are not used in fitting and illustrate the effectiveness of our field selection criteria. When deriving the Poisson error, the magnitude zero-point corrections provided in Tables \ref{tab:tab12} and \ref{tab:tab13} are used without any uncertainty in order to isolate the Poisson error. We independently shift each data point according to a normal distribution with a mean of 1 and standard deviation given by $\mathrm{1/\sqrt{N}}$, the fractional Poisson statistical uncertainty. Then a spline is fit to the perturbed data and the IGL value is obtained using the same integration process described above in Equations \ref{eqn:eqn1} \& \ref{eqn:eqn2}. This process is repeated 10,000 times where the median, 17th and 83rd percentiles of the corresponding distribution are used to define a percentage uncertainty due to Poisson statistical error only.
 
 %We independently shift each data point by its 1$\sigma$ uncertainty, 

\subsection{Cosmic Variance Errors}
\label{Section_5.2}
Calculating a cosmic variance error for our HST number counts is non-trivial because the area changes with depth as seen in Table \ref{tab:tab5}. As such, we calculate a fractional CV error for each SKYSURF galaxy count bin. This calculation is done using the \textsc{cosvararea} function in the \textsc{celestial} R software package\footnote{https://github.com/asgr/celestial/}. The function takes a number of independent sight lines, area, and redshift range to calculate a volume and return the CV error as a percentage using the formulation from \cite{Driver_2010}. We assume each SKYSURF field is fully independent for the purposes of calculating the number of independent sight lines. We obtain the redshift range from the DEVILS redshift catalogs in the D03 and D10 fields. We take the 17th and 83rd percentile of the redshift distribution for objects from AB 18-24 mag as the inputs for \textsc{cosvararea}. This is done separately for each band in the \textit{ugrizYJH} set given that the redshift distribution changes with wavelength. We obtain the CV uncertainty for both WAVES and DEVILS using the \textsc{cosvararea} function as well, where the value is constant regardless of magnitude given all magnitude bins cover the same area and number of independent sight lines.

Cosmic variance is a systematic uncertainty that affects each survey independently. We do not include the CV uncertainty in the weight assigned to each data point in the spline fit and maintain the weights as $\mathrm{1/\sigma_{Pois}}$. To properly address CV as a systematic error, we shift each entire set of number counts up/down according to their fractional cosmic variance errors again following a normal distribution with a mean of 1 and standard deviation of $\mathrm{\sigma_{CV}}$. As SKYSURF number counts are collated based on exposure time, the CV uncertainty at the faint end is larger than at the bright end, which is reflected in the relatively large spread of splines at the faint end in Figure \ref{fig:f606w_igl}. This process is repeated 10,000 times where the median, 17th and 83rd percentiles of the resultant distribution are used to define a percentage uncertainty due to CV error.

%We again perform Monte-Carlo analysis by shifting the number counts up/down according to their fractional cosmic variance errors again following a normal distribution with a mean of 1 and standard deviation of $\mathrm{\sigma_{CV}}$. As noted, the data from each survey (WAVES, DEVILS, and SKYSURF) are shifted systematically and independently for each data set. An error weighted spline is fit to the data and integrated. This process is repeated 10,000 times where the median, 17th and 83rd percentiles of the resultant distribution are used to define a percentage uncertainty due to CV error.

\begin{figure}
\vspace*{0.000cm}
\includegraphics[width=9cm, height = 9cm]{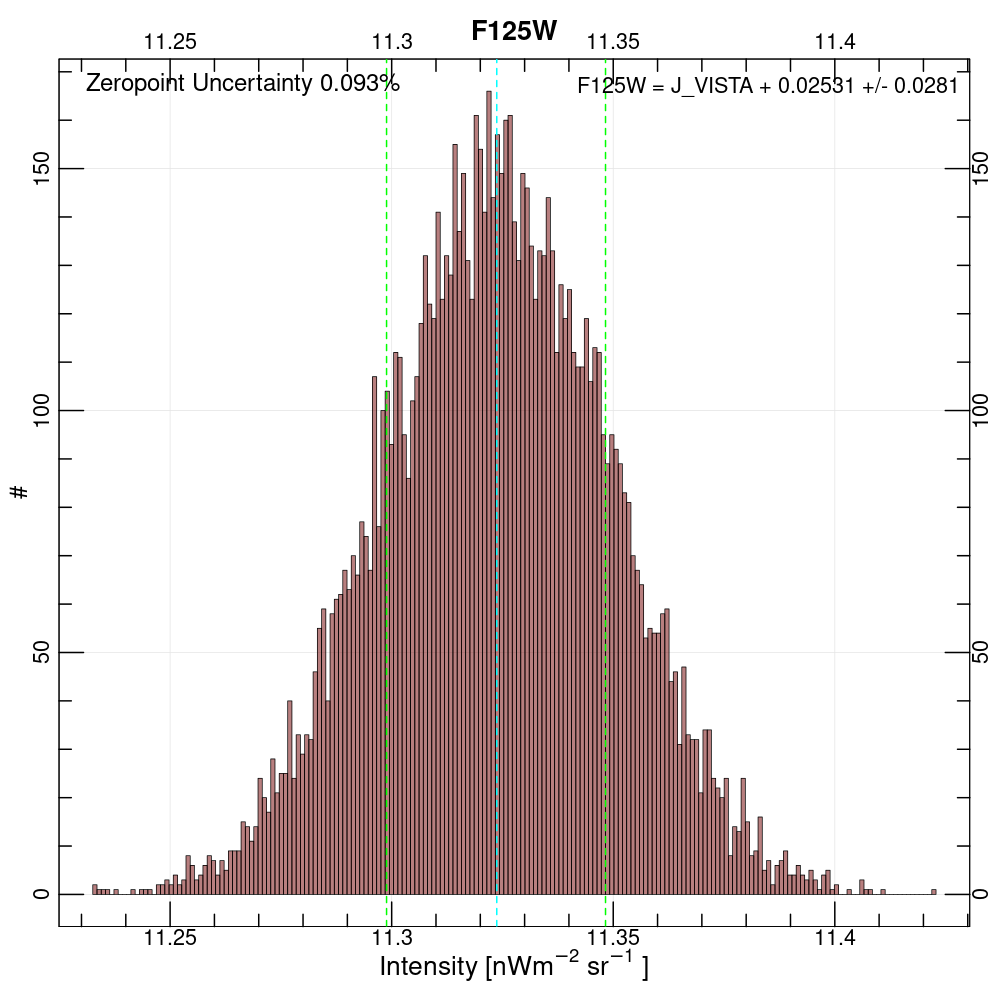}
\caption{ A distribution of 10,000 IGL values obtained from implementing the zero point corrections derived in Section \ref{section_5.3}. The median value is
marked by the vertical blue line with the 17th and 83rd percentiles used to drive the error shown in green. The uncertainty introduced by this process is quoted in the top left, while the magnitude zero point correction between WFC3IR F125W and VISTA-J is displayed in the top right.
\label{fig:zpt}}
\end{figure}

\begin{figure}
\vspace*{0.000cm}
\includegraphics[width=9cm, height = 9cm]{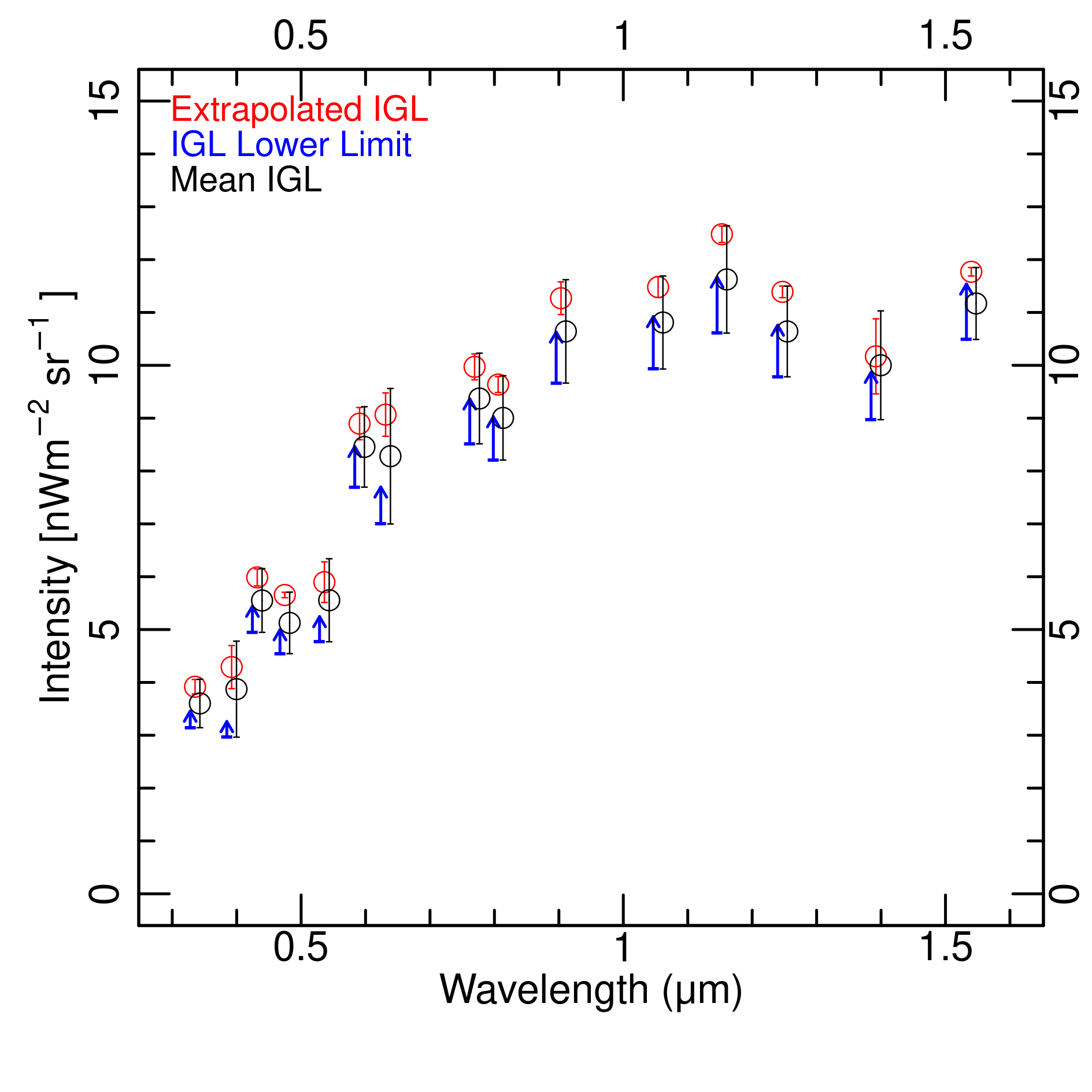}
\caption{ Our measurements of the IGL displayed on their own. Extrapolated IGL measurements are shown in red with lower limits shown as upward facing arrows. We also show the mean between the extrapolated IGL and lower limits with the respective errors spanning the possible 1$\sigma$ range of uncertainties on both measurements. The mean IGL data points are shifted slightly to the right for readability and listed in Table \ref{tab:tab7}. The lower limits (blue arrows) have also been shifted slightly to the left for readability}. We comment on discontinuities and discrepancies between our measurements in Section \ref{sec:sec_5.6}.
\label{fig:skysurf_only}
\end{figure}

\subsection{Magnitude Zero Point Corrections}
\label{section_5.3}
Magnitude zero point (ZPT) errors will also affect the data systematically. We ZPT correct number count data from WAVES and DEVILS onto the HST filter system which produces a systematic magnitude offset for both surveys. We calculate the color correction terms using the ProSpect \textsc{filterTranBands} function. As input, the function takes a filter, input filter, reference filter, and redshift range.  \textsc{ProSpect} comes with a long list of pre-calculated apparent magnitudes in each filter set which are used as inputs for the following equation \\

   $\mathrm{mag_{out} = mag_{in} + \alpha\times(mag_{in} - mag_{ref}) + \beta \pm \sigma}$   (2) \\

\label{eqn:eqn3}
where $\alpha$, $\beta$, and $\sigma$ are terms returned by \textsc{filterTranBands}. To determine the best solution for $\alpha$ and $\beta$, \& $\sigma$, the quantity $\mathrm{2\alpha^2} +\beta^2 + \sigma^2$ is minimised for all possible linear combinations of the reference and input facility filters, see \cite{Robotham_2020b}. $\mathrm{mag_{in}}$ is the apparent magnitude from the input filter, $\mathrm{mag_{ref}}$ is the magnitude for a reference filter, and $\mathrm{mag_{out}}$ is the desired magnitude in the target filter. For example, when correcting WAVES Y-band data to HST's WFC3IR F105W filter, the target, input, and reference filters are WFC3IR F105W, Y VISTA, and H VISTA, respectively. Using 1,000 of the \textsc{ProSpect} provided magnitudes, the final zero-point correction values and their uncertainties are provided in Tables \ref{tab:tab12} and \ref{tab:tab13}.

Since we merge number counts from filters that exist in both ACSWFC and WFC3UVIS such as F606W, we take the mean of the two correction values and their respective uncertainties to correct ground based data to the HST filter system. To derive the magnitude zero point uncertainties, we shift the WAVES and DEVILS number count magnitudes according to a normal distribution with a mean given by the magnitude offset and the standard deviation as the magnitude offset error. We do this 10,000 times to obtain a spread of IGL values and use the 17th and 83rd percentiles to obtain an uncertainty, see Figure \ref{fig:zpt}. The magnitude zero point corrections and their uncertainties are listed in the appendix.

 \subsection{Fitting Error}
 \label{sec:sec_5.4}

 For all of the spline fitting in this work, we use a spline with 10 degrees of freedom by default. We choose to introduce a fitting error by fitting the data with splines of order 8-12 and taking the best fit IGL value for each order of spline. This is consistent with what has been done in previous works, see \cite{Driver_2016, Koushan_2021}. To stop the spline fits from diverging at the bright we ignore any galaxy number count bins with 50 or fewer galaxies. See Figure \ref{fig:gebl} where bright galaxy number count bins in WAVES-g band (green points) could introduce diverging spline fits. For the F140W (Figure \ref{fig:jh_EBL}, bottom) filter where no ground-based data is available we simply ignore the brightest two magnitude bins which otherwise cause the splines to diverge.

\subsection{Dependencies on Ecliptic and Galactic Latitude}
\label{sec:sec_5.5}
We also test our counts for selection biases that may arise based on ecliptic latitude or galactic latitude. HST's standard processing steps and our own have performed sky subtractions, but we verify that when splitting our sample by ecliptic latitude, there are no significant changes. As Zodiacal light intensity depends strongly on ecliptic latitude, if sky subtraction imposes a systematic error it would be reflected in the IGL measurements when isolating SKYSURF fields in different ecliptic latitude bands. For both galactic and ecliptic latitude, we test the robustness of our sample by measuring the IGL in $\ang{20}\  \&\ \ang{15}$ wide bands, respectively. As the small number of fields in different regions can introduce divergent integrations, we elect to use lower limits by integrating across the completeness limits of our data as was done in Table \ref{tab:tab7}.

\begin{figure}
\vspace*{0.000cm}

\includegraphics[width=0.5\textwidth]{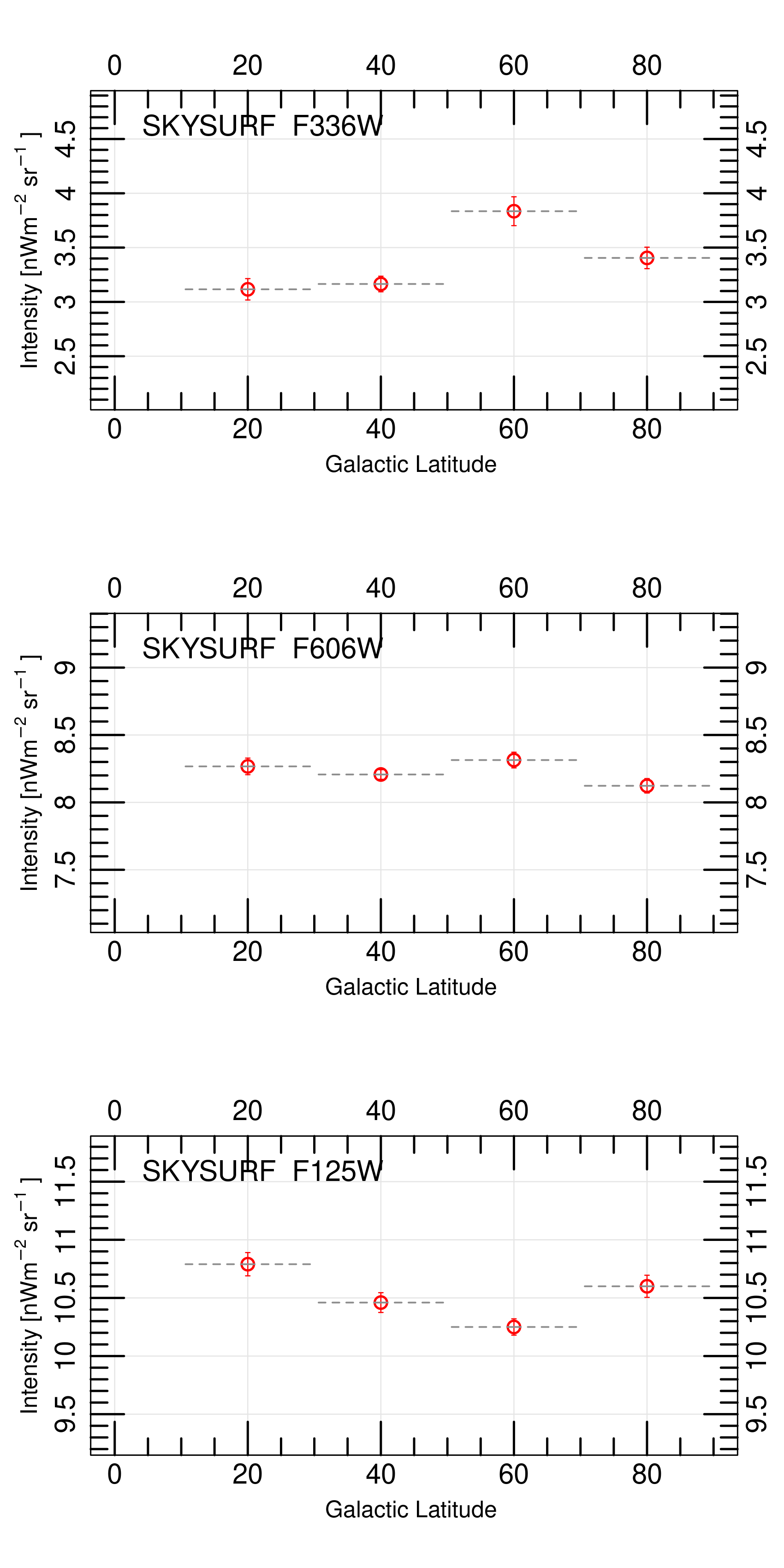}
\centering
\caption{ IGL measurements taken in galactic latitude bands from 10-30, 30-50, 50-70, and 70-90 degrees. Where extrapolations begin diverge errors are significantly higher.
\label{fig:latitude}
}
\end{figure}

 SKYSURF excluded all fields within \ang{10} of the galactic plane. We further test our IGL measurements by evaluating the IGL in \ang{20} wide galactic latitude bands from 10-30, 30-50, 50-70, and 70-90 degrees, respectively, Figure \ref{fig:latitude} The width of each galactic latitude selection is denoted by the dashed grey line through each point. During this process, we continue to include Poisson and Cosmic Variance, and magnitude zero-point uncertainties to account for the decreased number of fields. We find no significant dependencies on galactic latitude as shown in Figure \ref{fig:latitude}. This indicates that galactic extinction corrections do not introduce a systematic error at lower galactic latitudes. The number density of stars also increases at low galactic latitude, and thus we can deduce that our star-galaxy separation does not introduce significant uncertainty to our IGL measurements.

\begin{figure}
\vspace*{0.000cm}

\includegraphics[width=0.5\textwidth]{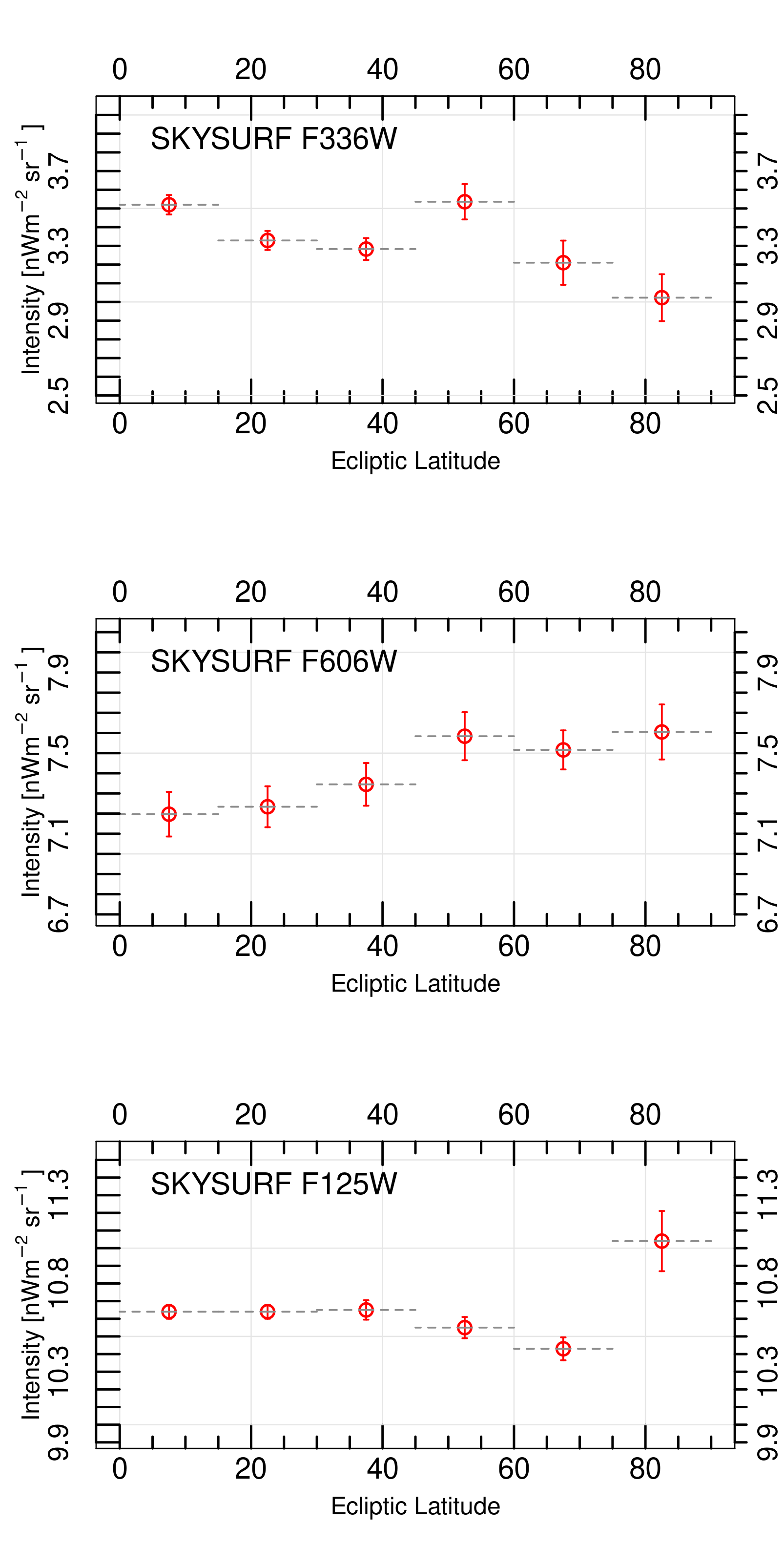}
\centering
\caption{Lower limit SKYSURF IGL values for three HST filters measured in $\ang{15}$ wide bands for different ecliptic latitude ranges.
\label{fig:ecl_latitude}
}
\end{figure}

SKYSURF did not make any selections based on ecliptic latitude, and thus we break our tests into \ang{15} wide intervals starting from zero. The intensity of the Zodical light varies significantly with ecliptic latitude and in principle could bias photometry. However, HST's standard processing pipeline performs initial sky subtraction during calibration \citep{Windhorst_2022}. We perform additional sky subtraction as part of our mosaic building and source detection. In Figure \ref{fig:ecl_latitude} we show IGL measurements in different \ang{15} wide bands. We find no consistent trend with ecliptic latitude and can conclude that HST's and our own Zodical foreground subtraction is sufficient to provide consistent photometry.

\subsection{Further Discussion Of Uncertainties and Biases}
\label{sec:sec_5.6}
We present our final IGL measurements below in Figure \ref{fig:skysurf_only}, where extrapolated IGL data is shown in red, complete with errors, with lower limits displayed as upward facing blue arrows. At most wavelengths where our faint number counts are well constrained, the lower limits on the IGL within the 1$\sigma$ uncertainties on the extrapolated IGL, suggesting that contribution to the optical IGL from galaxies fainter than AB 28.5 mag is small, ($\mathrm{< 2.5\%}$, see \cite{Windhorst_2022}. Where there is a significant difference, we attribute it to a poorly constrained slope in the faint number counts.

We attribute discontinuities in the IGL, such as the jump between the F555W and F606W at $\approx$0.54 \& 0.59 $\mu$m, to the choice of ground-based data sets used in conjunction with HST. The F555W filter falls between the g and r band data from DEVILS and WAVES and given its limited area coverage the IGL measurement for F555W is sensitive to the ground-based data chosen to accompany it. We elect to use g-band given it provides a lower uncertainty. Ideally, v-band data would be used alongside the F555W data though without access to wide area v-band number counts we are forced to use g-band data and accept a much larger uncertainty. Thus the significant increase in the IGL between $\approx$0.54 \& 0.59 $\mu$m can be attributed to the choice of reference data and should not be taken as an indication of any underlying physical process.

We test the assumption that our errors can be added in quadrature to avoid imprecision in our final error budget. We performed all of our error analysis techniques on the data simultaneously, and as expected the results are consistent with adding independent errors in quadrature. Across all bands, the mean contributions from zero-point, cosmic variance, Poisson, and fitting errors are $\approx$ 2.3, 2.0, 0.98, and 1.35 \%, respectively. Though the most significant on average, zero-point correction uncertainties can be attributed to the choice to use 19 HST filters at 15 different wavelengths to sample the COB at more wavelengths.  

 As seen in Tables \ref{tab:tab6} \& \ref{tab:tab7}, HST filters such as the F475W and F625W at pivot wavelengths of 0.475 and 0.631, $\mu$m, respectively, are designed to overlap with ground based \textit{g}\&\textit{r} band facilities. As such the filter transform errors are very small. However, in the final SKYSURF database these filters cover significantly less area and thus their CV uncertainties are much larger. This is why we chose to zero-point correct ground based reference data sets to the HST filter system as this avoided us having to choose which 8 HST filters to use with the \textit{ugrizYJH} filter set from WAVES \& DEVILS. Cosmic variance is unavoidable, and one of the main goals of SKYSURF was to reduce CV errors in faint galaxy number counts by sampling a large area and number of independent sight lines when compared to previous surveys at similar depths. We note the small, but consistent CV uncertainty introduced by WAVES at bright magnitudes. Despite covering $\mathrm{\approx 1068 deg^2}$, the CV error from WAVES is still 
 $\mathrm{\approx 1.8\%}$ and is in part responsible for the cosmic variance error floor where it is used. For the F140W, at a pivot wavelengths of 1.39 $\mu$m, only SKYSURF data is used and thus the cosmic variance uncertainty is lower than 1.5\%. However, without the bright counts provided by WAVES, this introduces significant Poisson statistical error.

\begin{figure}
\vspace*{0.000cm}
\includegraphics[width=8.2cm, height = 8.2cm]{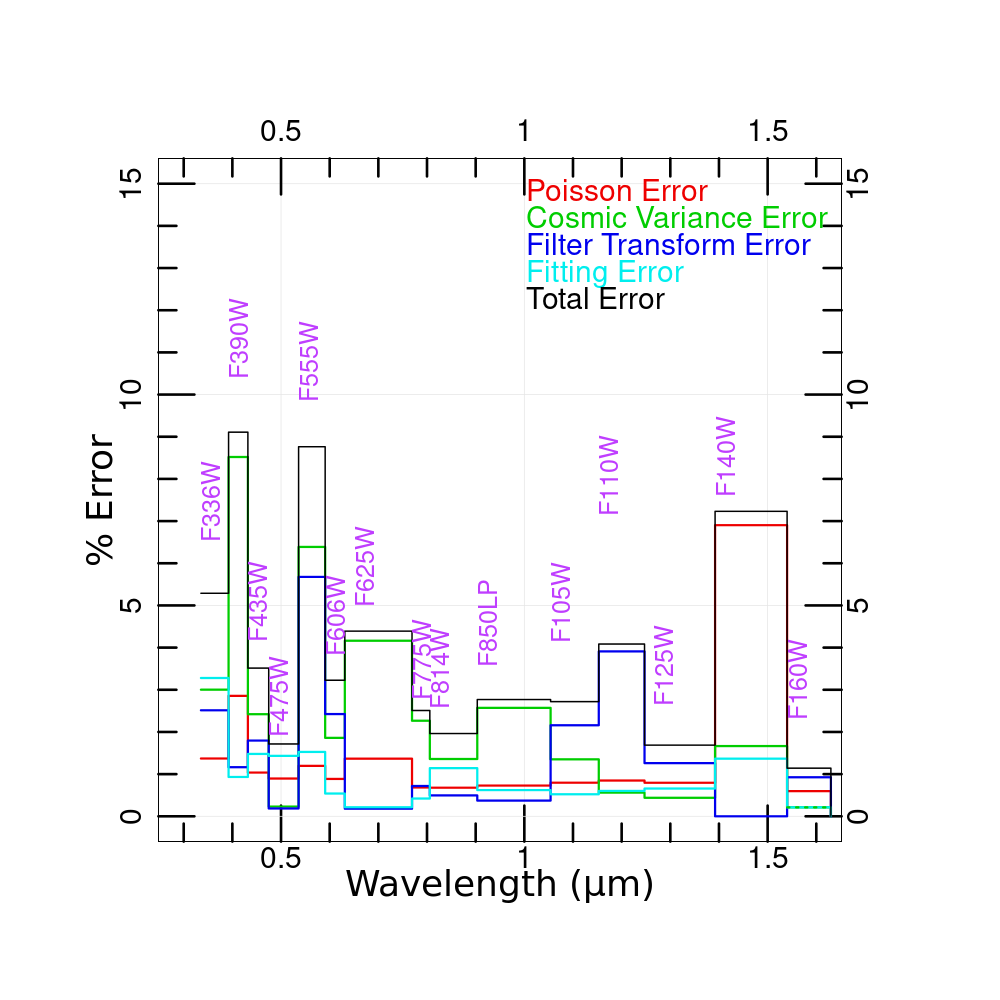}

\caption{ A visual display of the final error budget for the extrapolated IGL values in SKYSURF with ground-based data corrected to the HST filter system. These values are quoted in Table \ref{tab:tab6}.
\label{fig:hst_err}}
\end{figure}

 While the F140W SKYSURF IGL measurement offers a glimpse of the advantages provided by relying on a single facility to provide galaxy number counts, it has the largest uncertainty almost entirely due to Poisson statistical error. Without wide-area surveys to provide bright galaxy number counts, the small number of objects at bright magnitudes yield unavoidable statistical errors (see Figure \ref{fig:hst_err}). Similar to CV uncertainties, Poisson statistical errors can only be reduced by wider area surveys. Future data sets from missions such NASA's Nancy Grace Roman space telescope will cover a broad range of magnitudes and can produce IGL measurements without the need to combine galaxy number counts from different facilities, thus eliminating magnitude zero-point correction uncertainties entirely.

\begin{table*}
        \caption{The extrapolated final error IGL values along with the respective uncertainties displayed a percentage of the IGL. Errors are added in quadrature to calculate a final uncertainty. No ground-based reference data is used with the F140W filter and thus no zero-point error is included. See Figure \ref{fig:hst_err} for a visual depiction of the final error budget.
    \label{tab:tab6}}
    \centering
    \begin{tabular}{ccccccccc}
    \hline
        HST Filter & Pivot Wavelength & IGL Value   & Zero-Point  & CV & Poisson & Fitting Error & Total Error & \\ \hline
     &  $\mathrm{\mu m}$ & $\mathrm{nW m^{-2} sr^{-1}}$ &  \%  &  \% & \% & \% & \% \\ \hline
  F336W & 0.335 & 3.92 & 2.51 & 3.00 & 1.37 & 3.28 & 5.29 \\ 
  F390W & 0.392 & 4.29 & 1.16 & 8.52 & 2.86 & 0.93 & 9.11 \\ 
  F435W & 0.432 & 5.99 & 1.80 & 2.42 & 1.04 & 1.48 & 3.51 \\ 
  F475W & 0.475 & 5.65 & 0.19 & 0.23 & 0.90 & 1.43 & 1.72 \\ 
  F555W & 0.536 & 5.90 & 5.68 & 6.39 & 1.20 & 1.53 & 8.76 \\ 
  F606W & 0.591 & 8.90 & 2.42 & 1.86 & 0.89 & 0.54 & 3.23 \\ 
  F625W & 0.631 & 9.07 & 0.18 & 4.16 & 1.37 & 0.22 & 4.39 \\ 
  F775W & 0.769 & 9.97 & 0.72 & 2.27 & 0.68 & 0.42 & 2.51 \\ 
  F814W & 0.806 & 9.64 & 0.50 & 1.36 & 0.68 & 1.14 & 1.96 \\ 
  F850LP & 0.903 & 11.27 & 0.37 & 2.57 & 0.73 & 0.62 & 2.77 \\ 
  F105W & 1.054 & 11.48 & 2.16 & 1.35 & 0.80 & 0.52 & 2.72 \\ 
  F110W & 1.153 & 12.48 & 3.91 & 0.56 & 0.85 & 0.60 & 4.08 \\ 
  F125W & 1.247 & 11.39 & 1.26 & 0.44 & 0.80 & 0.66 & 1.69 \\ 
  F140W & 1.392 & 10.17 & N/A & 1.67 & 6.90 & 1.37 & 7.23 \\ 
  F160W & 1.540 & 11.77 & 0.93 & 0.21 & 0.60 & 0.21 & 1.14 \\ 
       % 2.15 & 10.03  & 999 & 0.84 & 0.84 \\ \hline
    \end{tabular}
\end{table*}

\begin{table*}
    \caption{The IGL without extrapolation. Without ground-based data the F140W filter does not have counts brighter than AB 15 mag given that objects brighter than this were ignored for all purposes in SKYSURF. 
    \label{tab:tab7}}
    \centering
    \begin{tabular}{cccccccccc}
    \hline
        HST Filter & Pivot Wavelength & ABmag Range & IGL Value   & Zero-Point  & CV & Poisson & Fitting Error & Total Error  \\ \hline
       &   $\mathrm{\mu m}$ & ABmag & $\mathrm{nW m^{-2} sr^{-1}}$   &  \% & \% & \%  & \%  & \% \\ \hline
F336W & 0.335 & 15.25-27 & 3.14 & 3.03 & 0.78 & 0.50 & 0.17 & 3.17 \\ 
  F390W & 0.392 & 15.25-27 & 2.96 & 1.49 & 0.83 & 0.51 & 0.10 & 1.78 \\ 
  F435W & 0.432 & 15.25-28 & 4.95 & 1.91 & 0.70 & 0.40 & 0.07 & 2.07 \\ 
  F475W & 0.475 & 15.25-26.5 & 4.54 & 0.20 & 0.17 & 0.36 & 0.03 & 0.44 \\ 
  F555W & 0.536 & 15.25-27.5 & 4.77 & 5.95 & 1.51 & 0.47 & 0.03 & 6.16 \\ 
  F606W & 0.591 & 15.25-28.5 & 7.70 & 2.47 & 0.44 & 0.36 & 0.06 & 2.53 \\ 
  F625W & 0.631 & 15.25-27 & 7.00 & 0.19 & 0.32 & 0.37 & 0.02 & 0.53 \\ 
  F775W & 0.769 & 15.25-28.5 & 8.51 & 0.77 & 0.74 & 0.42 & 0.09 & 1.15 \\ 
  F814W & 0.806 & 15.25-27 & 8.21 & 0.53 & 0.12 & 0.35 & 0.02 & 0.65 \\ 
  F850LP & 0.903 & 15.25-27.5 & 9.66 & 0.37 & 0.73 & 0.39 & 0.02 & 0.91 \\ 
  F105W & 1.054 & 15.25-27.5 & 9.93 & 2.10 & 0.42 & 0.43 & 0.02 & 2.19 \\ 
  F110W & 1.153 & 15.25-26 & 10.61 & 3.89 & 0.22 & 0.49 & 0.00 & 3.92 \\ 
  F125W & 1.247 & 15.25-26 & 9.78 & 1.26 & 0.19 & 0.48 & 0.04 & 1.36 \\ 
  F140W & 1.392 & 15.25-27 & 8.97 & 0.00 & 0.65 & 1.79 & 0.01 & 1.91 \\ 
  F160W & 1.540 & 15.25-26.5 & 10.49 & 0.91 & 0.14 & 0.41 & 0.04 & 1.00 \\ 
    \end{tabular}
\end{table*}

\begin{table}
  \caption{ The mean IGL values for the HST filter set shown in black on Figure \ref{fig:skysurf_only}. Errors are not quoted as a percentage given the asymmetrical nature of the error bars.
    \label{tab:tab8}}
    \centering
    \begin{tabular}{ccc}
    \hline
       Filter & Pivot Wavelength & IGL Value  \\ \hline
        & $\mathrm{\mu m}$ & $\mathrm{nW m^{-2} sr^{-1}}$   \\ \hline
F336W  &  0.335  & $ 3.60^{+0.46}_{-0.51}$   \\ 
F390W  &  0.392  & $ 3.87^{+0.91}_{-0.94}$   \\ 
F435W  &  0.432  & $ 5.55^{+0.6}_{-0.65}$   \\ 
F475W  &  0.475  & $ 5.13^{+0.58}_{-0.60}$   \\ 
F555W  &  0.536  & $ 5.55^{+0.79}_{-0.88}$   \\ 
F606W  &  0.591  & $ 8.46^{+0.76}_{-0.81}$   \\ 
F625W  &  0.631  & $ 8.28^{+1.28}_{-1.32}$   \\ 
F775W  &  0.769  & $ 9.37^{+0.86}_{-0.94}$   \\ 
F814W  &  0.806  & $ 9.01^{+0.80}_{-0.83}$   \\ 
F850LP  &  0.903  & $ 10.64^{+0.98}_{-1.07}$   \\ 
F105W  &  1.05  & $ 10.81^{+0.88}_{-0.96}$   \\ 
F110W  &  1.15  & $ 11.62^{+1.02}_{-1.13}$   \\ 
F125W  &  1.25  & $ 10.64^{+0.86}_{-0.92}$   \\ 
F140W  &  1.39  & $ 10.00^{+1.03}_{-1.2}$   \\ 
F160W  &  1.54  & $ 11.17^{+0.68}_{-0.73}$   \\ \hline
    \end{tabular}
\end{table}

%\subsection{The Extrapolated IGL and Galaxy Number Counts}

%We now discuss the motivation for and implications of our extrapolated measurements from spline fits. In the absence of yet fainter number count data, extrapolating the slope of the number counts and thus IGL beyond our magnitude completeness limits is necessary to remain fully empirical. However, there is also observable and theoretical evidence to believe that the slope of the number counts should not turn upwards again beyond the completeness limits reached by HST.

%The logarithmic slope of the number counts must drop below the critical value of 0.4 in order for the EBL to be finite \citep{Madau_2000}. There are also cosmological constraints on the contribution of high redshift populations to number counts and thus the EBL. Presented as a surface brightness per unit area, the EBL thus contains information on the total number of galaxies per unit comoving volume. Both the standard '7-3-7' and Planck cosmologies predict that  comoving volume increases rapidly until z$\approx$2.5, after which it once again begins to decrease. When coupled with cosmological dimming, the faint end of galaxy number counts is not extremely sensitive to the nature of populations at high redshift \citep{Manzoni2025}. Thus, any population that would significantly affect the number counts and IGL would have to be at lower redshifts while not driving the slope of the counts above the critical slope of 0.4.

\section{Comparison With Previous COB Studies}
\label{Section_6}

\subsection{Comparison to Previous IGL Studies}
\begin{figure}
\vspace*{0.000cm}
\includegraphics[width=8cm, height = 16cm]{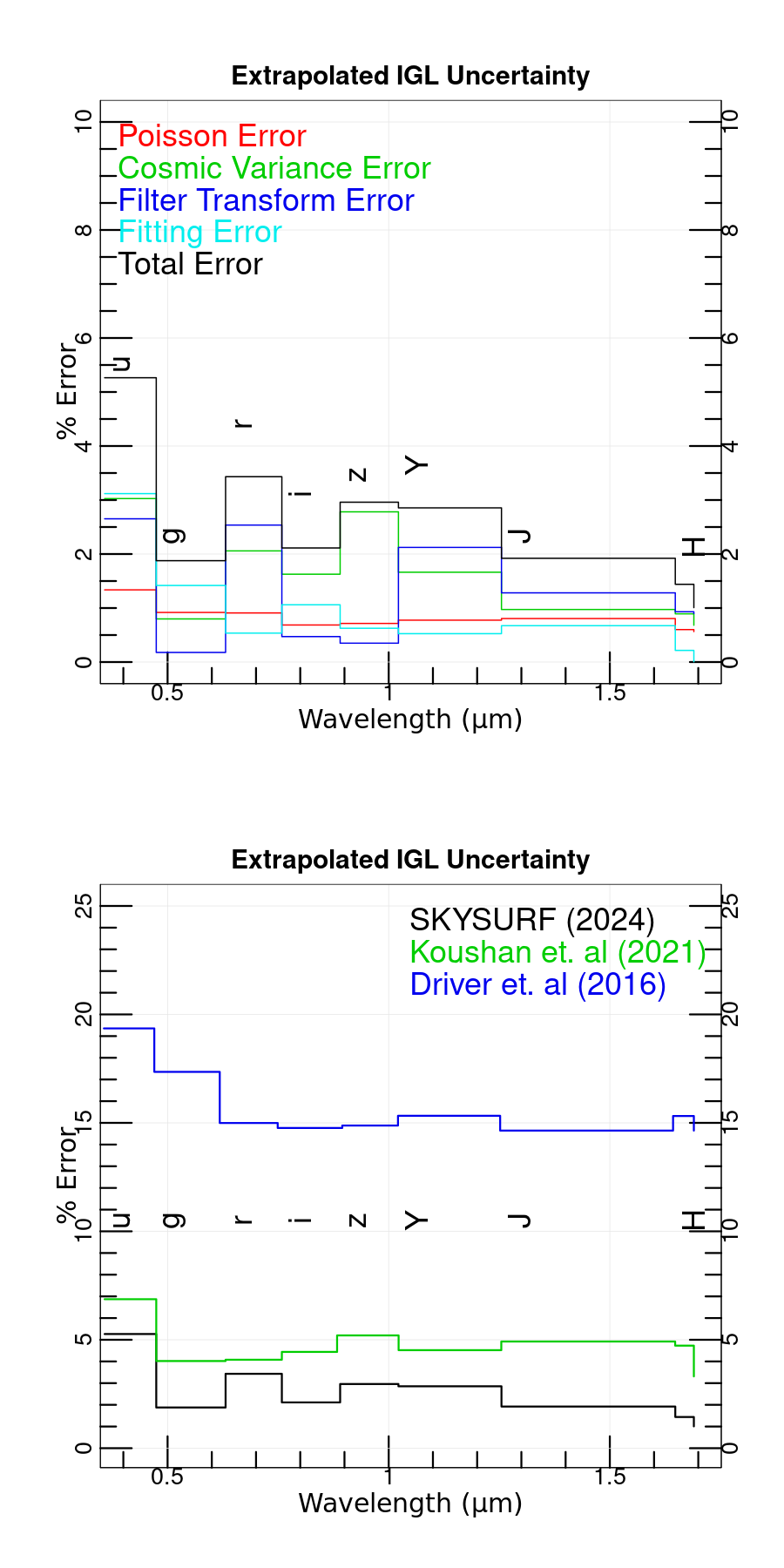}
\caption{ (Top) The same information as displayed in Figure \ref{fig:hst_err} where our HST data has been corrected to the \textit{ugrizYJH} filter system with the final error obtained by adding uncertainties in quadrature. (Bottom) Our final error budget compared to previous IGL work.
\label{fig:fig_vst_err}}
\end{figure}

Here we compare our IGL measurements to previous studies. To make a fair comparison, when comparing to previous IGL works we magnitude zero-point correct our HST number counts to the ground-based \textit{ugrizYJH} filter system. Over the last decade studies of the IGL have consistently improved due in large part to the accumulation of galaxy number count data reducing cosmic variance uncertainties. Beginning with \cite{Driver_2016}, the compilation of number counts from different existing surveys produced IGL measurements with uncertainties between $\mathrm{\approx 13-21\%}$. Cosmic variance contributed a dominant $\approx$ 10\% while zero-point uncertainties remained steady at $\approx$ 4.5\%. These errors arise from the smaller area coverage of the Galaxy and Mass Assembly Survey (GAMA) \citep{Driver_2010, Liske_2015} relative to WAVES and a significant number of data sets brought to a common magnitude zero-point. In \cite{Koushan_2021}, the introduction of DEVILS \citep{Davies_2021} increased area coverage at faint magnitudes and thus reduced the number of supplementary data sets required, which reduced cosmic variance and zero-point uncertainties to bring the IGL uncertainty down to $\approx$5\%. Our extrapolated IGL uncertainties are compared to previous IGL studies in Figure \ref{fig:fig_vst_err} (bottom).

We also compare our IGL measurements to Carter et al. 2025 (submitted), whose work serves as a sister paper to this work and the other capstone paper of the SKYSURF project. Their work introduces a novel technique for measuring the IGL. They measure total galaxy flux in individual fields for galaxies between AB 18-25 mag and extrapolate beyond that range using the ratio between the IGL contained within these bounds and extrapolation to $\mathrm{\pm \infty}$ from this work. Given both works originate from the same HST data set, we expect our measurements to be consistent. 

Within the respective errors, we find our measurements to be consistent with those of Carter et al. 2025 (submitted), though our IGL values are found to be slightly higher. This can be attributed to \textsc{ProFound's} well documented tendency to extract more flux from faint sources compared to \textsc{SourceExtractor} \cite{Robotham_2018}. In particular, the flux extraction difference between the two tools is primarily a result of the segment dilation approach employed by \textsc{ProFound}. The mechanics of the dilation approach are discussed extensively in \cite{Robotham_2018}. Both extended low surface brightness and faint sources require more dilation steps until the flux extraction converges.

\begin{figure*}
\vspace*{0.000cm}
\includegraphics[width=1\textwidth]{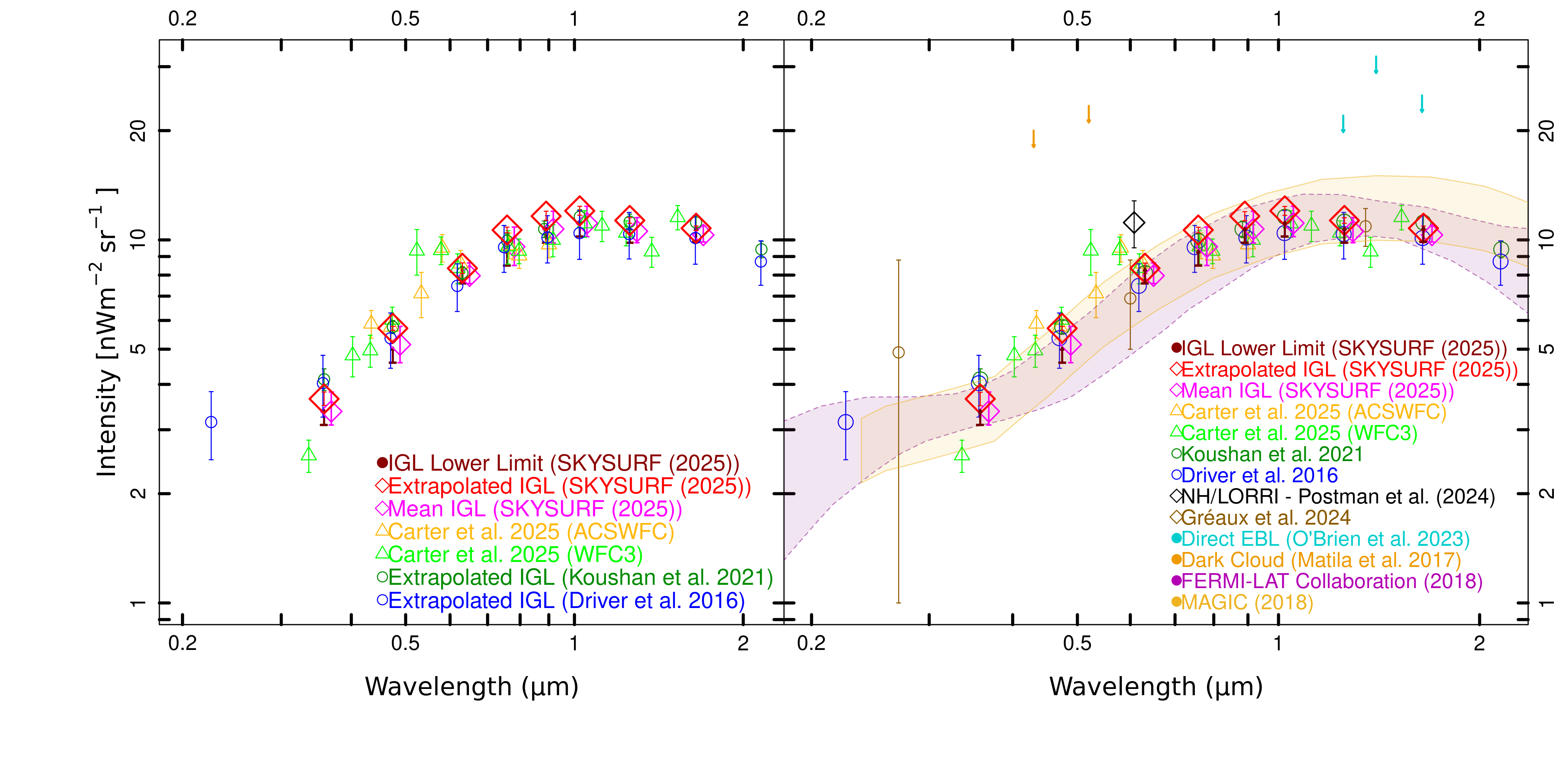}
\centering
\caption{ (Left) A comparison of the final IGL measurements in Carter et al. 2025 (submitted) \citet{Driver_2016, Koushan_2021} and this work. In Carter et al. 2025 (submitted), the IGL measurements from the WFC3 and ACSWFC instruments are treated separately. (Right) Our IGL measurements placed in context with broader COB measurements. The mean IGL values (magenta diamonds) are shifted slightly to the right for readability.  
\label{fig:vst_IGL}
}
\end{figure*}

\begin{figure}
\vspace*{0.000cm}
\includegraphics[width=9.0cm, height = 6.0cm]{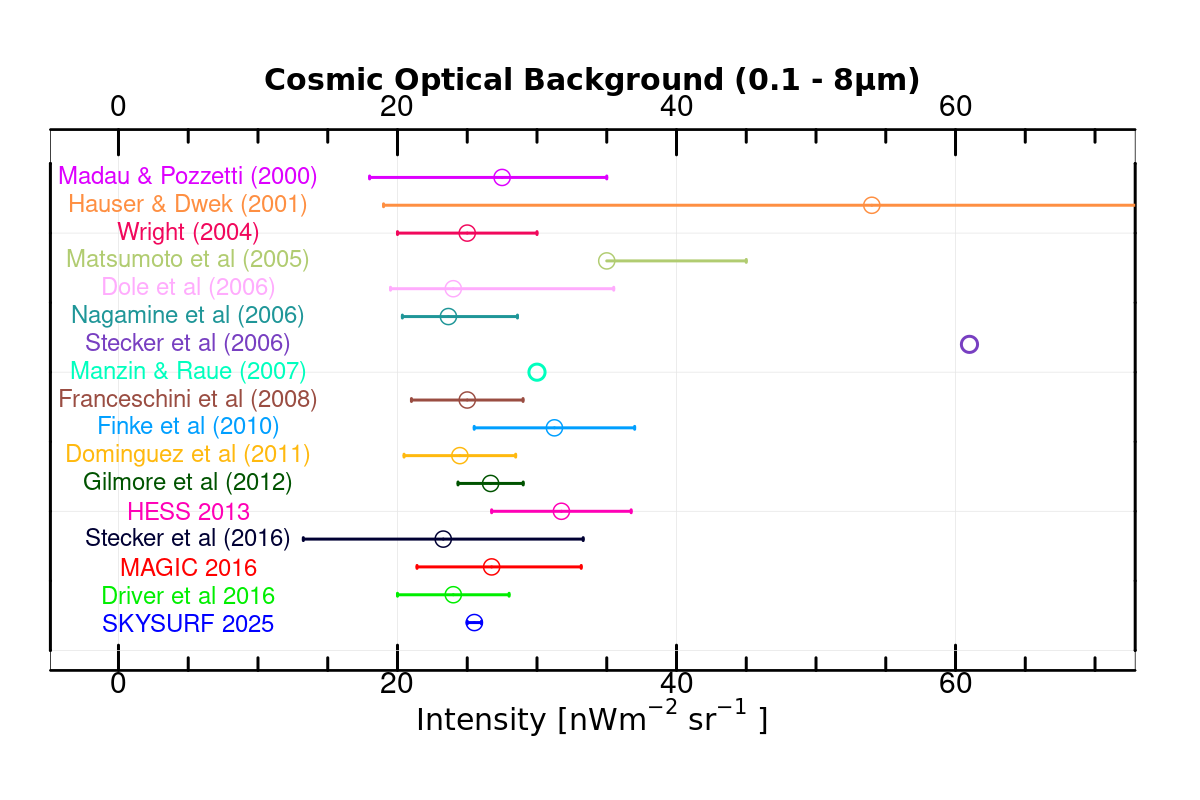}

\caption{Our measurement of the cosmic optical background compared with previous works \citet{Madau_2000, Hauser_2001, Wright_2004, Matsumoto_2005, Dole_2006, Nagamine_2006, Stecker_2006, Raue_2007, Franceschini_2008, Finke_2010, Dominguez_2011, Gilmore_2012, HESS_2013, Stecker_2016, Driver_2016, Ahnen_2016}. Error bars are shown where provided.}
\label{fig:cobhist}
\end{figure}

\begin{table*}
    \caption{The extrapolated final IGL values along with the respective uncertainties displayed as a percentage of IGL values. Errors are added in quadrature to calculate a final uncertainty in the IGL. In this table we have zero-point corrected a set of 7 HST filters to the ground-based \textit{ugrizYJH} filter set used in WAVES. The chosen HST filter is listed alongside the photometric band. See Figure \ref{fig:fig_vst_err} for a visual depiction of the final error budget when using the WAVES filter set. 
    \label{tab:tab9}}
    \centering
    \begin{tabular}{cccccccc}
    \hline
       Band/HST Filter & Pivot Wavelength & IGL Value   & Zero-Point  & CV & Poisson & Fitting Error & Total Error  \\ \hline
       & $\mathrm{\mu m}$ & $\mathrm{nW m^{-2} sr^{-1}}$ &  \%  &  \% & \% & \% & \%  \\ \hline

  u / F336W & 0.3577 & 3.65 & 2.65 & 3.03 & 1.34 & 3.08 & 5.24 \\ 
  g / F475W & 0.4744 & 5.71 & 0.18 & 0.80 & 0.92 & 1.34 & 1.82 \\ 
  r / F606W & 0.6312 & 8.35 & 2.54 & 2.06 & 0.91 & 0.63 & 3.45 \\ 
  i / F814W & 0.7584 & 10.65 & 0.47 & 1.63 & 0.69 & 0.33 & 1.86 \\ 
  z / F850LP & 0.8902 & 11.63 & 0.35 & 2.78 & 0.71 & 0.65 & 2.96 \\ 
  Y / F105W & 1.0220 & 12.02 & 2.12 & 1.66 & 0.78 & 0.54 & 2.86 \\ 
  J / F125W & 1.2550 & 11.31 & 1.28 & 0.97 & 0.81 & 0.66 & 1.92 \\ 
  H / F160W & 1.6480 & 10.77 & 0.93 & 0.89 & 0.60 & 0.23 & 1.44 \\ 

    \end{tabular}

\end{table*}

\begin{table*}
    \caption{The IGL without extrapolation with our 7 chosen HST filters zero-point corrected to the WAVES \textit{ugrizYJH} filter system. 
    \label{tab:tab10}}
    \centering
    \begin{tabular}{cccccccccc}
    \hline
       Band/HST Filter & ABmag Range & Pivot Wavelength & IGL Value & Zero-Point & CV & Poisson & Fitting Error  & Total Error  \\ \hline
       & ABmag & $\mathrm{\mu m}$ & $\mathrm{nW m^{-2} sr^{-1}}$ &  \%  &  \% & \% & \% & \%   \\ \hline
 u / F336W & 15.25-27.0 & 0.3577 & 3.09 & 2.96 & 1.14 & 0.50 & 0.18 & 4.48 \\  
  g / F475W & 15.25-26.5 & 0.4744 & 4.59 & 0.18 & 0.82 & 0.33 & 0.03 & 1.68 \\ 
  r / F606W & 15.25-28.5 & 0.6312 & 7.58 & 2.27 & 0.98 & 0.34 & 0.08 & 2.55 \\ 
  i / F814W & 15.25-27.0 & 0.7584 & 8.50 & 0.44 & 0.76 & 0.28 & 0.01 & 1.40 \\ 
  z / F850LP & 15.25-27.5 & 0.8902 & 9.81 & 0.32 & 1.10 & 0.44 & 0.02 & 1.38 \\
  Y / F105W & 15.25-27.5 & 1.022 & 10.21 & 1.91 & 0.96 & 0.50 & 0.01 & 2.26 \\  
  J / F125W & 15.25-26.0 & 1.255 & 9.82 & 1.17 & 0.82 & 0.47 & 0.03 & 1.65 \\ 
  H / F160W & 15.25-26.5 & 1.648 & 9.87 & 0.90 & 0.86 & 0.39 & 0.07 & 1.32 \\ 
      \hline

    \end{tabular}

\end{table*}

\begin{table*}
    \caption{ The mean IGL values for the \textit{ugrizYJH} filter set shown in pink on Figure \ref{fig:vst_IGL}. Errors are not quoted as a percentage given the asymmetrical nature of the error bars.
    \label{tab:tab11}}
    \centering
    \begin{tabular}{ccc}
    \hline
       Band/HST Filter & Pivot Wavelength & IGL Value  \\ \hline
        & $\mathrm{\mu m}$ & $\mathrm{nW m^{-2} sr^{-1}}$   \\ \hline
    u / F336W  &  0.358  & $ 3.37^{+0.44}_{-0.32}$   \\ 
    g / F475W  &  0.474  & $ 5.15^{+0.63}_{-0.6}$   \\ 
    r / F606W  &  0.631  & $ 7.97^{+0.67}_{-0.47}$   \\ 
    i / F814W  &  0.758  & $ 9.57^{+1.27}_{-1.14}$   \\ 
    z / F850LP  &  0.89  & $ 10.72^{+1.28}_{-1.03}$   \\ 
    Y / F105W  &  1.02  & $ 11.12^{+1.26}_{-1.01}$   \\ 
    J / F125W  &  1.25  & $ 10.57^{+0.95}_{-0.84}$   \\ 
    H / F160W  &  1.65  & $ 10.32^{+0.61}_{-0.55}$   \\ 
    \end{tabular}
\end{table*}

The $\approx$ 1.2\% error imposed by CV in WAVES in commination with comparable magnitude zero-point uncertainties introduced by using a suite of HST filters limits our improvements over \cite{Koushan_2021}. Our most precise measurements come from HST filters where we cover significant area and magnitude zero-point corrections are small such as the F814W, F125W, and F160W filters.
 
From Figure \ref{fig:vst_IGL} we see that extrapolated IGL measurements are consistently above that of previous works while still within each other's 1$\sigma$ uncertainties. Our higher IGL values could most likely be attributed to \textsc{ProFound}'s tendency to extract slightly more flux from faint objects than \textsc{SourceExtractor} \citep{Bertin_1996} as found in \cite{Robotham_2018}. The HST data used in \cite{Driver_2016, Koushan_2021} was processed using \textsc{SourceExtractor}. In this work, all photometry for DEVILS, WAVES, and HST is provided by \textsc{ProFound}. Nevertheless all measurements are consistent within their quoted errors.

\subsection{Comparison to VHE and Direct Measurements of the COB}

Finally, we compare our IGL measurements to different measurements of the COB in Figure \ref{fig:vst_IGL} (right). Our measurements are consistent with estimates of the COB provided by VHE studies \citep{Desai_2017, Fermi_LAT_2018, Gréaux_2024}. VHE studies provide a probe of the COB independent of the systematic uncertainties discussed in Section \ref{section_5.0}. Our extrapolated IGL value at $\mathrm{0.6 \mu m}$ is also consistent with the latest direct measurements from New Horizons presented in \cite{Lauer_2024}. Values of $\mathrm{8.90 \pm 0.29 nW m^{-2} sr^{-1}}$, $\mathrm{11.16 \pm 1.65 nW m^{-2} sr^{-1}}$, and $\mathrm{6.9 \pm 1.9 nW m^{-2} sr^{-1}}$ from our IGL measurements, the direct measurements of \cite{Lauer_2024}, and VHE measurements of \cite{Gréaux_2024} no longer present statistically significant disagreement. Therefore, refining measurements of the COB will provide a better understanding of the remaining foreground flux that constitutes the difference between measurements of the IGL and direct measurements. As the COB and IGL become equivalent, there is merit in further reducing uncertainties in IGL the to $\mathrm{\lessapprox} 1\%$ in order to serve as constraint on models.

With significant diffuse and/or unresolved sources of the COB ruled out, the IGL contains all of the information necessary to reconstruct the history of UV-NIR photon production since recombination. Thus the COB can be used as a powerful constraint on models of galaxy formation and evolution as measurements become more precise. Previous works such as \citep{Koushan_2021} demonstrated that the IGL could be used as a constraint on the CSFH. At the time, measurement precision meant that only the normalization of existing CSFH models could be tested. With more precise measurements the shape of the CSFH and models of more complex processes within galaxies can be evaluated using the COB as a constraint. 

Finally, we display our measurement of the integrated COB ($\mathrm{0.1-8 \mu m}$) alongside many different historical measurements, Figure \ref{fig:cobhist}. To obtain our measurement, we fit a spline with 10 degrees of freedom to the extrapolated IGL data from this work, \cite{Driver_2016}, and \cite{Koushan_2021} between $\mathrm{0.1-8 \mu m}$. We then perform the same Monte-Carlo analysis as was done for the IGL measurements discussed in Section \ref{Section_5.1} and integrate 10,000 times. The median, 17th, and 83rd percentiles serve as our measurement of the COB and its 1$\sigma$ uncertainties. We obtain a COB value of $\mathrm{25.15 \pm 0.49 nW m^{-2} sr^{-1}}$, an uncertainty of $\approx$ 2\%.

\subsection{Prospects For Future Studies of the EBL}

\begin{figure*}
\vspace*{0.000cm}
\includegraphics[width=15cm, height = 10cm]{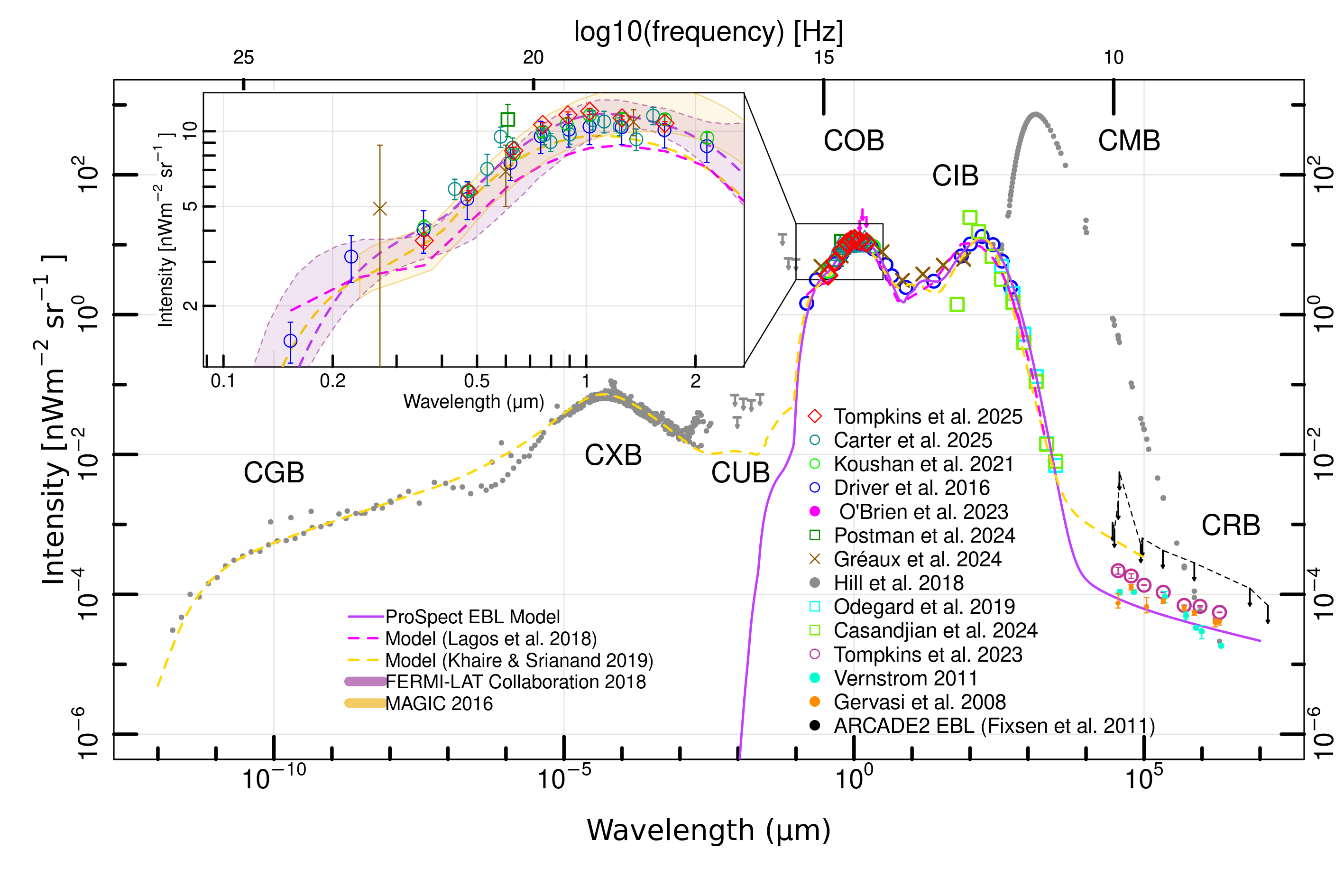}
\caption{ Our final extrapolated EBL measurements placed in context with the current status of EBL studies across all wavelengths. We show data from \citet{Gervasi_2008, Vernstrom_2011, Fixsen_2011, Ahnen_2016, Driver_2016, Hill_2018, Odegard_2019, Koushan_2021, O'Brien_2023, Tompkins_2023, Gréaux_2024}; Carter et al. 2025 (submitted) and models generated by \citet{Lagos_2018, Khaire_2019}. We show a simple EBL model generated by \textsc{ProSpect} using the currently in-development \textsc{ProSpectEBL} function. Future work on EBL modelling using \textsc{ProSpect} will be discussed in Tompkins et al. (in prep) \& Driver et al. (in prep).
\label{fig:fig_all_ebl}}
\end{figure*}

We display our COB measurements in context with the current status of the EBL across all wavelengths in Figure \ref{fig:fig_all_ebl}. Measurements of the COB have finally converged and thus studies of the COB have the potential to become used for precise galactic evolution studies. Current and upcoming facilities such as Euclid, Roman, and SphereX will have the capability to survey much larger areas of sky compared to existing data sets. Euclid has already surveyed a larger area than will ever be accessible to HST. As discussed in Section \ref{section_5.0}, uncertainties in measurements of the IGL are due to statistical errors caused by limited survey volumes and the need to combine galaxy number counts from multiple facilities. Euclid and the Vera C. Rubin Observatory in particular have the ability to address both of these issues by surveying vast areas at comparable depths to all but the deepest HST fields. 

Future surveys will avoid the existing requirement to combine galaxy number counts from different facilities, eliminating magnitude zero-point corrections. The area coverage will also reduce both Poisson and cosmic variance uncertainties to unprecedented levels. However, the HST filter set in combination with existing \textit{ugrizYJH} data provides a finer sampling of the COB which peaks at $\mathrm{\approx 1.2 \mu m}$.

Studies of the EBL also have potential to be improved outside of the range covered in this work. The successor to SKYSURF, SKYSURFIR (Windhorst et al. (in prep), plans to use archival JWST NIRcam to perform a similar analysis to what has been done in this work. In combination with SphereX, the same level of precision should be accessible for the NIR EBL at $\mathrm{\approx 0.9-4.4 \mu m}$. 

Outside of the optical and NIR regimes, next generation facilities such as the square kilometre array (SKA) have the potential to unlock new wavelength regimes to precisely study the EBL. The CRB still suffers from the disagreement between upper limits established by direct detection techniques \citep{Fixsen_2011} and the integration of source counts \citep{Gervasi_2008, Vernstrom_2011, Tompkins_2023}. The SKA has the potential to rule out significant populations of undetected faint or clustered sources, which had been proposed as a cause of previous disagreements in studies of the COB, as a cause for the current status of radio EBL measurements.

The COB encodes the CSFH and using the semi-analytical modelling software \textsc{ProSpect} \cite{Robotham_2020}. In Figure \ref{fig:fig_all_ebl} we show that using a prescribed cosmic star formation history we are able predict the COB. In combination with the CSFH, producing an accurate model requires some understanding of ongoing evolutionary processes within galaxies such as a stellar initial mass function (IMF), evolving metallicity history, and canonical model for dust attenuation. We plan to use our COB measurements in combination with the CRB measurements of \cite{Tompkins_2023} to reverse this process and predict the CSFH in future work. As such, we only provide a brief discussion of EBL modelling and the CSFH. We adopt the BC03 stellar libraries \cite{Bruzual_2003} and dust attenuation templates of \cite{Dale_2014}. We also maintain a constant \cite{Chabrier_2003} IMF throughout. In this demonstration, our closed box metallicity evolution begins at a metallicity value of $10^{-4}$ and reaches solar metallicity at z=0. To generate mock SEDs, we keep most \textsc{ProSpectSED} function parameters at default values. For a detailed discussion of all parameters see the relevant documentation. We produce EBL predictions using these inputs via the summation of the appropriately redshifted ProSpect SEDs starting with a look-back time of 13.5 GYR and ending at a look-back time of $10^{-5}$ GYR.

\section{Summary}
In this work we present the conclusions of the SKYSURF HST legacy archival program in which we produce the most precise measurements of the IGL to date. We produced over 16,000 mosaics starting with calibration level 3 image FLC and FLT image products. We have processed a significant amount of HST data using the same processing pipeline to produce consistent mosaics and object catalogs. Across $\sim$\,16.8 $deg^2$ of area we reach completeness limits of AB 24-28.5 mag in different exposure time ranges. By rigorously filtering for problematic fields in biased HST data, we are able to produce reliable galaxy number counts across 17 of 22 HST filters from which mosaics were constructed. We use our HST number counts alongside WAVES and DEVILS to measure the IGL at 17 different wavelengths between 0.3-1.6 $\mu$m by magnitude zero-point correcting the WAVES and DEVILS number counts to the HST filter set. We extensively discuss the significant sources of error affecting the IGL, Poisson, cosmic variance, and magnitude zero point uncertainties. We test our sample for systematic biases which could arise as a function of either galactic or ecliptic latitude. We then comment on the self-consistency of our HST IGL measurements in adjacent HST filters. 

We also select 8 HST filters corrected to the \textit{ugrizYJH} filter set used in WAVES to compare with existing literature which has done the same. In both cases, we report extrapolated IGL values and lower limits obtained by integrating across the magnitude range covered by the data. We then place our 8 IGL measurements on the \textit{ugrizYJH} filter set in context with previous IGL studies and measurements of the COB at large. We also quote measurements at the mean value between our lower limits and extrapolated IGL which serve as the most conservative result in this work.

\section{Acknowledgments}

This work was made possible by the following grants, HST programs AR-09955 and AR-15810 provided by NASA through the Space Telescope Science Institute,
which is operated by the Association of Universities for Research in Astronomy, Incorporated, under NASA contract NAS5-26555. Timothy Carleton is thankful for support from the Beus Center for Cosmic Foundations. Finally, we thank our reviewer whose comments and suggestions strengthened the presentation and robustness of our results.

\section{Software}

\textsc{R:} \cite{R_software} \\
\textsc{astropy:} \cite{Astropy_2022} \\
\textsc{ProFound} \cite{Robotham_2018} \\
\textsc{ProSpect} \cite{Robotham_2020} \\
\textsc{ProPane} \cite{Robotham_2023} \\

\section{Data Availability}

In accordance with the SKYSURF grant conditions we make our source catalogs publicly available upon request to the corresponding author. The catalogs contain the entire output of the \textsc{ProFound} source detection code, for a description of \textsc{ProFound} columns see the \textsc{ProFound} documentation. Here we describe the added columns.\\
\textbf{Starflag} - Flag specifying if an object is a identified as a star. A value of 0 specifies an object as a galaxy and a value of 1 as a star.\\
\textbf{Secondary Flag} - A flag specifying whether the object is used in the production of counts. A value of 0 is a good object identified as a galaxy, while 1 is given to stars, 2 to suspected artifacts, 4 to given to objects falling within exclusion radii around stars such as diffraction spikes, and a value of 3 is given to objects brighter than AB 15 mag which are not used and have their surroundings masked.\\
\textbf{Exptime} - The stacked exposure time at the centre of each object.\\
\textbf{wht} - The weightmap value, or number of good pixels in the stack, at the centre of each object. Objects with a value of zero are possible noise and not included. \\
\textbf{Region} - An integer, identifying which exposure time bin, eg, 200-400 (1), 400-800 (2), an object falls into, used for keeping track of areas.\\
\textbf{Area} - The total unmasked area of each mosaic.\\
\textbf{sub area} - The total area of unmasked pixels in the associated exposure time range.\\
\textbf{fraction} - The fraction of area covered by the corresponding exposure time range.\\
\textbf{visit} - ID flag given to specify which mosaic an object comes from.\\
\textbf{corr mag} - Galactic extinction corrected magnitudes.

\bibliographystyle{mnras}
\bibliography{references.bib}

\section{Appendix}

\begin{figure*}

\caption{ The 4-panel galaxy number count and IGL figures for all filter sets used in this work. 
\label{fig:uebl}}

\vspace*{0.000cm}
\includegraphics[width=0.675\linewidth]{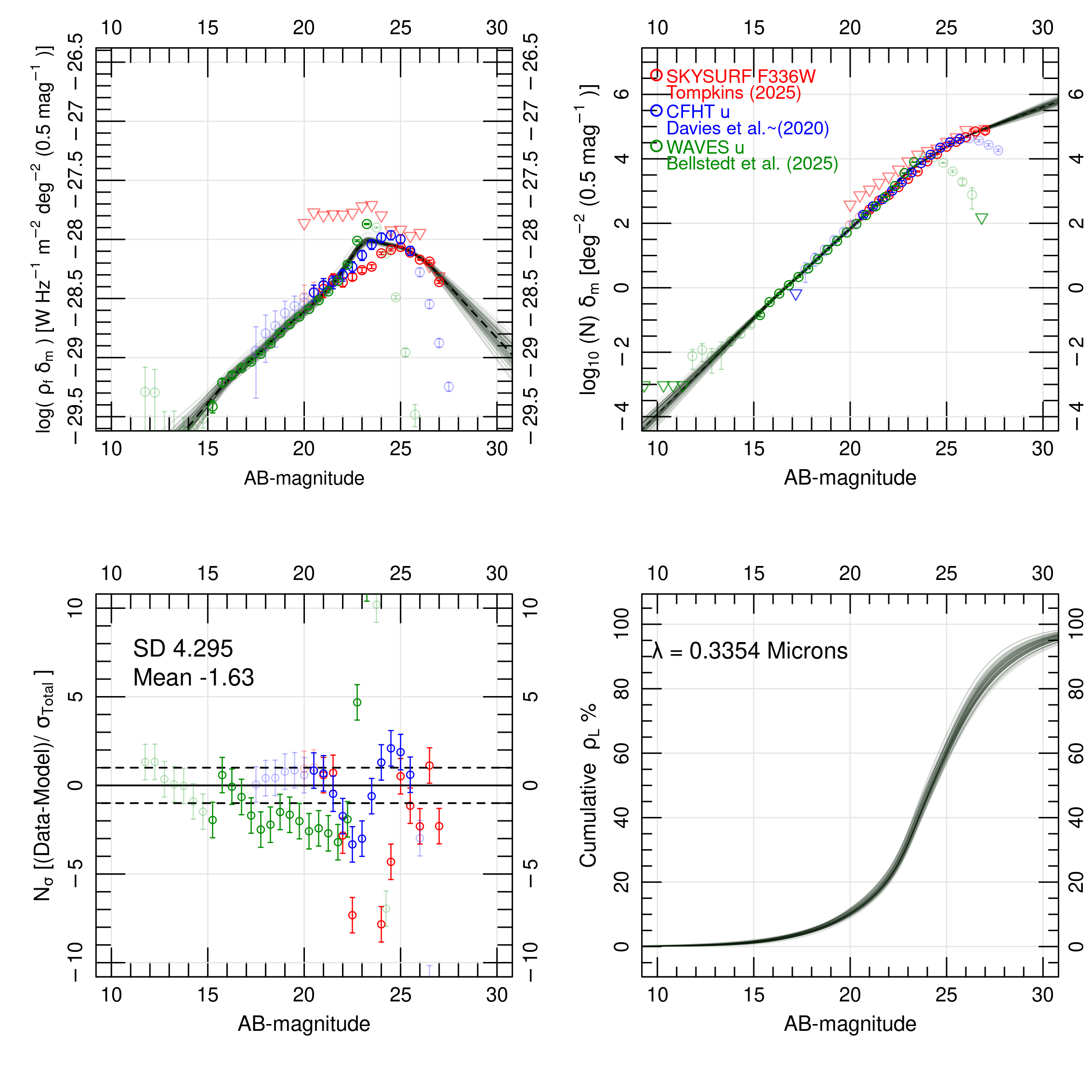}
\includegraphics[width=0.675\linewidth]{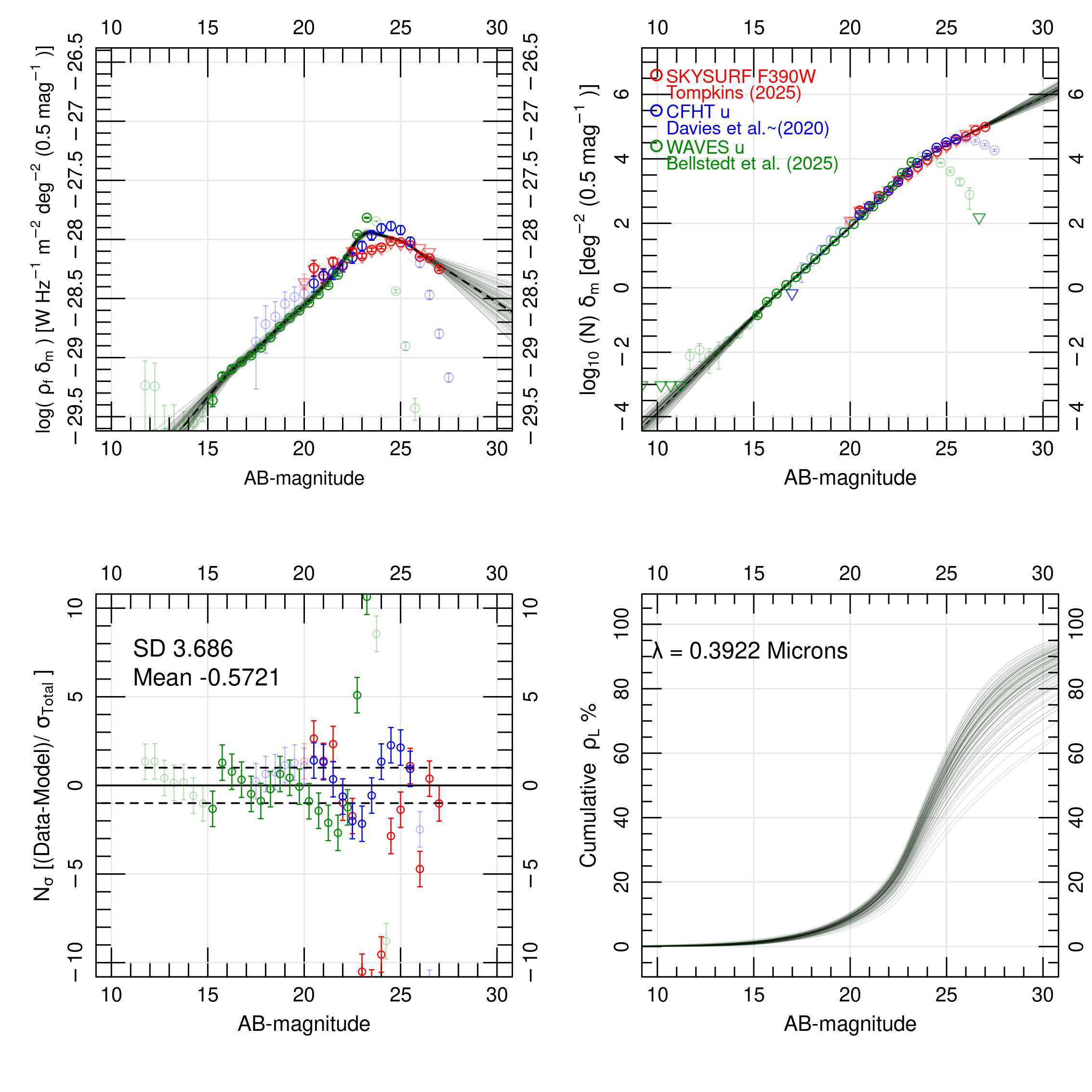}

\end{figure*}

\begin{figure*}

\caption{ The 4-panel galaxy number count and IGL figures for all filter sets used in this work. 
\label{fig:gebl}}

\includegraphics[width=0.675\linewidth]{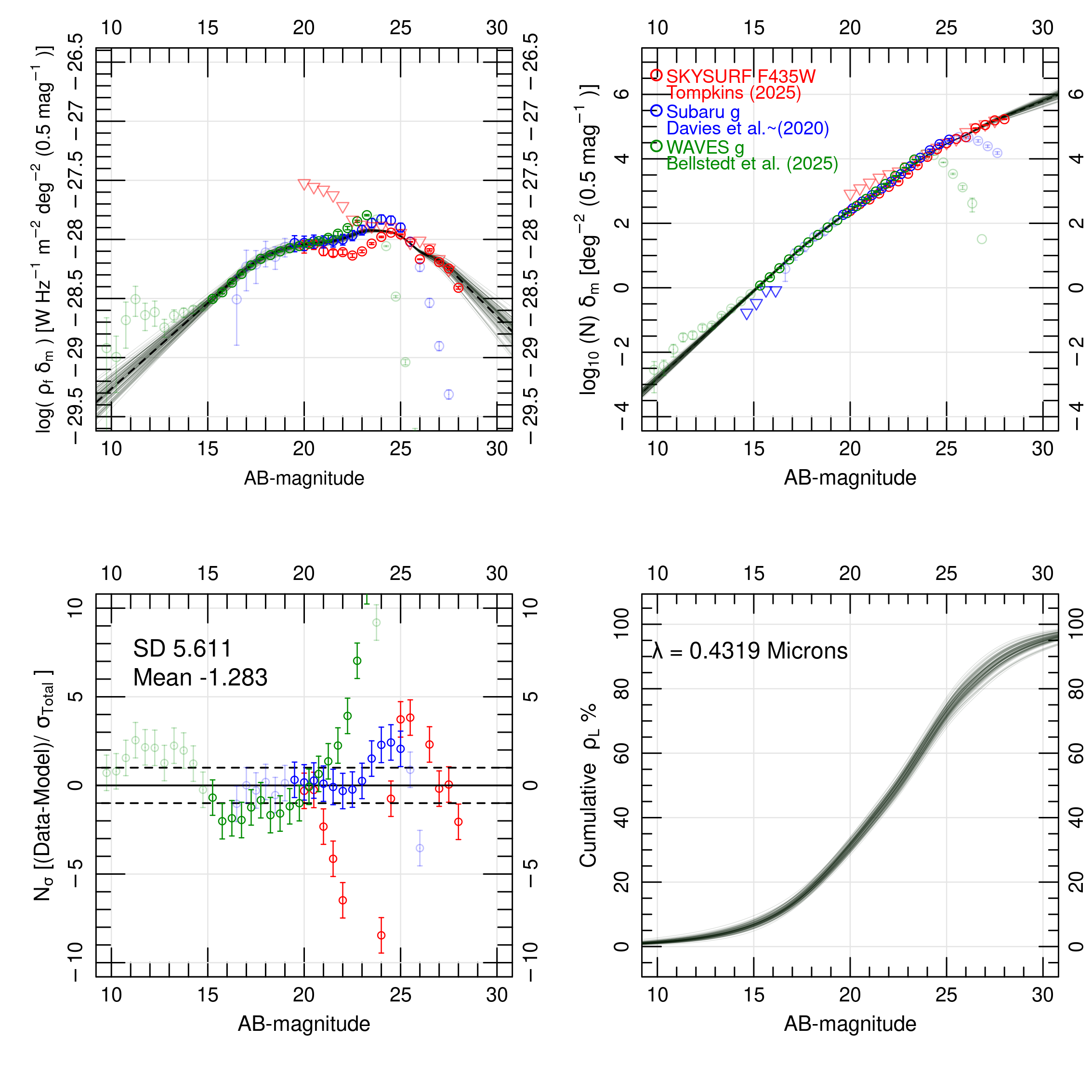}
\includegraphics[width=0.675\linewidth]{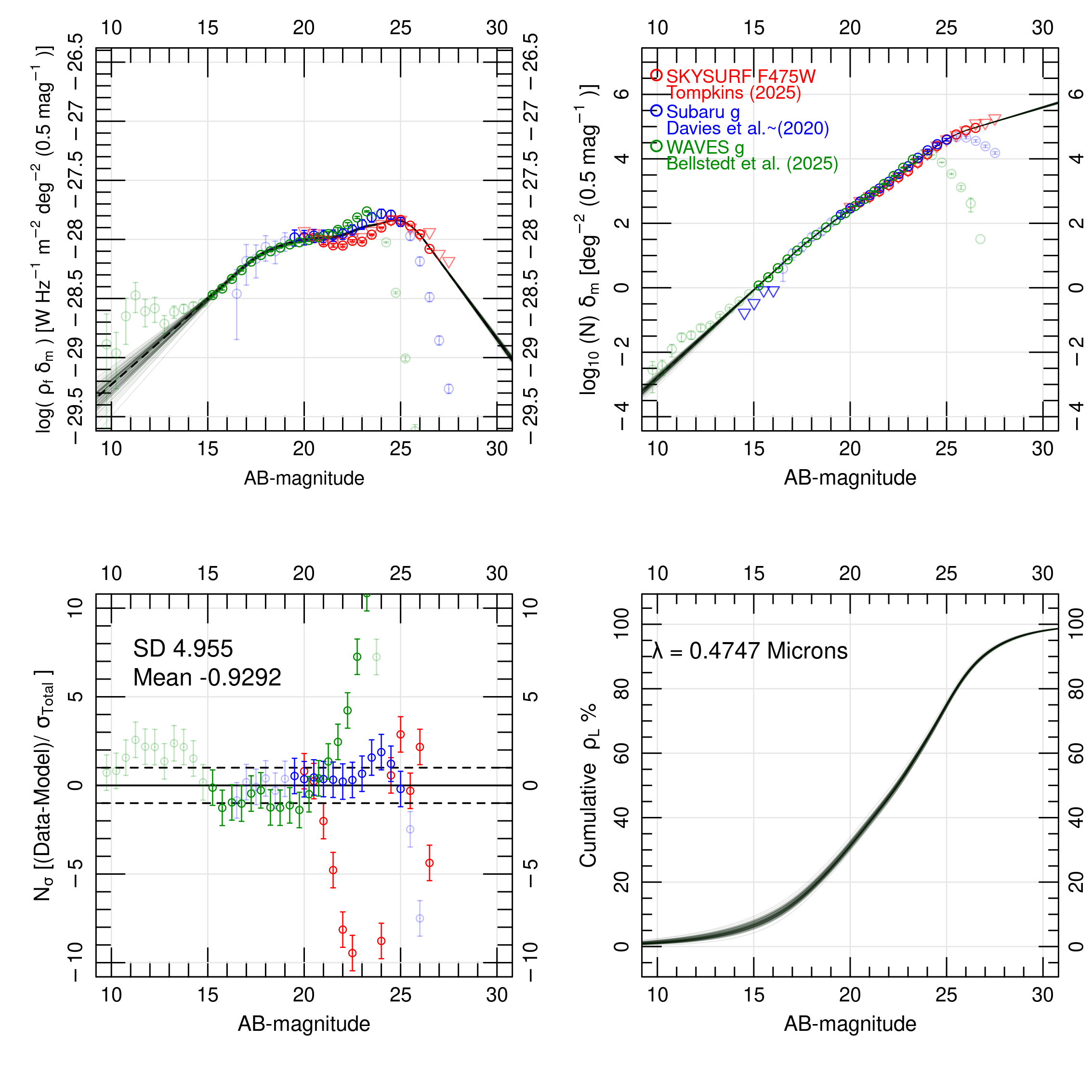}

\end{figure*}

\begin{figure*}

\caption{ The 4-panel galaxy number count and IGL figures for all filter sets used in this work.
\label{fig:r_ebl}}

\vspace*{0.000cm}
\includegraphics[width=0.675\linewidth]{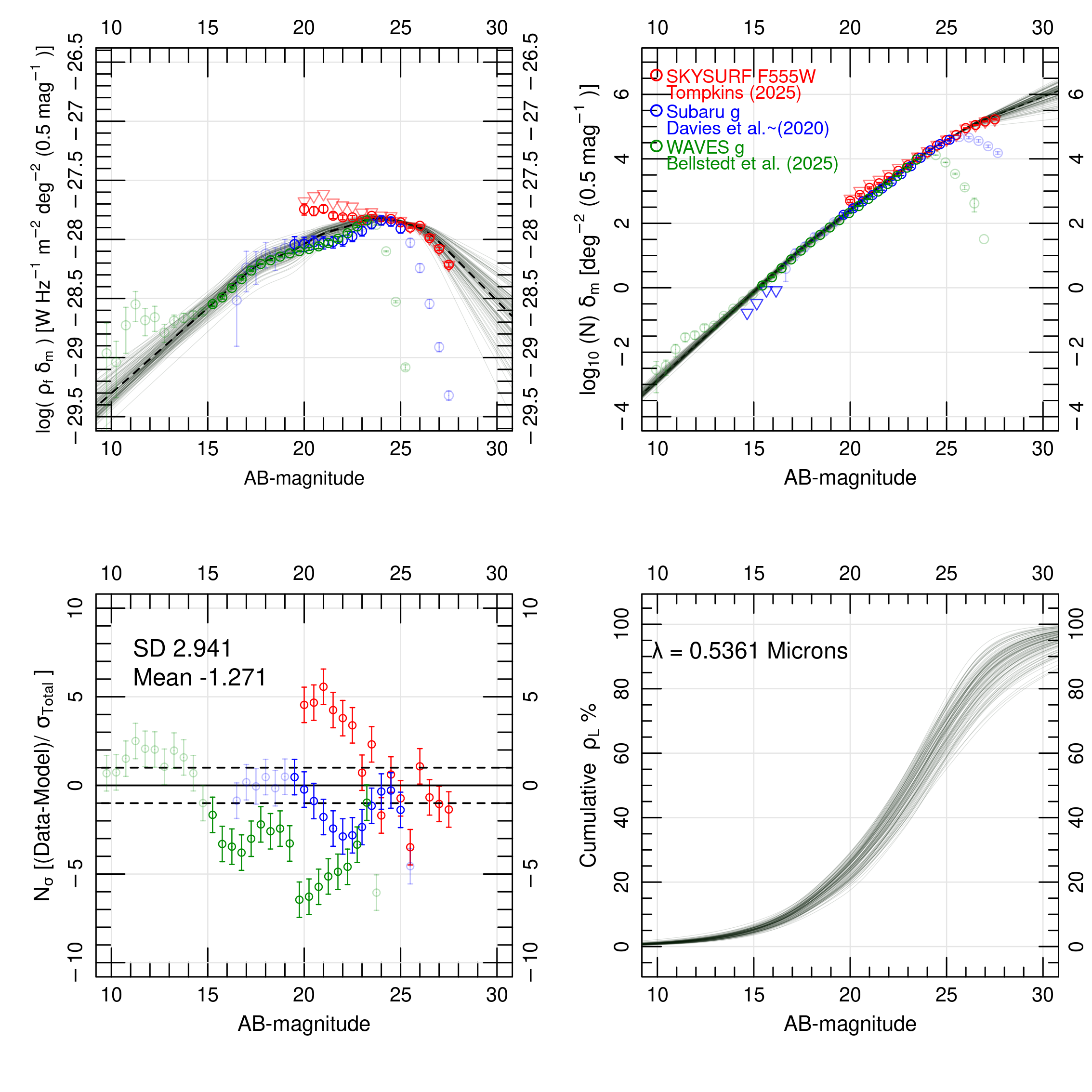}
\includegraphics[width=0.675\linewidth]{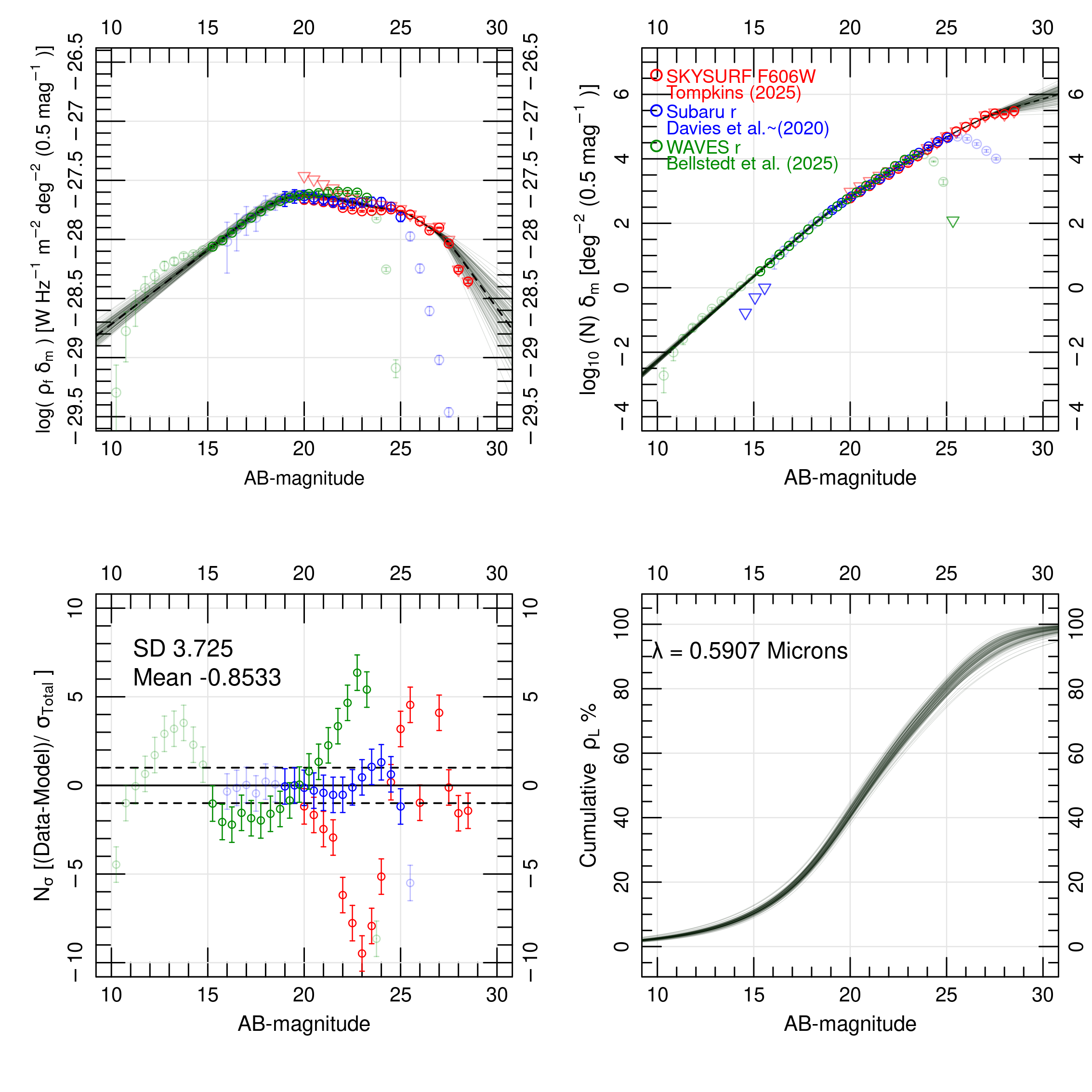}

\end{figure*}

\begin{figure*}

\caption{ The 4-panel galaxy number count and IGL figures for all filter sets used in this work.
\label{fig:ri_ebl}}

\includegraphics[width=0.675\linewidth]{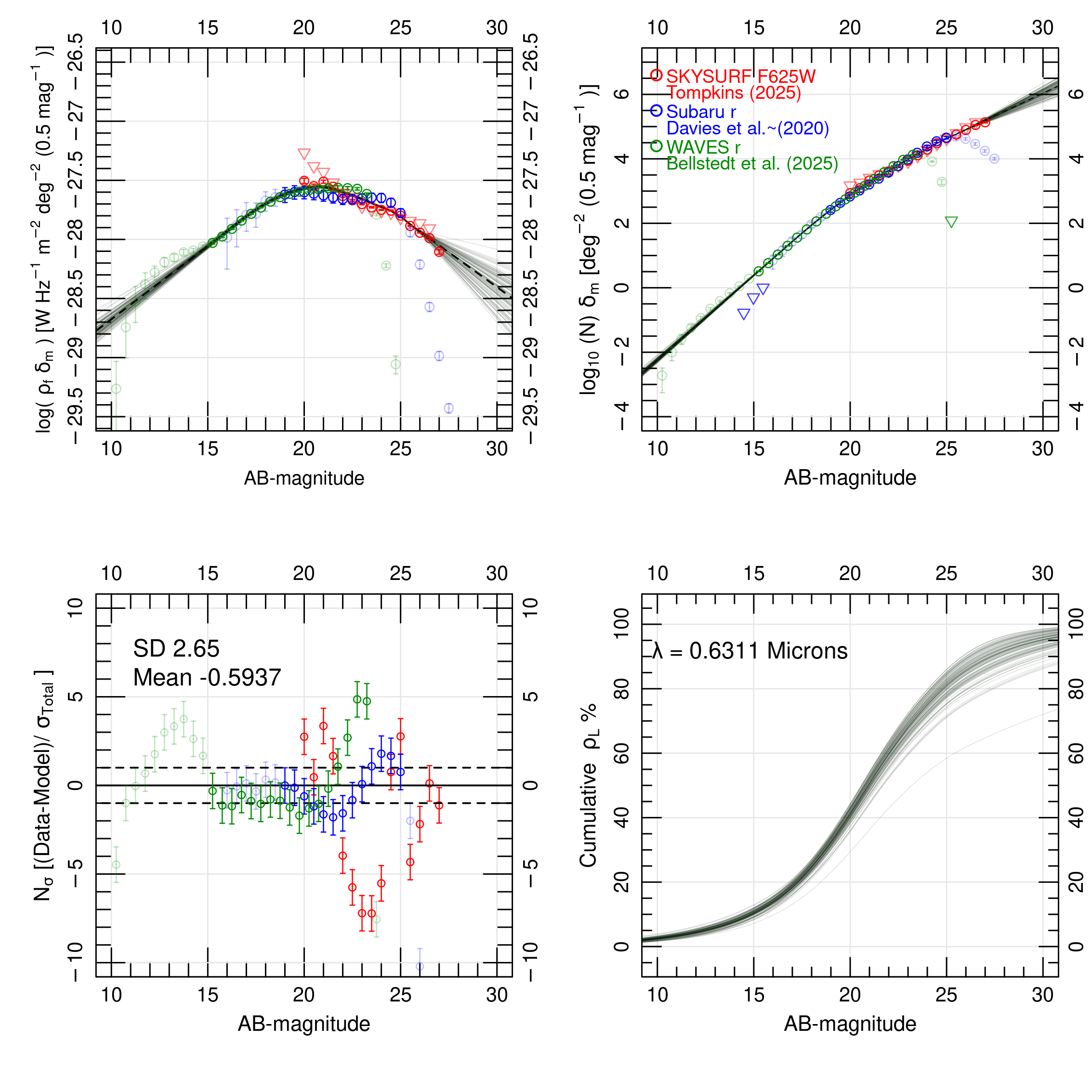}
\includegraphics[width=0.675\linewidth]{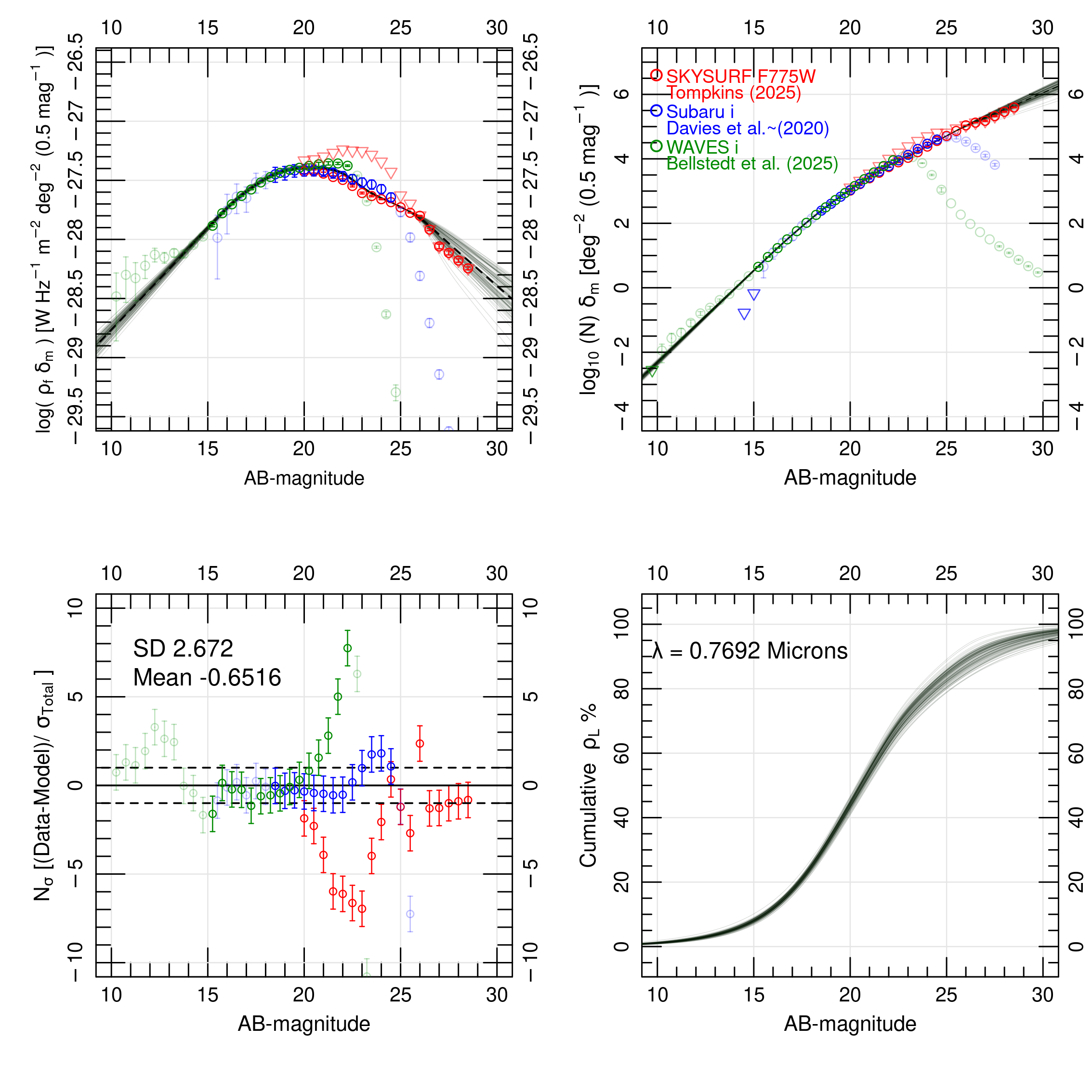}

\end{figure*}

\begin{figure*}

\caption{ The 4-panel galaxy number count and IGL figures for all filter sets used in this work.
\label{fig:iz_EBL}}

\vspace*{0.000cm}

\includegraphics[width=0.675\linewidth]{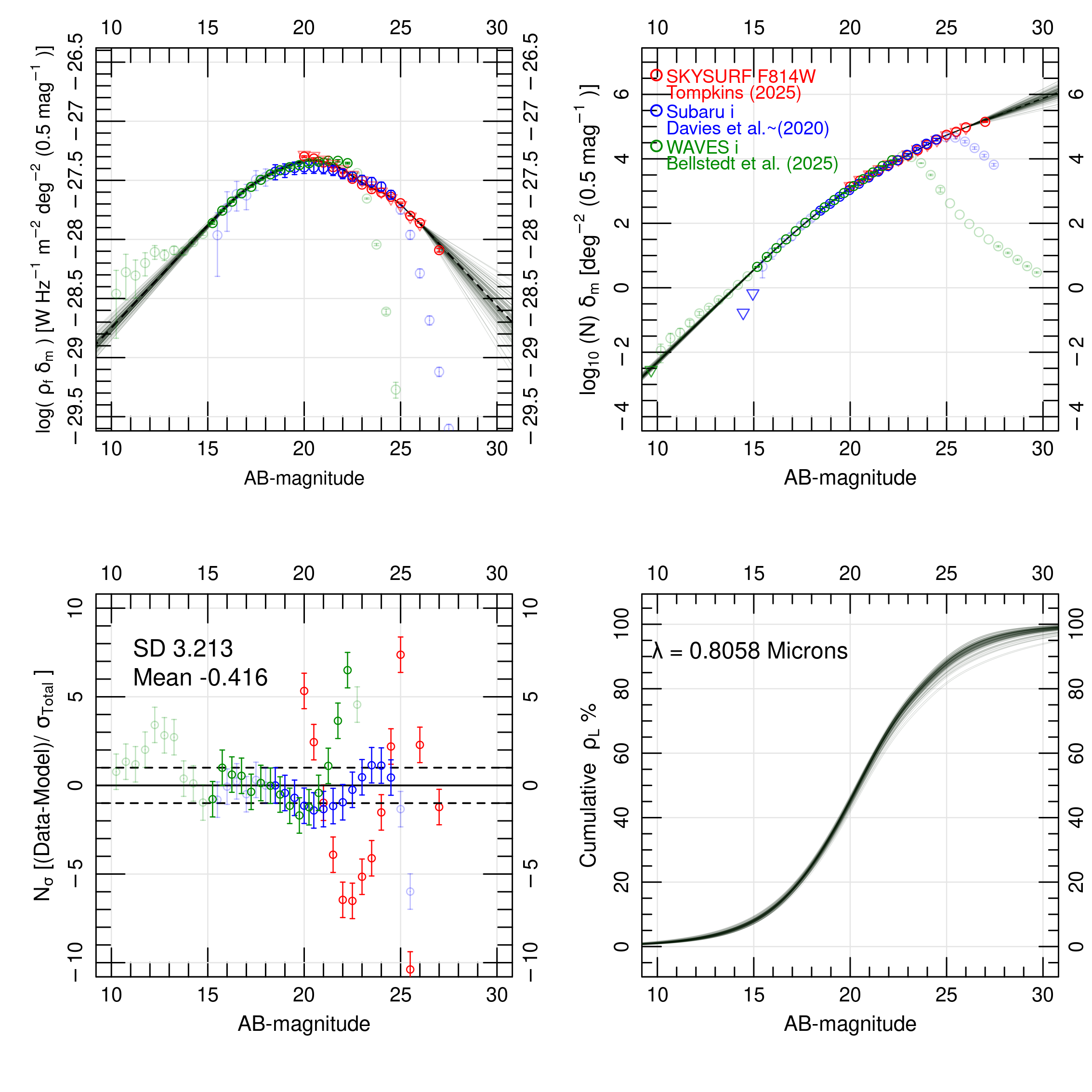}

\includegraphics[width=0.675\linewidth]{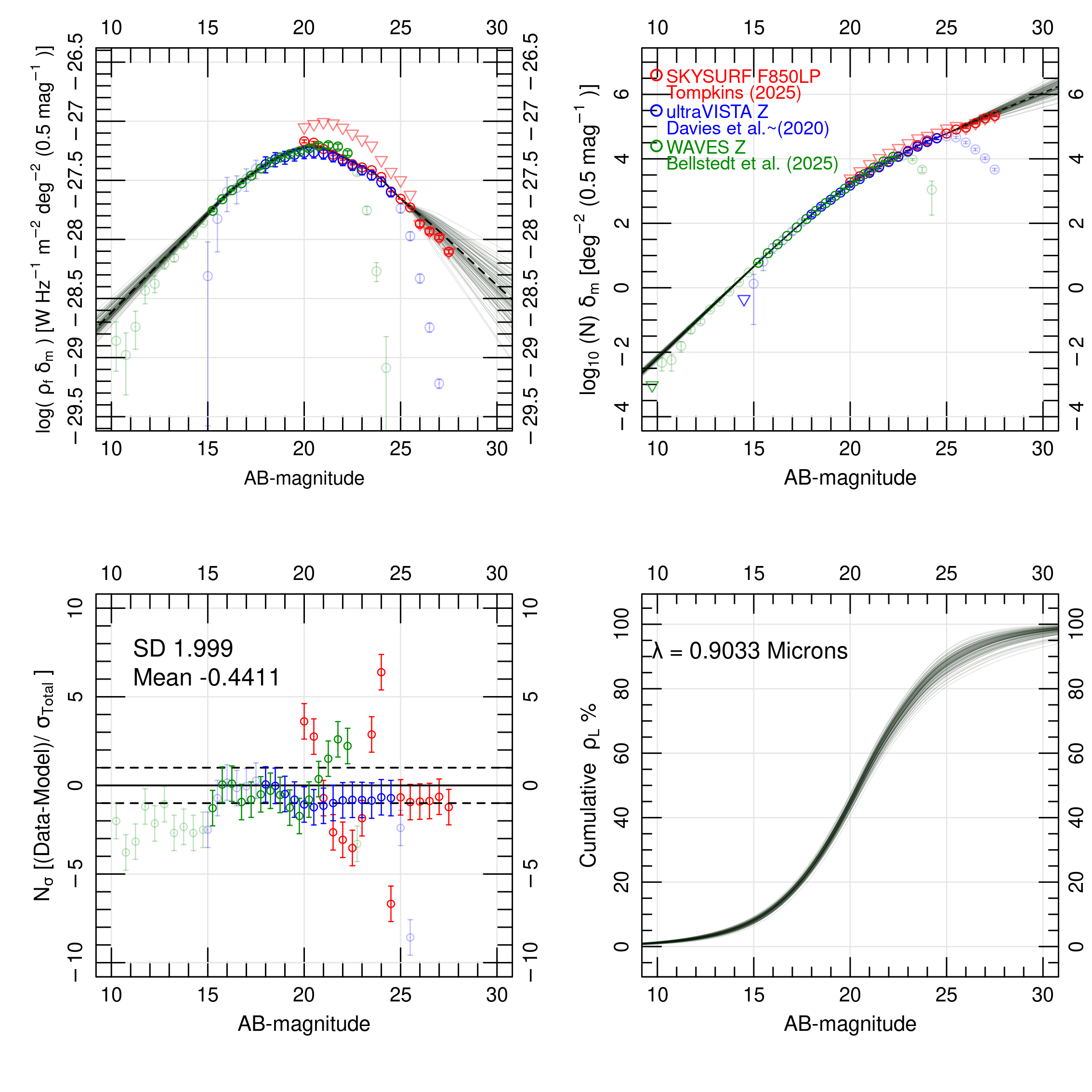}

\end{figure*}

\begin{figure*}

\caption{ The 4-panel galaxy number count and IGL figures for all filter sets used in this work.
\label{fig:y_EBL}}

\includegraphics[width=0.675\linewidth]{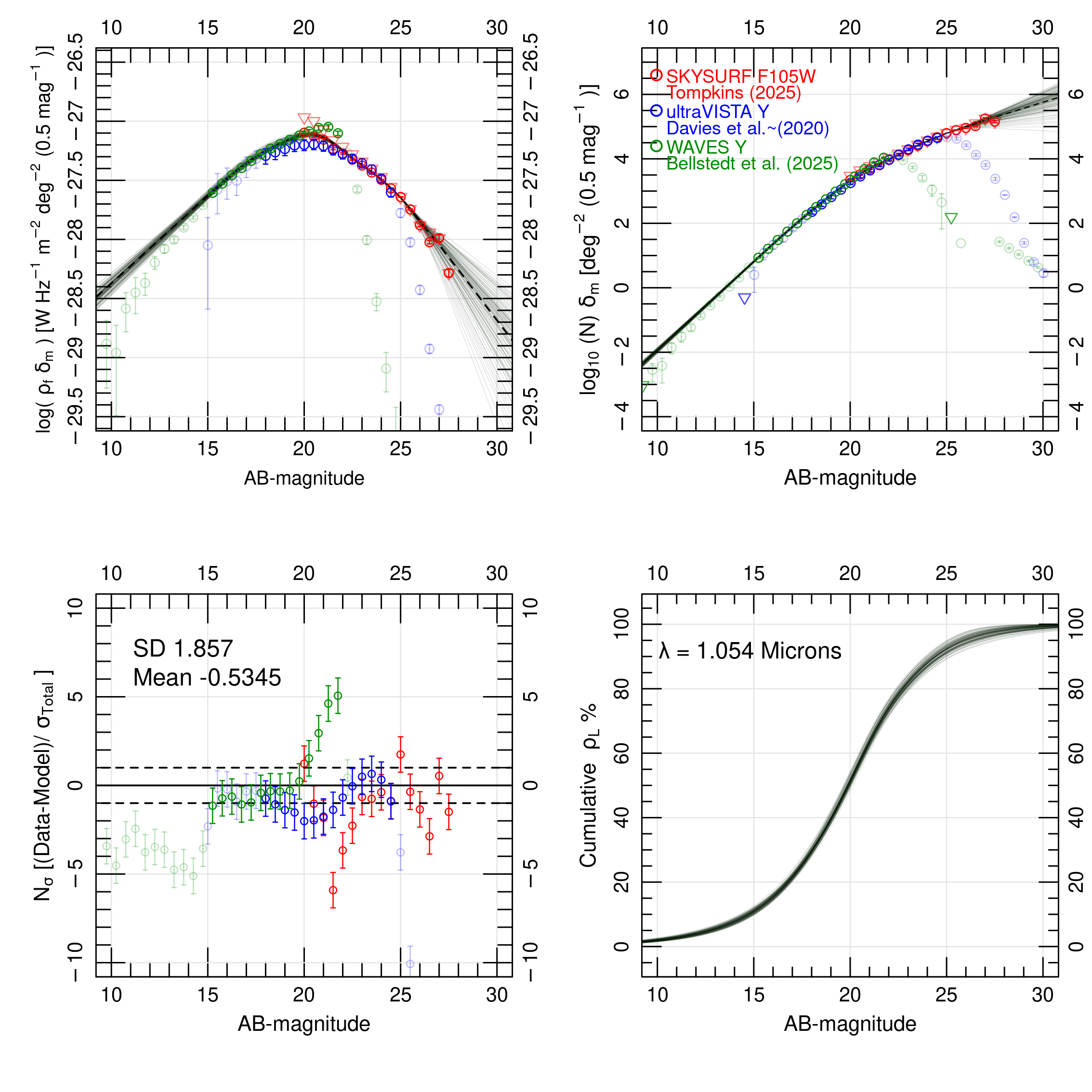}

\includegraphics[width=0.675\linewidth]{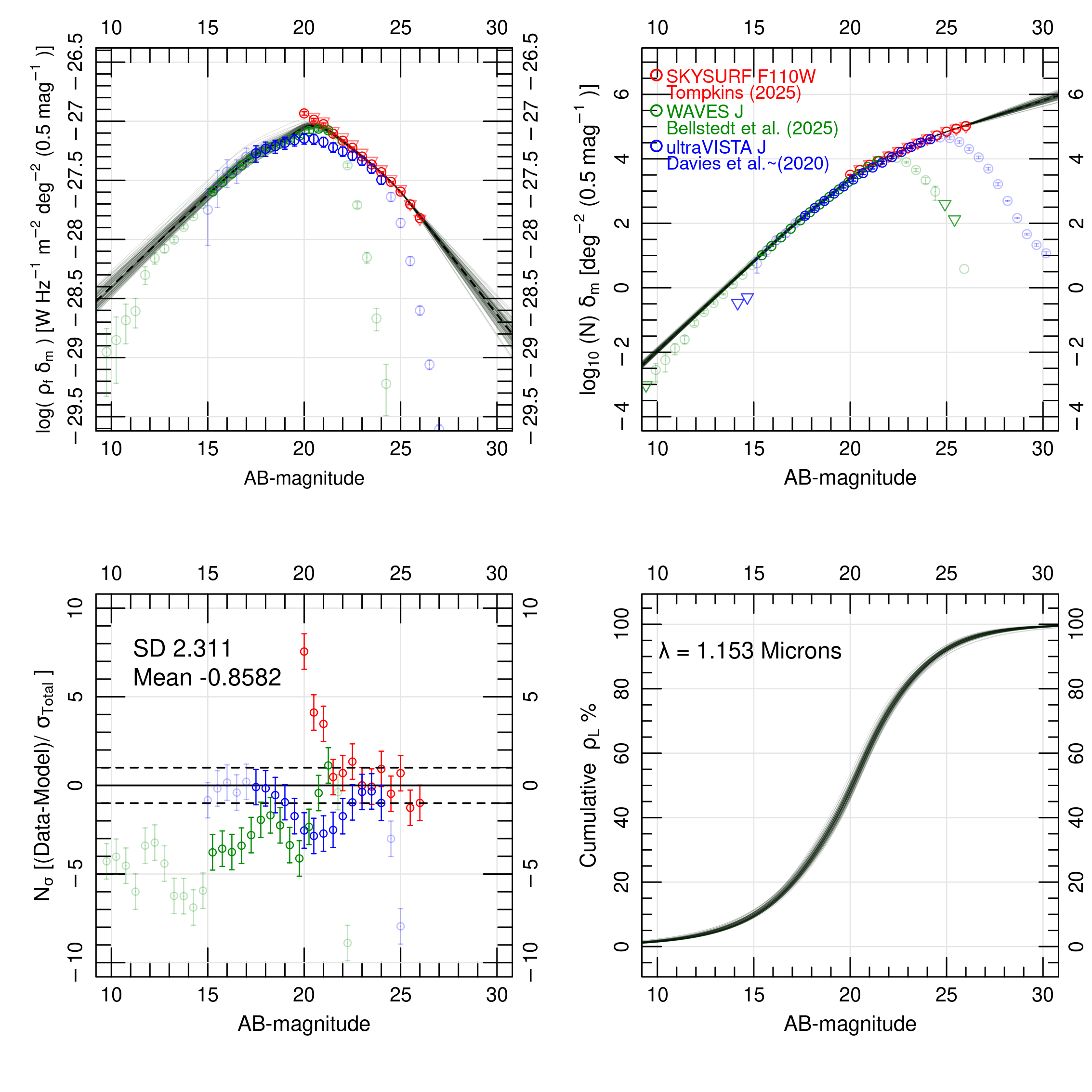}

\end{figure*}

\begin{figure*}

\caption{ The 4-panel galaxy number count and IGL figures for all filter sets used in this work. No secondary data is used with the F140W filter.
\label{fig:jh_EBL}}

\vspace*{0.000cm}
\includegraphics[width=0.675\linewidth]{F125W_1.247_4_panel_.png}
\includegraphics[width=0.675\linewidth]{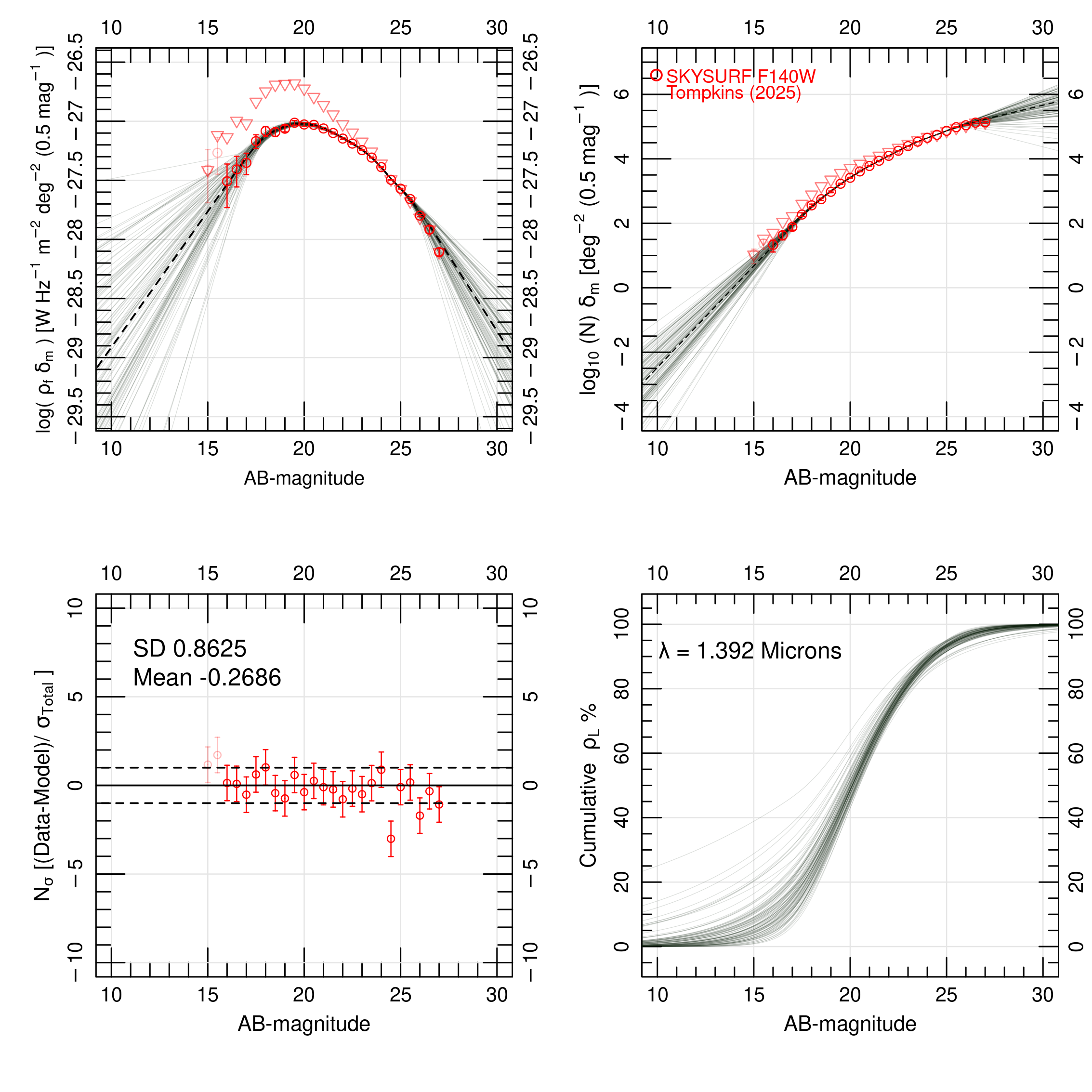}
\end{figure*}

In the appendix, we show the 4-panel figures with the galaxy number counts, the galaxy number counts adjusted to show the
contribution of each magnitude bin to the IGL, and the cumulative sum of the IGL for all HST filters in Figures \ref{fig:uebl}, \ref{fig:gebl}, \ref{fig:r_ebl}, \ref{fig:iz_EBL}, \ref{fig:y_EBL}, \ref{fig:jh_EBL}. We list the magnitude zero-point correction information for all filters in Tables \ref{tab:tab12} \& \ref{tab:tab13}. Finally we display the star, galaxy, and artifact separation boundaries for all filters in Figures \ref{fig:fig_blue_ele}, \ref{fig:fig_more_ele},
\ref{fig:fig28}, \& \ref{fig:fig_rest_ele}.

\begin{figure*}
\caption{ The 4-panel galaxy number count and IGL figures for all filter sets used in this work.
\label{fig:h_EBL}}

\includegraphics[width=0.675\linewidth]{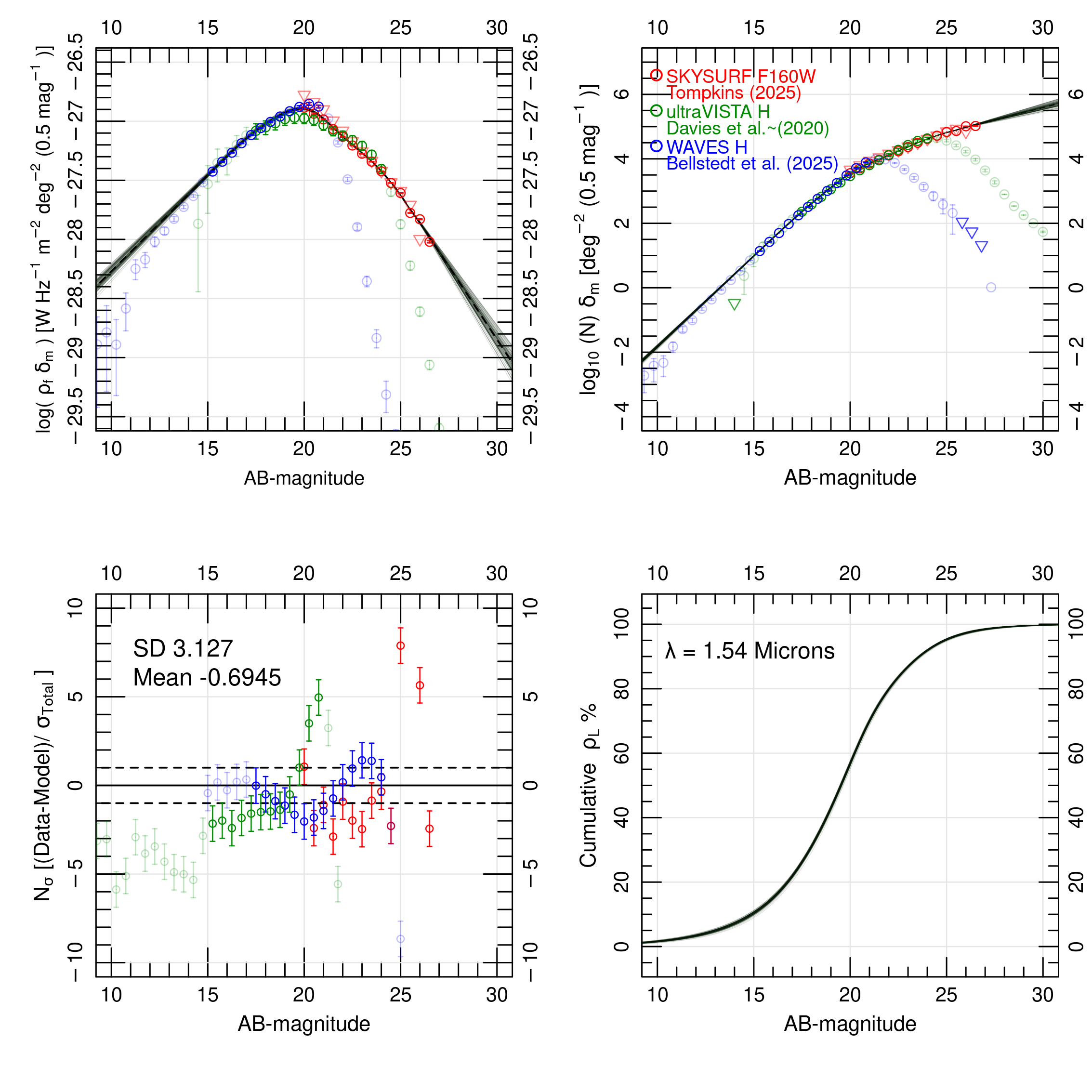}
\end{figure*}

\begin{table*}
    \centering
    \caption{The magnitude zero point corrections used to move the WAVES number counts to the HST filter system. The \textsc{ProSpect} \textsc{filterTranBands} function takes in a target filter, input filter, and reference filter.}
      \label{tab:tab12}
    \begin{tabular}{cccccc}
    \hline
        Target Filter & Input Filter & Reference Filter & band & ABMag Offset & ABMag Offset Err \\ \hline
        ACSWFC F435W & g VST & r VST & g & 0.008085 & 0.042613 \\ %\hline
        ACSWFC F475W & g VST & r VST & g & 0.001512 & 0.004756 \\ %\hline
        ACSWFC F555W & g VST & r VST & g & 0.190506 & 0.15061 \\ %\hline
        ACSWFC F606W & r VST & i VST & r & 0.06081 & 0.04948 \\ %\hline
        ACSWFC F625W & r VST & i VST & r & 0.0002331 & 0.003608 \\ %\hline
        ACSWFC F775W & i VST & z VST & i & -0.008711 & 0.01549 \\ %\hline
        ACSWFC F814W & i VST & z VST & i & -0.02324 & 0.0124 \\ %\hline
        ACSWFC F850LP & z VST & Y VISTA & Z & -0.0000293 & 0.008342 \\ %\hline
        WFC3UVIS F336W & u VST & g VST & u & 0.07004 & 0.08185 \\ %\hline
        WFC3UVIS F390W & u VST & g VST & u & -0.062654 & 0.04275 \\ %\hline
        WFC3UVIS F606W & r VST & i VST & r & 0.06442 & 0.05412 \\ %\hline
        WFC3UVIS F814W & i VST & z VST & i & -0.02372 & 0.01074 \\ %\hline
        WFC3IR F105W & Y VISTA & J VISTA & Y & 0.02315 & 0.0455 \\ %\hline
        WFC3IR F110W & J VISTA & H VST & J & 0.16356 & 0.09030 \\ %\hline
        WFC3IR F125W & J VISTA & H VISTA & J & 0.03092 & 0.0281 \\ %\hline
        WFC3IR F160W & H VISTA & Ks VISTA & H & 0.03837 & 0.01932 \\ \hline
    \end{tabular}
\end{table*}

\begin{table*}
    \centering
    \caption{The same information displayed in Table \ref{tab:tab12} but used for moving the DEVILS number counts to the HST filter set.}
      \label{tab:tab13}
    \begin{tabular}{cccccc}
    \hline
        Target Filter & Input Filter & Reference Filter & band & ABMag Offset & ABMag Offset Err \\ \hline
        ACSWFC F435W & g HSC & r HSC & g & 0.133864 & 0.046885 \\% \hline
        ACSWFC F475W & g HSC & r HSC & g & 0.007533 & 0.005143 \\% \hline
        ACSWFC F555W & g HSC & r HSC & g & 0.15564 & 0.13648 \\% \hline
        ACSWFC F606W & r HSC & i HSC & r & 0.04741 & 0.03753 \\ %\hline
        ACSWFC F625W & r HSC & i HSC & r & -0.009079 & 0.01413 \\ %\hline
        ACSWFC F775W & i HSC & z HSC & i & 0.002311 & 0.006701 \\ %\hline
        ACSWFC F814W & i HSC & z HSC & i & -0.01921 & 0.008543 \\ %\hline
        ACSWFC F850LP & z HSC & Y VISTA & Z & 0.00847 & 0.01794 \\ %\hline
        WFC3UVIS F336W & u CFHT & g VST & u & 0.17658 & 0.12925 \\
        WFC3UVIS F390W & u CFHT & g VST & u & -0.01819 & 0.01667 \\
        WFC3UVIS F606W & r HSC & i HSC & r & 0.05058 & 0.04195 \\ %\hline
        WFC3UVIS F814W & i HSC & z HSC & i & -0.0194 & 0.01172 \\ %\hline
        WFC3IR F105W & Y VISTA & J VISTA & Y & 0.02315 & 0.0455 \\ %\hline
        WFC3IR F110W & J VISTA & H VST & J & 0.16356 & 0.09030 \\ %\hline
        WFC3IR F125W & J VISTA & H VISTA & J & 0.03092 & 0.0281 \\ %\hline
        WFC3IR F160W & H VISTA & Ks VISTA & H & 0.03837 & 0.01932 \\ %\hline
    \end{tabular}
\end{table*}

%%%

\begin{figure*}
    \centering % <-- added

\begin{subfigure}{0.5\textwidth}
  \includegraphics[width=\linewidth]{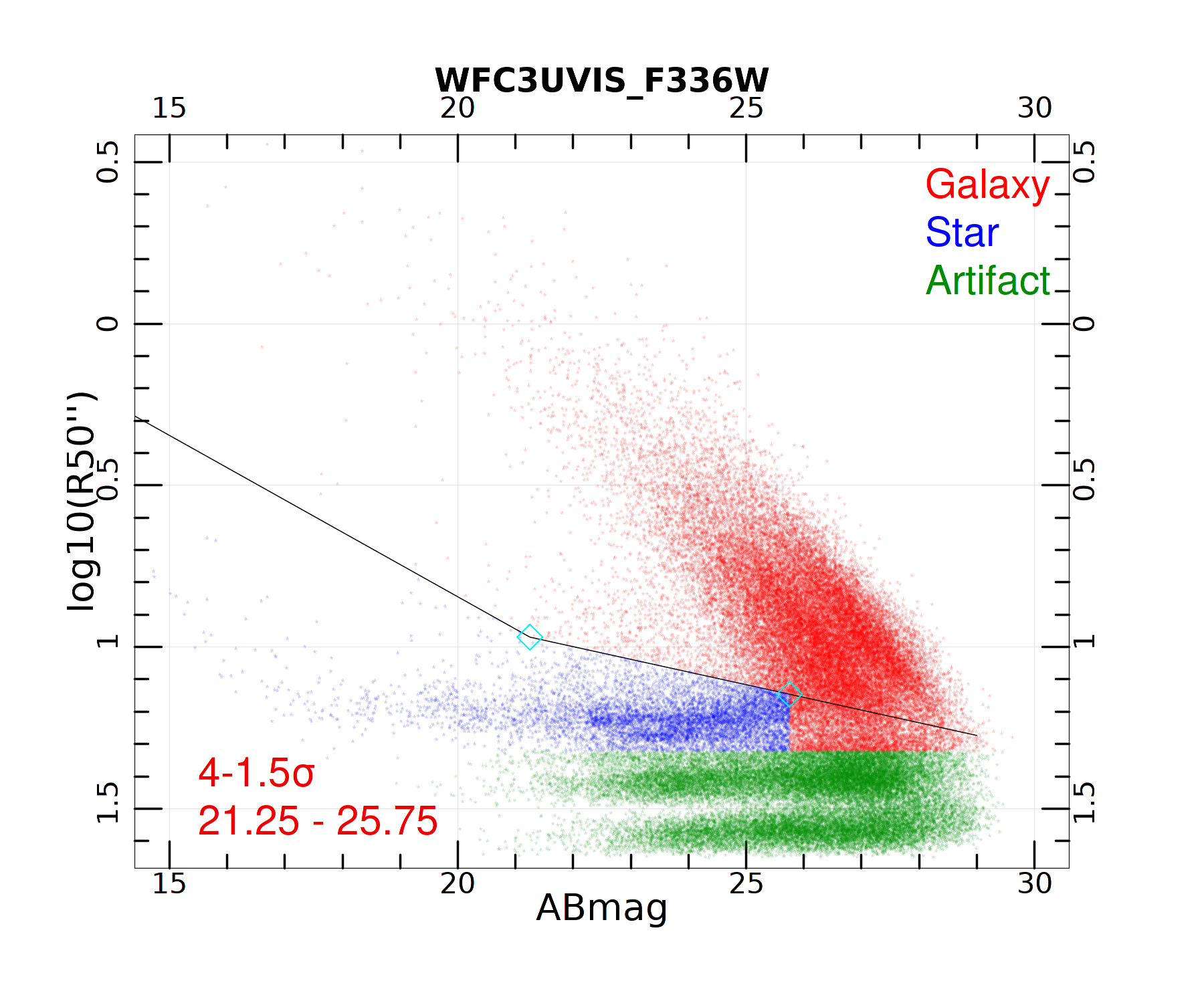}
  \caption{$\log_{10}$ vs R50 parameter space for WFC3UVIS F336W.}

\end{subfigure}\hfil % <-- added
\begin{subfigure}{0.5\textwidth}
  \includegraphics[width=\linewidth]{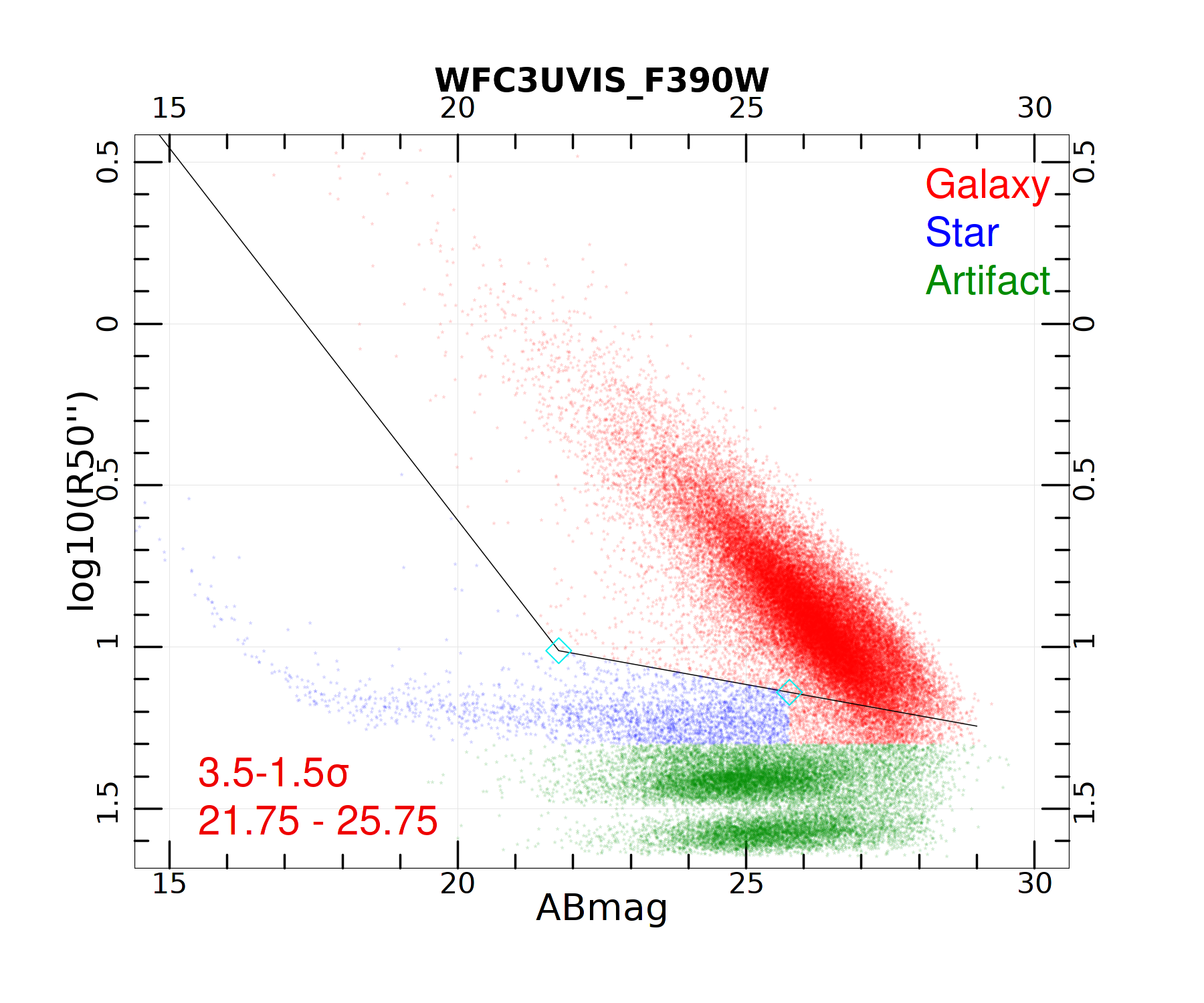}
  \caption{$\log_{10}$ vs R50 parameter space for WFC3UVIS F390W.}

\end{subfigure}\hfil % <-- added

\begin{subfigure}{0.5\textwidth}
  \includegraphics[width=\linewidth]{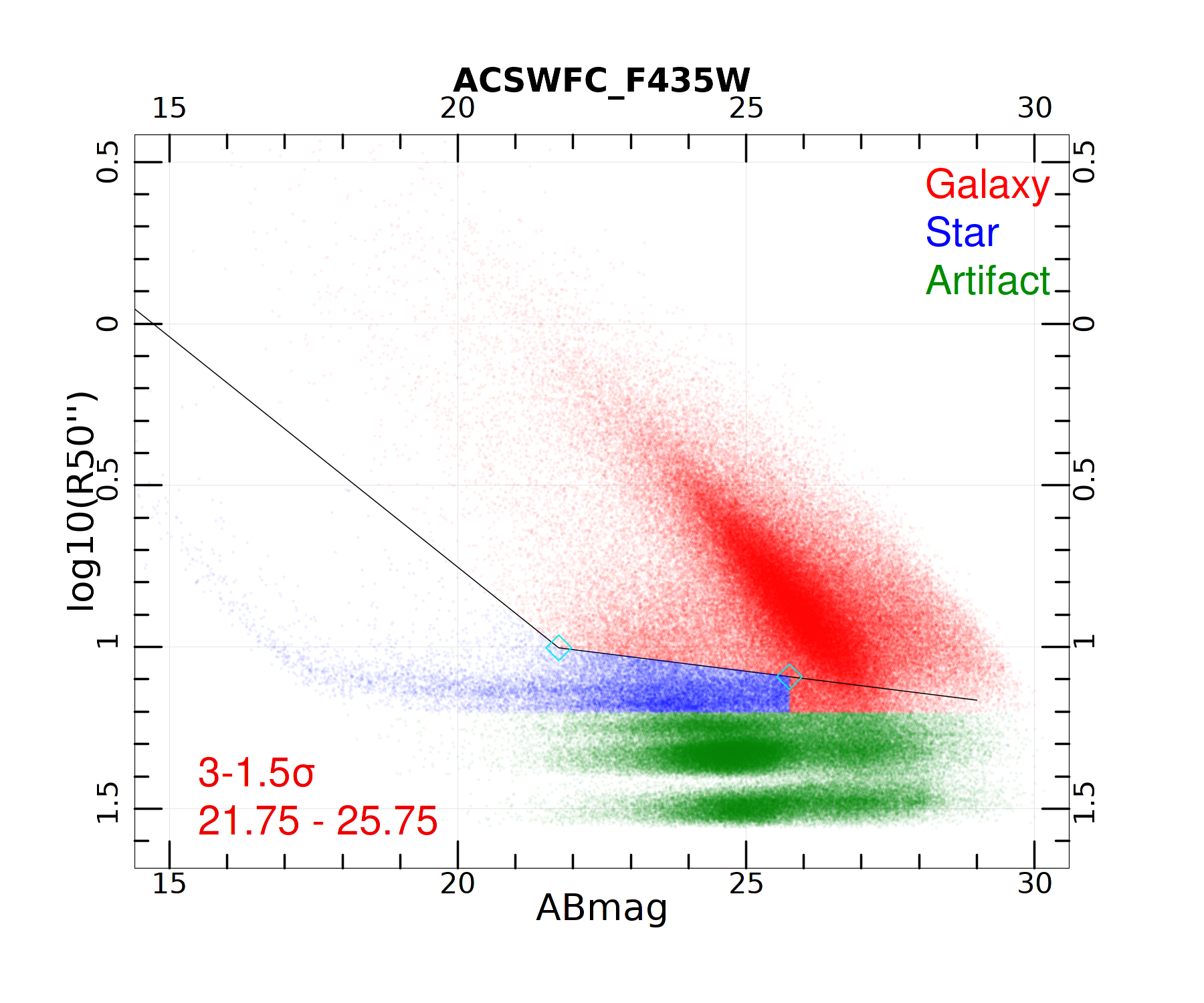}
  \caption{$\log_{10}$ vs R50 parameter space for ACSWFC F435W.}

\end{subfigure}\hfil % <-- added
\begin{subfigure}{0.5\textwidth}
  \includegraphics[width=\linewidth]{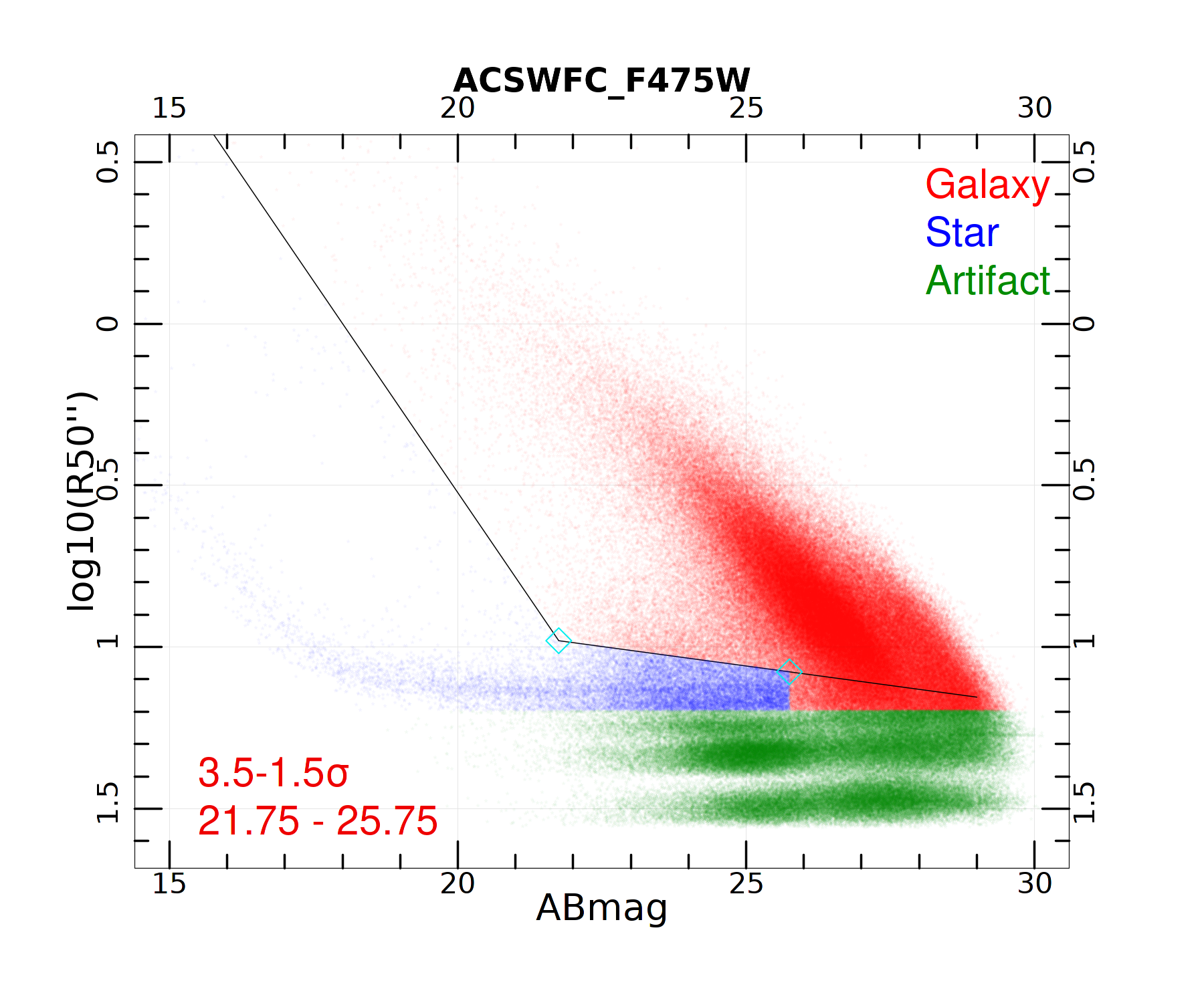}
  \caption{$\log_{10}$ vs R50 parameter space for ACSWFC F475W.}

\end{subfigure}\hfil % <-- added
  \caption{$\log_{10}$ vs R50 parameter space for the other 16 filters used in this work.}
\label{fig:fig_blue_ele}
\end{figure*}

\begin{figure*}

\begin{subfigure}{0.5\textwidth}
  \includegraphics[width=\linewidth]{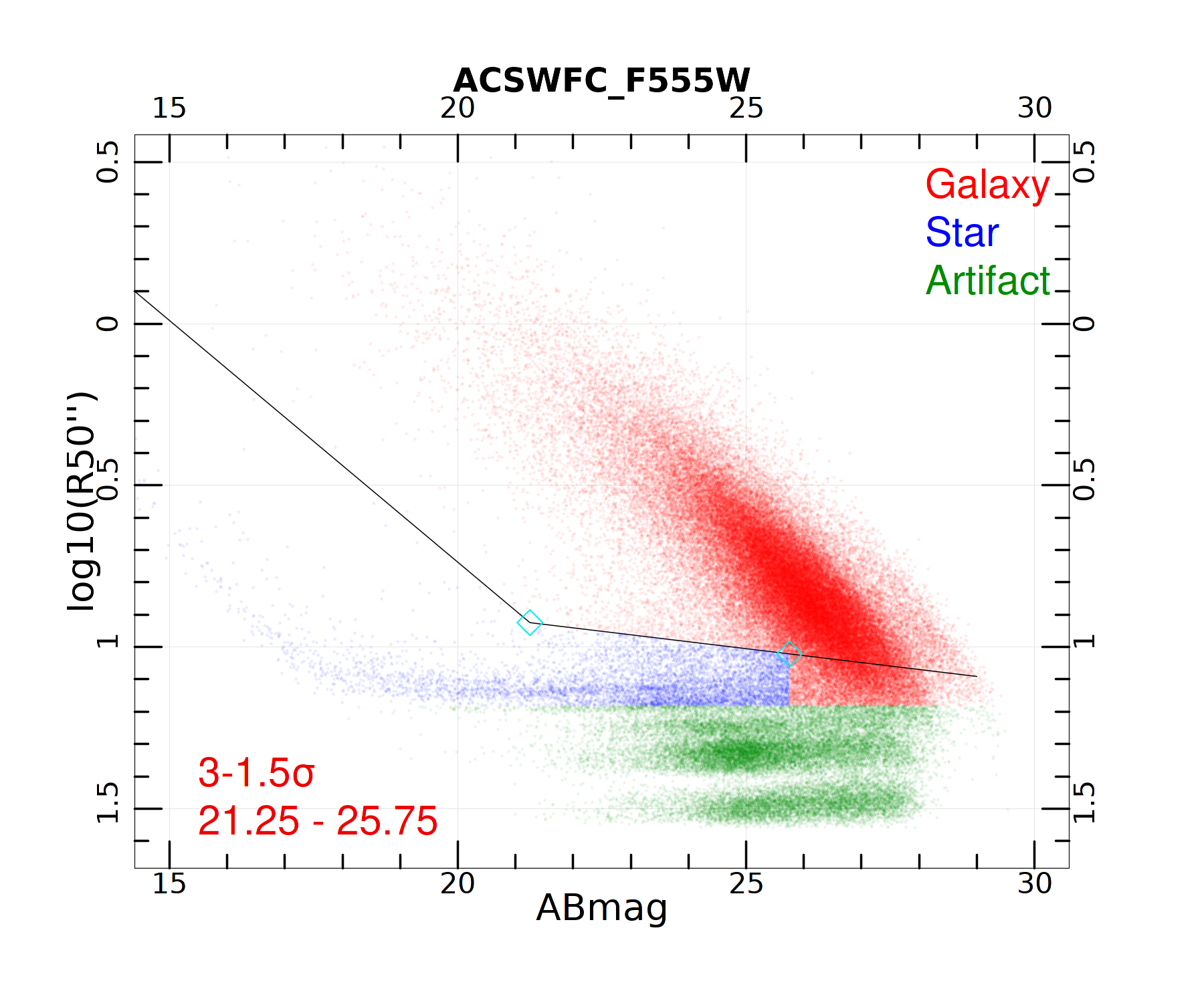}
  \caption{$\log_{10}$ vs R50 parameter space for ACSWFC F555W.}

\end{subfigure}\hfil % <-- added
\begin{subfigure}{0.5\textwidth}
  \includegraphics[width=\linewidth]{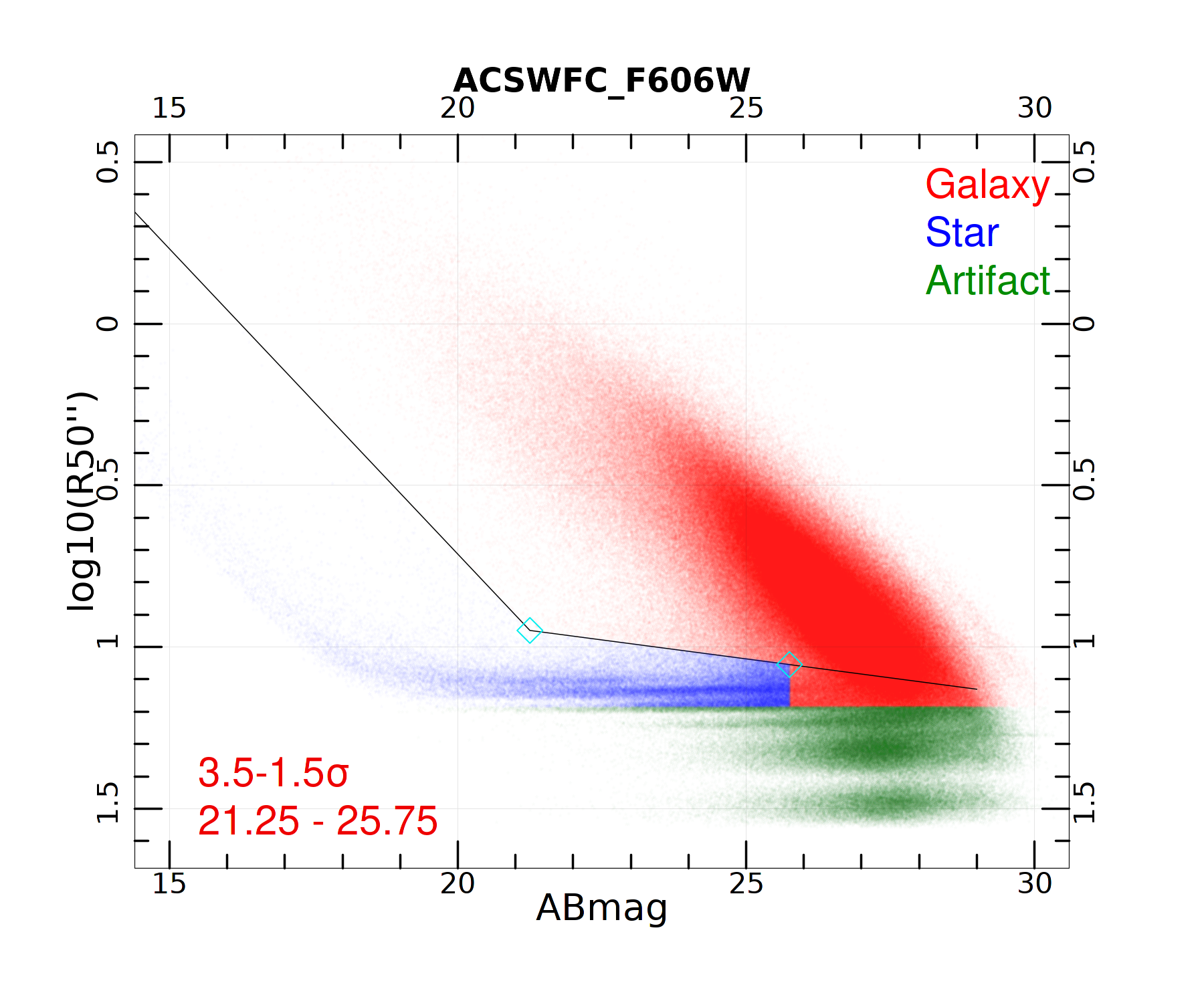}
  \caption{$\log_{10}$ vs R50 parameter space for ACSWFC F606W.}

\end{subfigure}\hfil % <-- added
  \caption{$\log_{10}$ vs R50 parameter space for the other 16 filters used in this work.}

\begin{subfigure}{0.5\textwidth}
  \includegraphics[width=\linewidth]{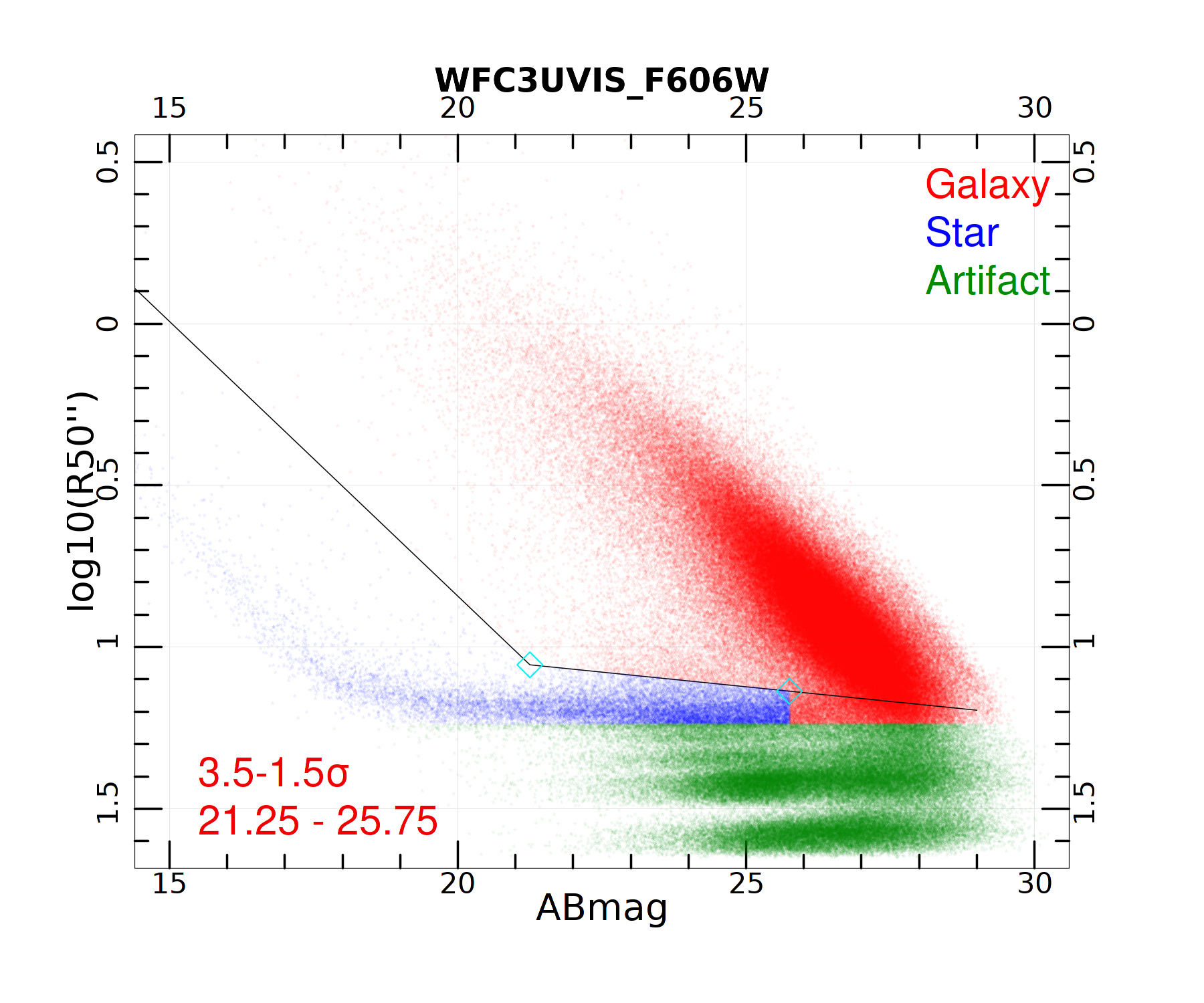}
  \caption{$\log_{10}$ vs R50 parameter space for WFC3UVIS F606W.}

\end{subfigure}\hfil % <-- added
\begin{subfigure}{0.5\textwidth}
  \includegraphics[width=\linewidth]{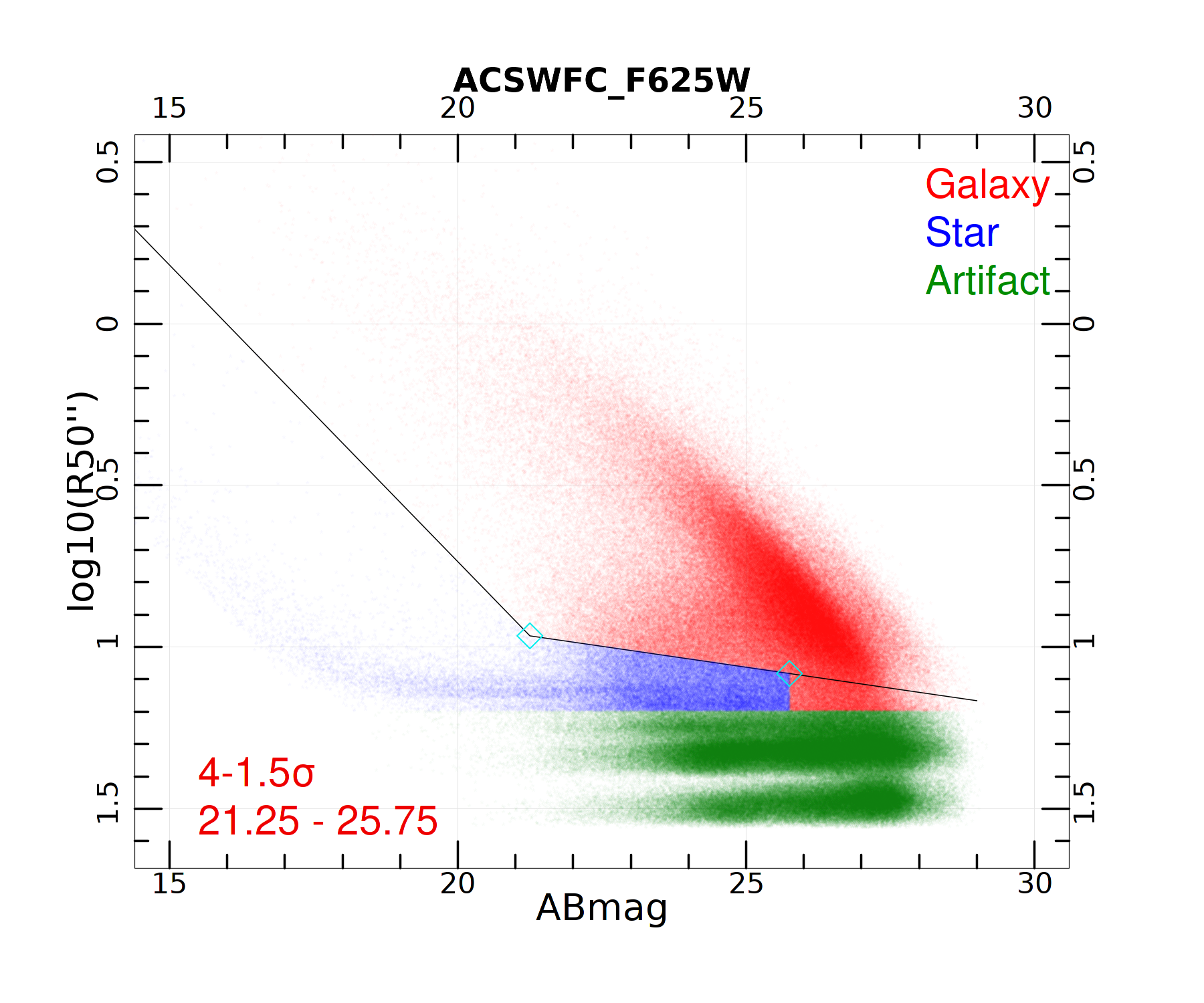}
  \caption{$\log_{10}$ vs R50 parameter space for ACSWFC F625W.}

\end{subfigure}\hfil % <-- added

\medskip

 %   \centering % <-- added
\begin{subfigure}{0.5\textwidth}
  \includegraphics[width=\linewidth]{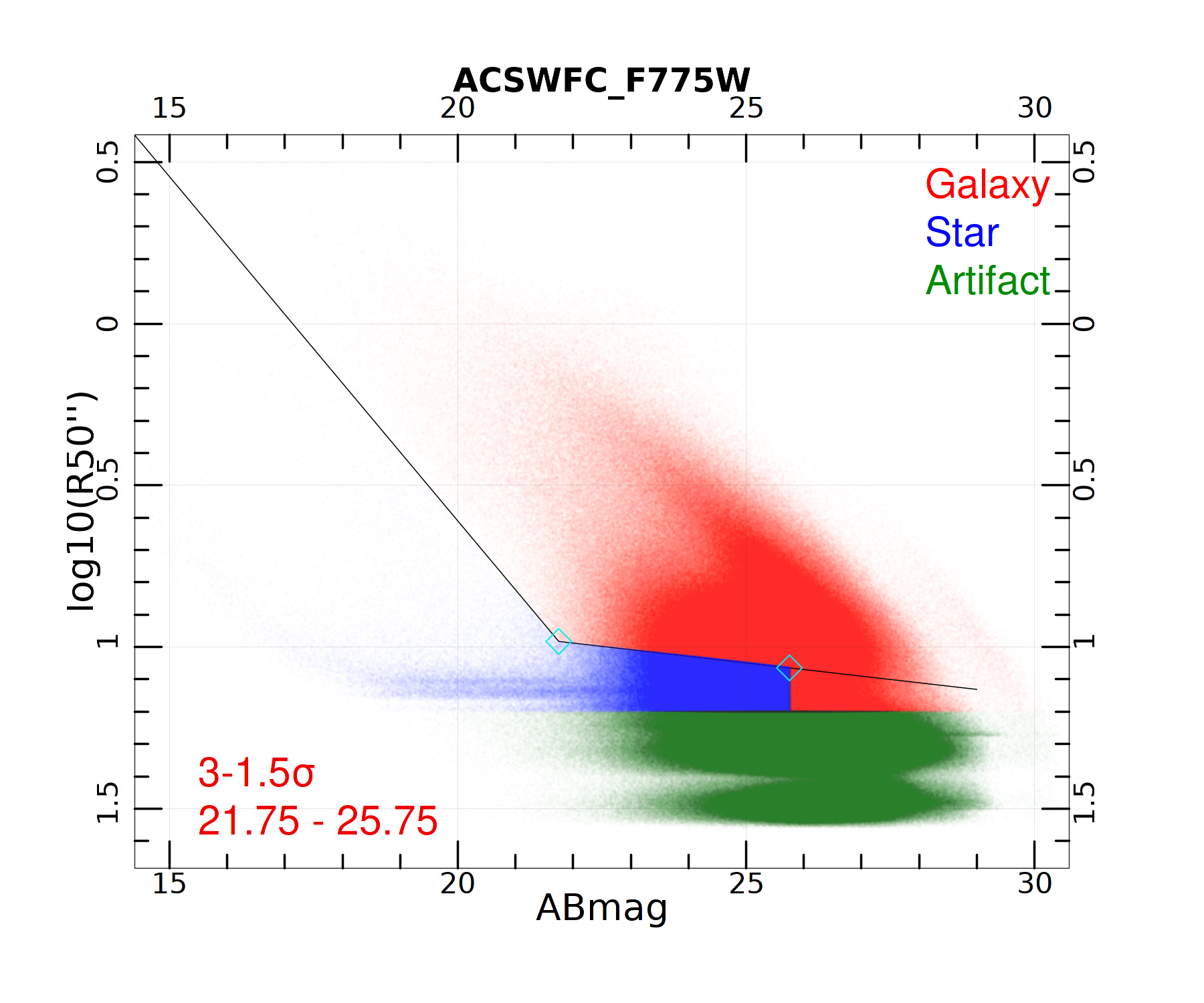}
  \caption{$\log_{10}$ vs R50 parameter space for ACSWFC F775W.}

\end{subfigure}\hfil % <-- added
\begin{subfigure}{0.5\textwidth}
  \includegraphics[width=\linewidth]{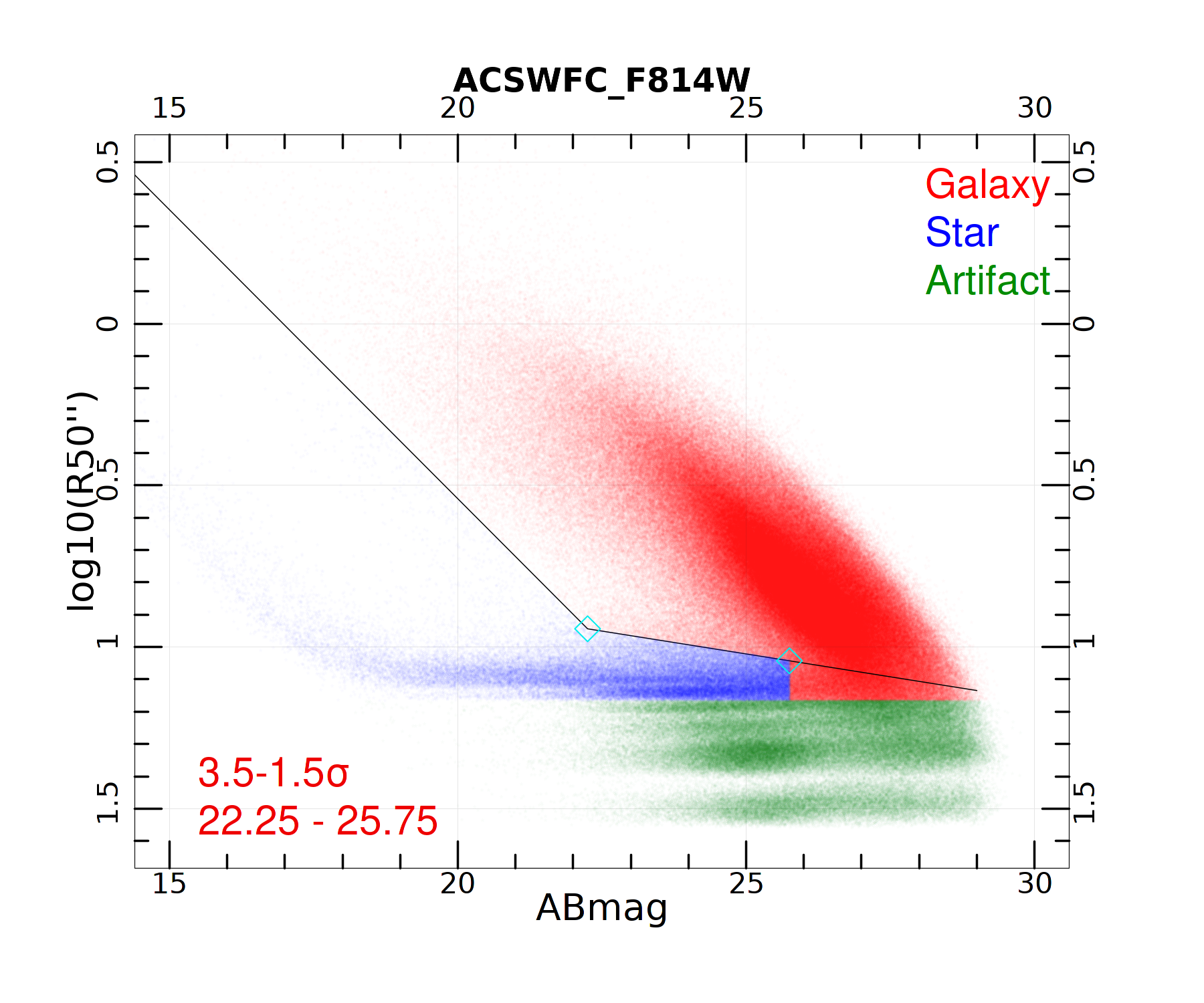}
  \caption{$\log_{10}$ vs R50 parameter space for ACSWFC F814W.}
 % \label{fig:2}
\end{subfigure}\hfil % <-- added
  \caption{$\log_{10}$ vs R50 parameter space for the other 16 filters used in this work.}
\label{fig:fig_more_ele}
\end{figure*}

\begin{figure*}

    %\centering % <-- added
\begin{subfigure}{0.5\textwidth}
  \includegraphics[width=\linewidth]{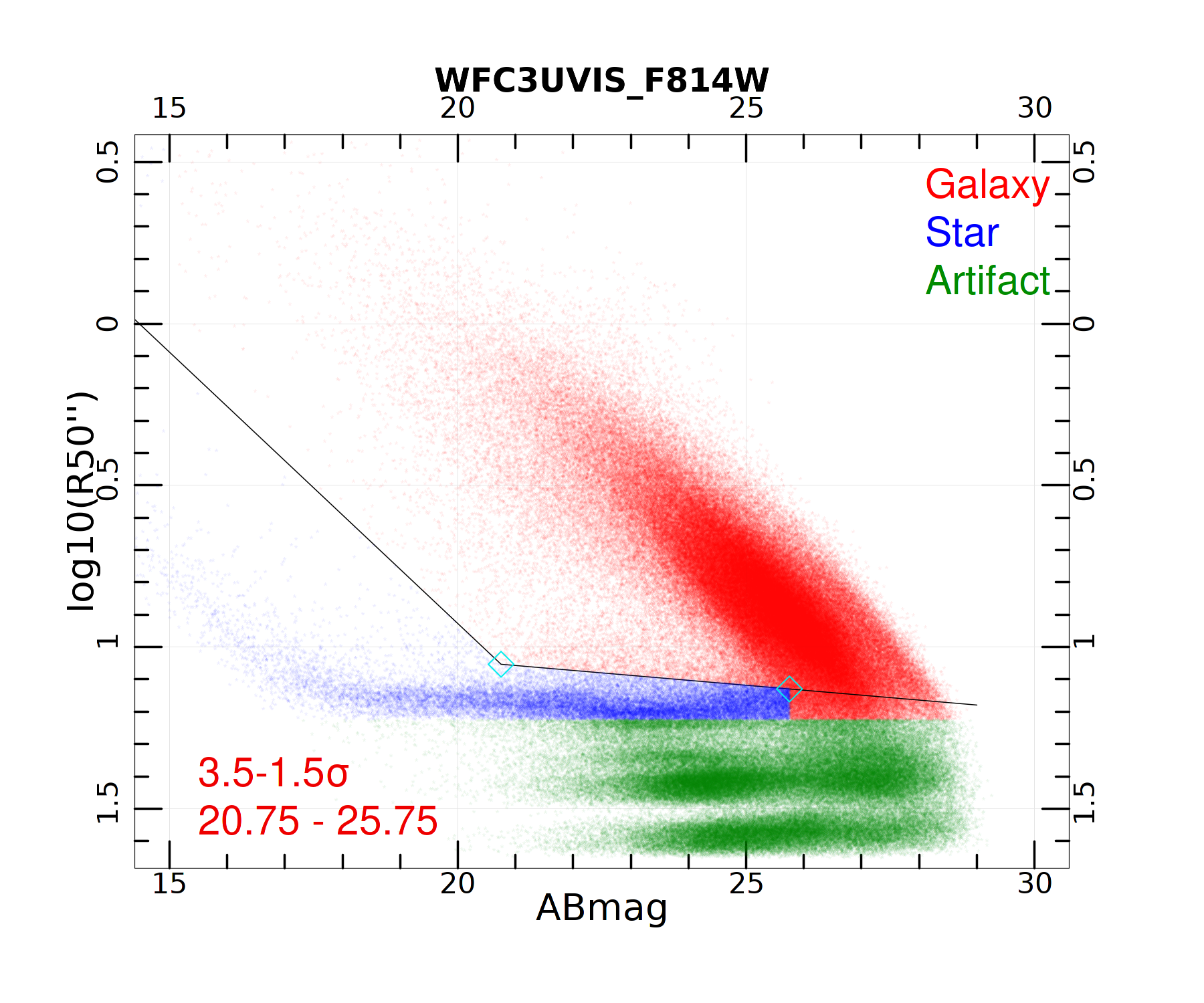}
  \caption{$\log_{10}$ vs R50 parameter space for WFC3UVIS F814W.}
  \label{fig:1}
\end{subfigure}\hfil % <-- added
\begin{subfigure}{0.5\textwidth}
  \includegraphics[width=\linewidth]{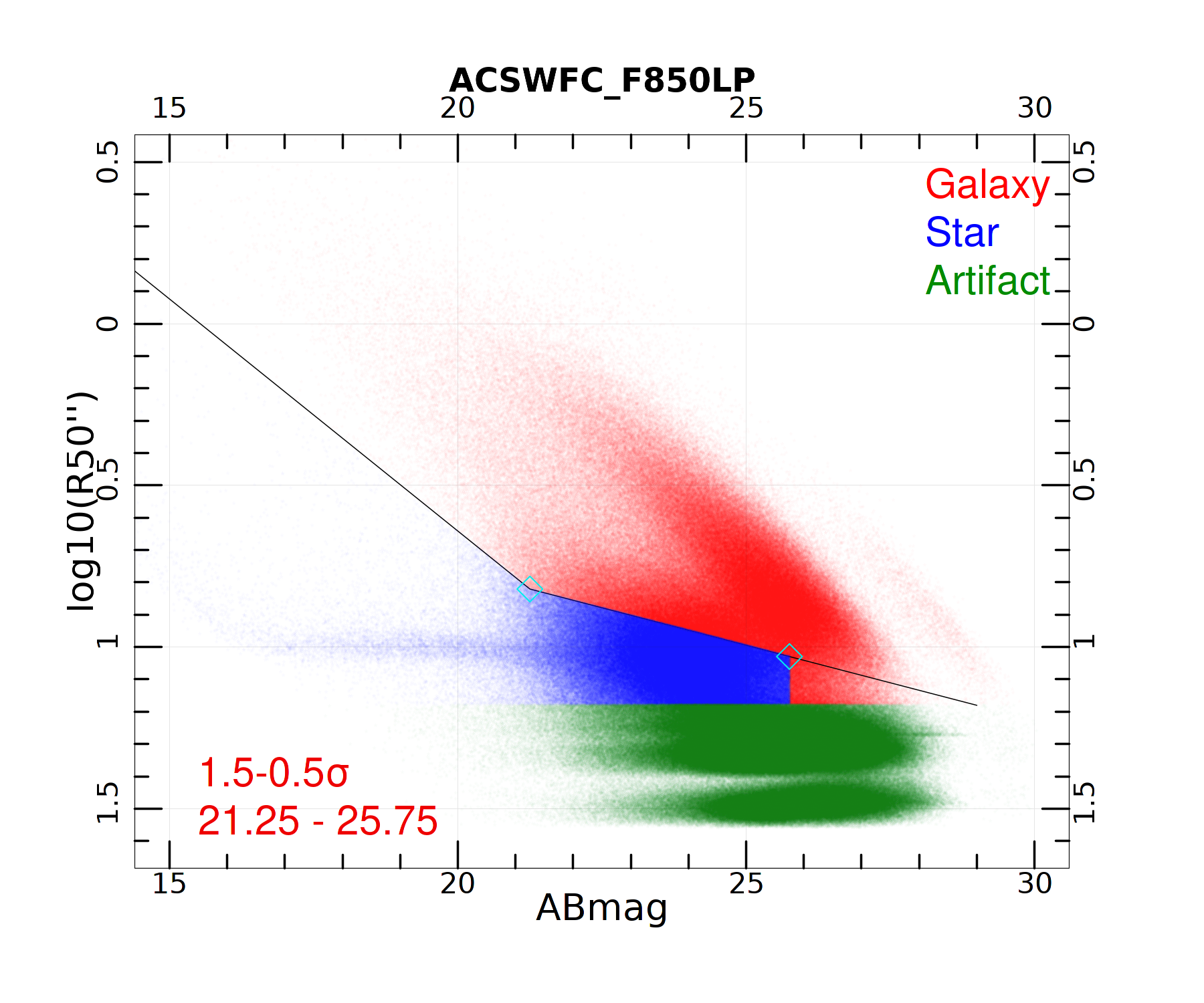}
  \caption{$\log_{10}$ vs R50 parameter space for ACSWFC F850LP.}

\end{subfigure}\hfil % <-- added

\begin{subfigure}{0.5\textwidth}
  \includegraphics[width=\linewidth]{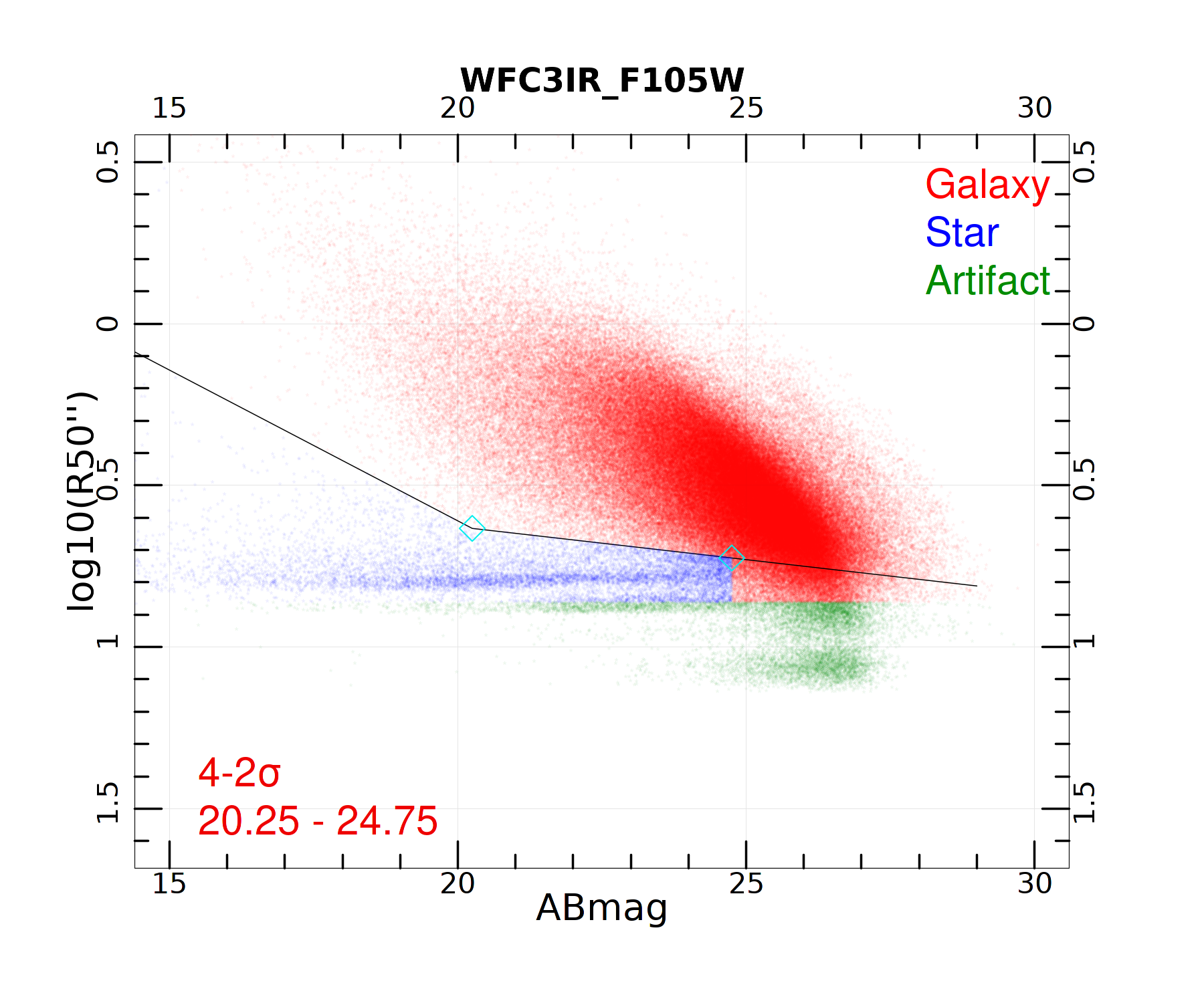}
  \caption{$\log_{10}$ vs R50 parameter space for WFC3IR F105W.}

\end{subfigure}\hfil % <-- added
\begin{subfigure}{0.5\textwidth}
  \includegraphics[width=\linewidth]{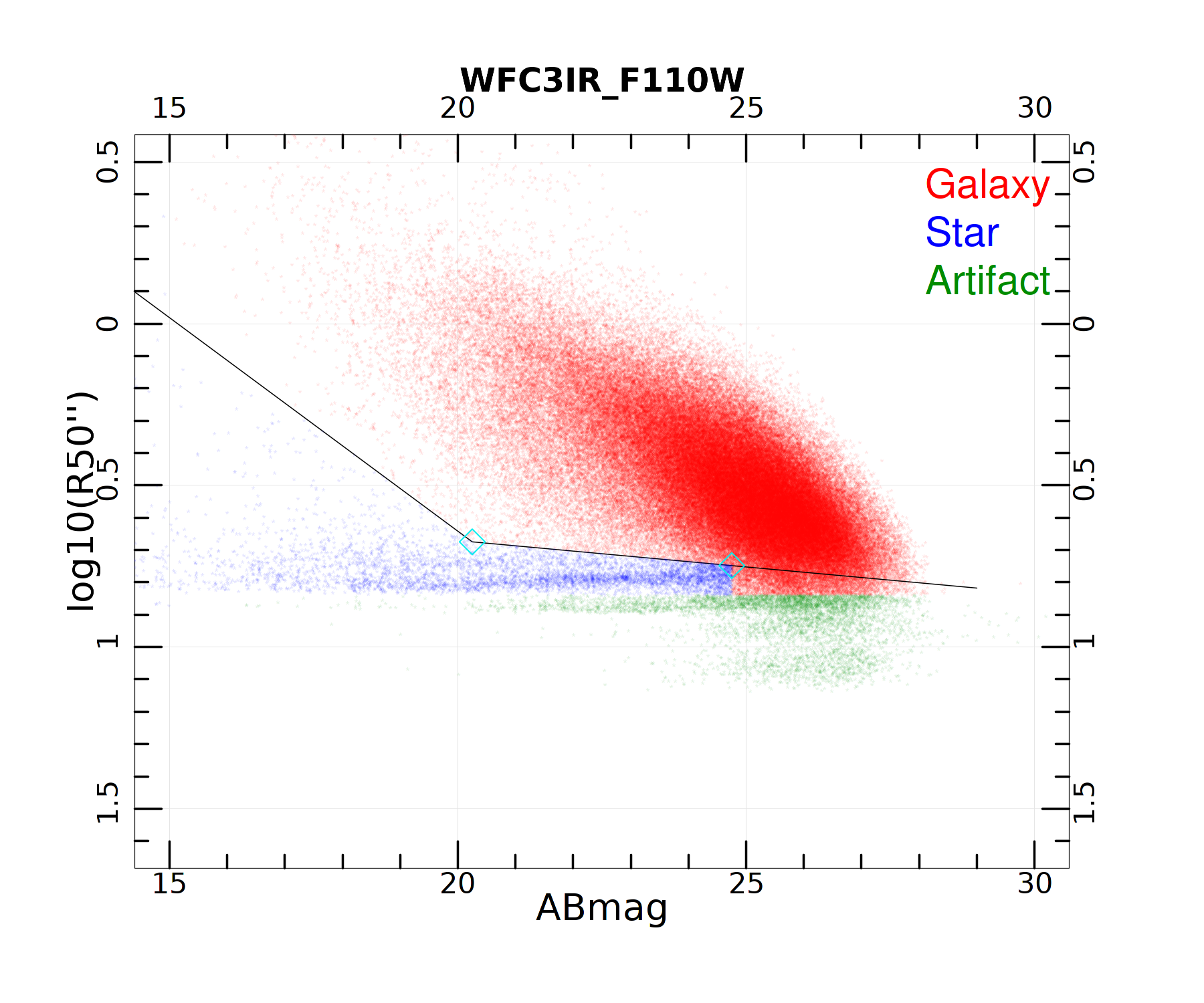}
  \caption{$\log_{10}$ vs R50 parameter space for WFC3IR F110W.}

\end{subfigure}\hfil % <-- added
  \caption{$\log_{10}$ vs R50 parameter space for the other 16 filters used in this work.}
\label{fig:fig28}
\end{figure*}
\begin{figure*}

\begin{subfigure}{0.5\textwidth}
  \includegraphics[width=\linewidth]{WFC3IR_F125W_view.png}
  \caption{$\log_{10}$ vs R50 parameter space for WFC3IR F125W.}

\end{subfigure}\hfil % <-- added
\begin{subfigure}{0.5\textwidth}
  \includegraphics[width=\linewidth]{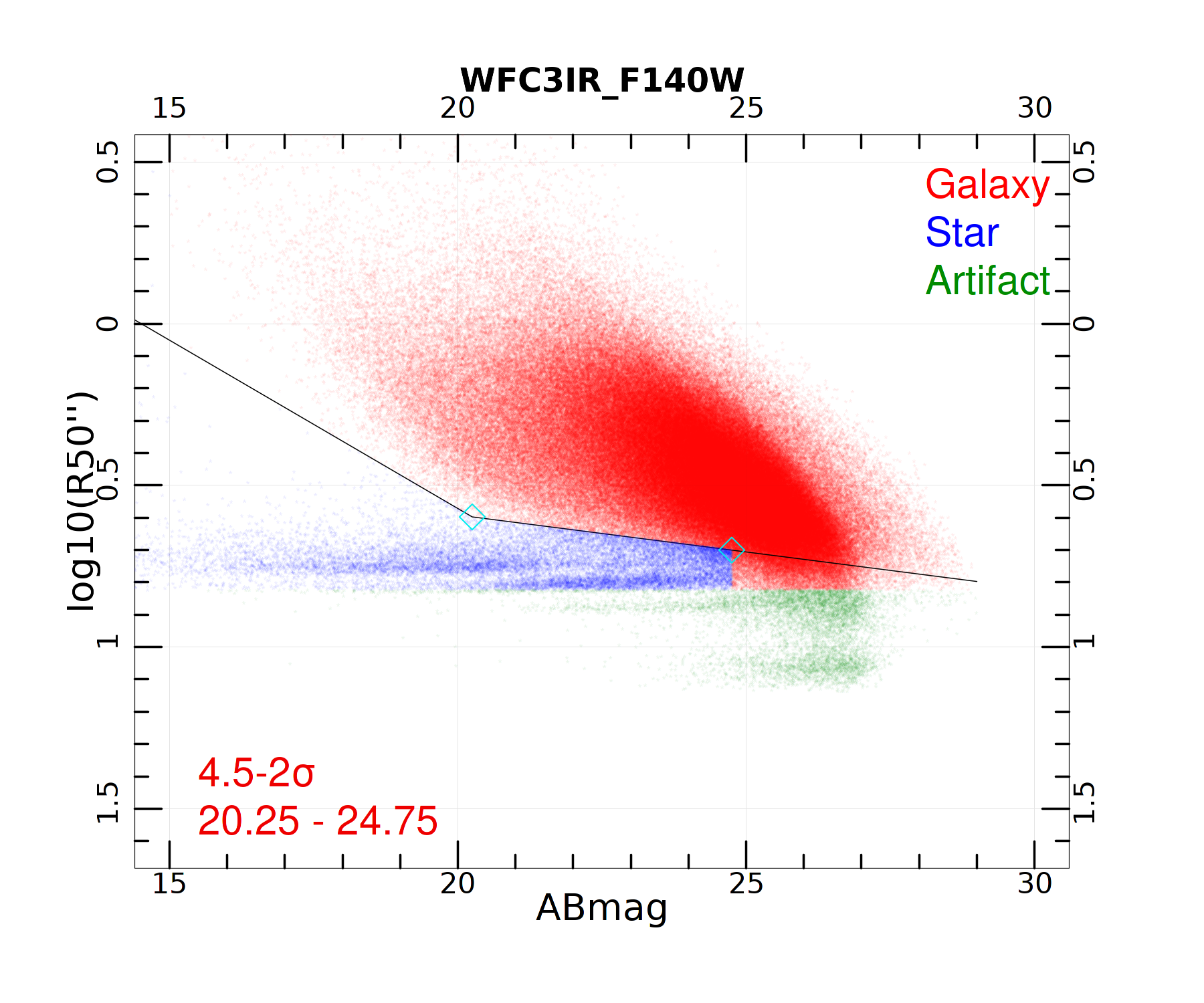}
  \caption{$\log_{10}$ vs R50 parameter space for WFC3IR F140W.}

\end{subfigure}\hfil % <-- added

\begin{subfigure}{0.5\textwidth}
  \includegraphics[width=\linewidth]{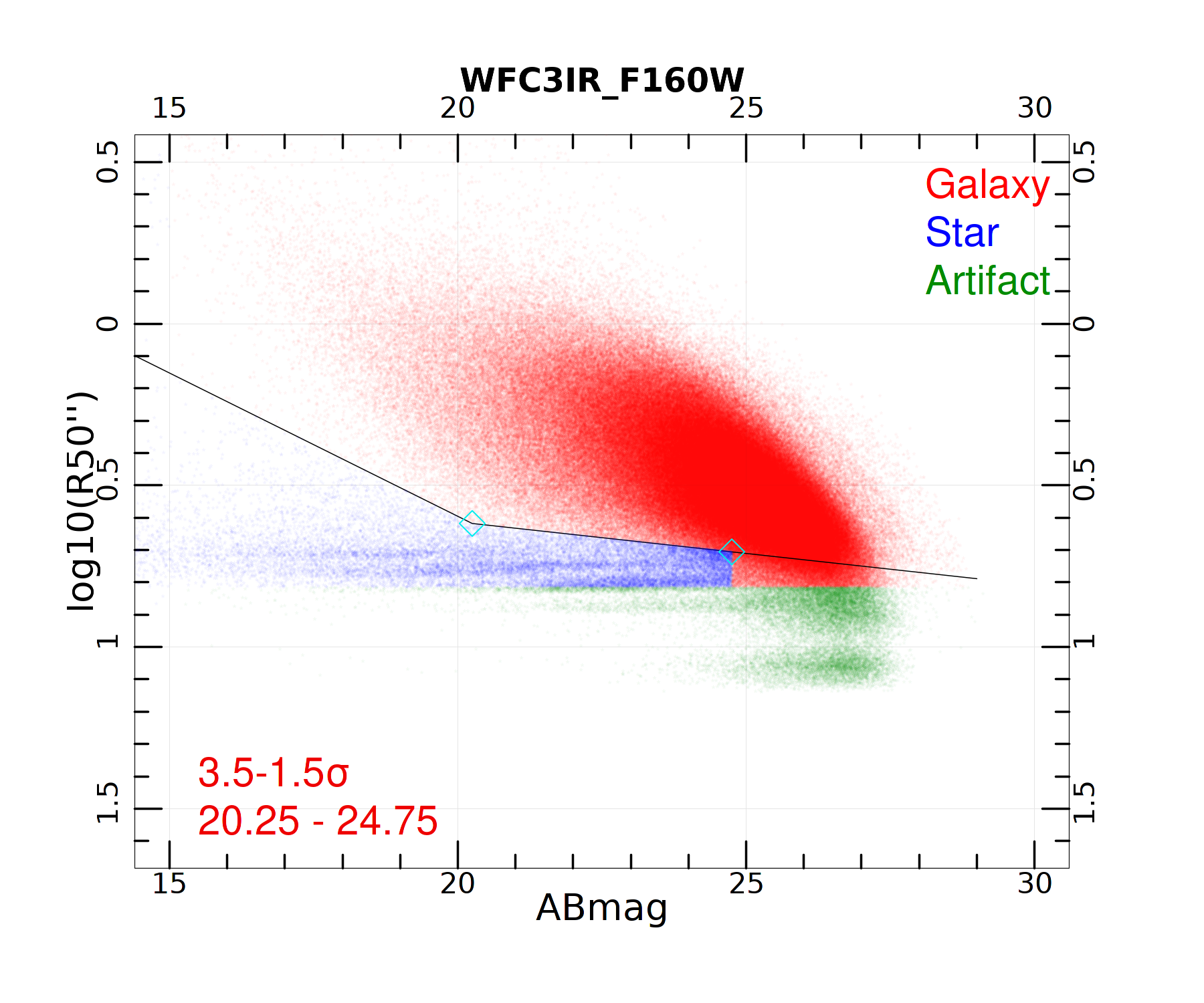}
  \caption{$\log_{10}$ vs R50 parameter space for WFC3IR F160W.}

\end{subfigure}\hfil % <-- added
% \begin{subfigure}{0.4\textwidth}
%   \includegraphics[width=\linewidth]{WFC3IR_F160W_view.png}
%   \caption{$\log_{10}$ vs R50 parameter space for WFC3IR F160W.}

% \end{subfigure}\hfil % <-- added

  \caption{$\log_{10}$ vs R50 parameter space for the other 16 filters used in this work.}
 \label{fig:fig_rest_ele}
\label{lastpage}
\end{figure*}

\clearpage

\begin{table*}
    \centering

        \caption{For convenience, here we quote the optical IGL values from \citet{Driver_2016} (Top) \& \citet{Koushan_2021} (Bottom). For a detailed breakdown of the error budget, see the respective works.}
    \begin{tabular}{cccc}
    \hline
       Band & Pivot Wavelength  & IGL Value   & Total Error \% \\ \hline
         --  & $\mathrm{\mu m}$ & $\mathrm{nW m^{-2} sr^{-1}}$ &    \%   \\ \hline
       % 0.337 & 999 & 999  & 999 & 999 & 999 \\ \hline
       % 0.396 & 999 & 999  & 999 & 999 & 999 \\ \hline
        FUV & 0.153  & 1.45 & 18.62  \\ 
        NUV & 0.225  & 3.15 & 21.27 \\ %\hline
        u &  0.356   & 4.03 & 19.35  \\ %\hline
        g & 0.470   & 5.36 & 17.35 \\ %\hline
        r & 0.618   & 7.47 & 14.99\\ %\hline
        i & 0.749   & 9.55 & 14.76  \\ %\hline
        z & 0.895   & 10.15 & 14.88 \\ %\hline
        Y & 1.021   & 10.44 & 15.33 \\ %\hline
        J & 1.252   & 10.38 & 14.64 \\ %\hline
        H & 1.643  & 10.12 & 15.31 \\ %\hline
        K & 2.150  & 8.72 & 13.99 \\ \hline
            &   &  & \\ \hline
        u &  0.3577   & 4.13 & 6.87  \\ %\hline
        g & 0.4744   & 5.76 & 4.02 \\ %\hline
        r & 0.6312   & 8.11 & 4.08 \\ %\hline
        i & 0.7584   & 9.94 & 4.44  \\ %\hline
        z & 0.8833   & 10.71 & 5.20 \\ %\hline
        Y & 1.0224   & 11.58 & 4.52 \\ %\hline
        J & 1.2546   & 11.22 & 4.92 \\ %\hline
        H & 1.6477  & 11.17 & 4.73 \\ %\hline
        Ks & 2.1549  & 9.42 & 5.21 \\ \hline
       % 2.15 & 10.03  & 999 & 0.84 & 0.84 \\ \hline
    \end{tabular}

    \label{tab:tab14}
\end{table*}

\end{document}